\documentclass[rmp,aps,showpacs,twocolumn,superscriptaddress]{revtex4}

\usepackage{mathpple}
\usepackage{epsfig}
\usepackage{bm}                 
\usepackage{amsmath}
\usepackage{amssymb}
\usepackage{dcolumn}
\usepackage{graphicx}
\usepackage{psfrag}

\def\unity{\mbox{\small 1} \!\! \mbox{1}}

\begin{document}
 

\title{Linear optical quantum computing}

\author{Pieter Kok} \email{pieter.kok@materials.ox.ac.uk}
\affiliation{Department of Materials, Oxford University, Oxford OX1 3PH, UK}
\affiliation{Hewlett-Packard Laboratories, Filton Road Stoke Gifford, 
Bristol BS34 8QZ, UK}

\author{W.J.\ Munro} 
\affiliation{Hewlett-Packard Laboratories, Filton Road Stoke Gifford, 
Bristol BS34 8QZ, UK}

\author{Kae Nemoto} 
\affiliation{National Institute of Informatics, 2-1-2 Hitotsubashi,
  Chiyoda-ku, Tokyo 101-8430, Japan} 

\author{T.C.\ Ralph} 
\affiliation{Centre for Quantum Computer Technology, University of
  Queensland, St.\ Lucia, Queensland 4072, Australia} 

\author{Jonathan P.\ Dowling} 
\affiliation{\mbox{Hearne Institute for Theoretical Physics, Department of
    Physics and Astronomy, LSU, Baton Rouge LA, 70803, USA}}  
\affiliation{Institute for Quantum Studies, Department of Physics, Texas A\&M
  University, 77843-4242, USA}   

\author{G.J.\ Milburn} 
\affiliation{Centre for Quantum Computer Technology, University of
  Queensland, St.\ Lucia, Queensland 4072, Australia}

\date{\today}

\begin{abstract}
 Linear optics with photon counting is a prominent candidate for practical
 quantum computing. The protocol by Knill, Laflamme, and Milburn [Nature {\bf
 409}, 46 (2001)] explicitly demonstrates that efficient scalable quantum
 computing with single photons, linear optical elements, and projective
 measurements is possible. Subsequently, several improvements on this protocol
 have  started to bridge the gap between theoretical scalability and practical
 implementation. We review the original theory and its improvements, and we
 give a few examples of experimental two-qubit gates. We discuss the use of
 realistic components, the errors they induce in the computation, and how
 these errors can be corrected.
\end{abstract}

\pacs{03.67.Hk, 03.65.Ta, 03.65.Ud}

\maketitle

\tableofcontents

\bigskip

\section{Quantum computing with light}

\noindent
Quantum computing has attracted much attention over the last ten to
fifteen years, partly because of its promise of super-fast factoring
and its potential for the efficient simulation of quantum
dynamics. There are many different architectures for quantum computers
based on many different physical systems. These include atom- and
ion-trap quantum computing, superconducting charge and flux qubits,
nuclear magnetic resonance, spin- and charge-based quantum dots,
nuclear spin quantum computing, and optical quantum computing (for a
recent overview see Spiller et al.\ \nocite{spiller05} 2005). All
these systems have their own advantages in quantum information
processing. However, even though there may now be a few front-runners, such
as ion-trap and superconducting quantum computing, no physical
implementation seems to have a clear edge over others at this
point. This is an indication that the technology is still in its
infancy. Quantum computing with linear quantum optics, the subject of
this review, has the advantage that the smallest unit of quantum
information (the photon) is potentially free from decoherence: The
quantum information stored in a photon tends to stay there. The
downside is  that photons do not naturally interact with each other,
and in order to apply two-qubit quantum gates such interactions are
essential. 

Therefore, if we are to construct an optical quantum computer, one way
or another we have to introduce an effective interaction between
photons. In section \ref{sec:kerr}, we review the use of so-called
large cross-Kerr nonlinearities to induce a single-photon controlled-NOT
operation. However, naturally occurring nonlinearities of this sort are
many orders of magnitude smaller than what is needed for our
purposes. An alternative way to induce an effective interaction
between photons is to make projective measurements with
photo-detectors. The difficulty with this technique is that such
optical quantum gates are probabilistic: More often than not, the gate
fails and destroys the information in the quantum computation. This can be
circumvented by using an exponential number of optical modes, but this is by
definition not scalable (see also section \ref{sec:kerr}).
In 2001, Knill, Laflamme, and Milburn (KLM 2001) \nocite{knill01}
constructed a protocol in which probabilistic two-photon gates are
teleported into a quantum circuit with high probability. Subsequent
error correction in the quantum circuit is used to bring the error
rate down to fault-tolerant levels. We describe the KLM protocol in
detail in section \ref{klm}. 

Initially, the KLM protocol was designed as a proof that linear optics and
projective measurements allow for scalable quantum computing in
principle. However, it subsequently spurred on new experiments in quantum
optics, demonstrating the operation of high-fidelity probabilistic  two-photon
gates. On the theoretical front, several improvements of the protocol were
proposed, leading to ever smaller overhead cost on the computation. A number of
these improvements are based on {\em cluster-state} quantum
computing, or the one-way quantum computer. Recently,
a circuit-based model was shown to have similar scaling properties as the
best-known cluster state model. In section \ref{sec:imp}, we describe the
several improvements to linear optical quantum information processing in
considerable detail, and in section \ref{sec:real}, we describe the issues
involved in the use of realistic components such as photon detectors, photon
sources and quantum memories. Given these realistic components, we discuss
loss tolerance and general error correction for Linear Optical Quantum
Computing (LOQC) in section \ref{sec:error}.

We will restrict our discussion to the theory of single-photon implementations
of quantum information processors, and we assume some familiarity with the
basic 
concepts of quantum computing. For an introduction to quantum computation and
quantum information, see e.g., Nielsen and Chuang (2000). For a review article
on optical quantum  information processing with continuous variables, see
Braunstein and Van Loock \nocite{vanloock05} (2005). In section
\ref{sec:outlook} we conclude with an outlook on other promising optical
quantum information processing techniques, such as photonic band-gap
structures, weak cross-Kerr nonlinearities, and hybrid matter-photon
systems. We start our review with a short introduction to linear optics,
$N$ port optical interferometers and circuits, and we define the different
versions of the optical qubit.

\subsection{Linear quantum optics}

\noindent
The basic building blocks of linear optics are beam splitters, half-
and quarter-wave plates, phase shifters, etc. In this section we will describe
these devices mathematically and establish the convention that is used
throughout the rest of the paper. 

\begin{figure}[t]
  \begin{center}
  \begin{psfrags}
       \epsfig{file=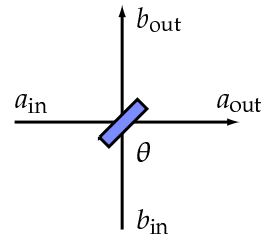}
  \end{psfrags}
  \end{center}
  \caption{The beam splitter with transmission amplitude $\cos\theta$.}
  \label{fig:bs}
\end{figure}

The quantum-mechanical plane-wave expansion of the electromagnetic
vector potential is usually expressed in terms of the annihilation
operators $\hat{a}_j(k)$ and their adjoints, the creation operators:
\begin{equation}\nonumber
 A^{\mu}(x,t) = \int \frac{d^3 k}{2\omega_k} \sum_{j=1,2}
 \epsilon^{\mu}_j(k)\, \hat{a}_j(k)\, e^{ikx-i\omega_k t} + \text{H.c.,} 
\end{equation} 
where $j$ indexes the polarisation in the Coulomb gauge and
$\epsilon^{\mu}_j$ is the corresponding polarisation vector. For the
moment we suppress the polarisation degree of freedom and consider
general properties of the creation and annihilation operators. They
bear their names because they act in a specific way on the Fock states
$|n\rangle$:  
\begin{equation}
 \hat{a} |n\rangle = \sqrt{n} |n-1\rangle \quad\text{and}\quad
 \hat{a}^{\dagger} |n\rangle = \sqrt{n+1} |n+1\rangle \; ,
\end{equation}
where we suppressed the $k$ dependence. It is straightforward to show
that $\hat{n}(k) \equiv \hat{a}^{\dagger}(k) \hat{a}(k)$ is the number
operator $\hat{n} |n\rangle = n |n\rangle$ for a given mode with
momentum $k$. The canonical commutation relations between $\hat{a}$
and $\hat{a}^{\dagger}$ are given by 
\begin{eqnarray}\label{eq:commutation}
 \left[\hat{a}(k),\hat{a}^{\dagger}(k')\right] &=& \delta(k-k') \cr
 \left[\hat{a}(k),\hat{a}(k')\right] &=& \left[\hat{a}^{\dagger}(k),
 \hat{a}^{\dagger}(k')\right] = 0\; .
\end{eqnarray}
In the rest of this review, we denote the information about the
spatial mode, $k$, by a subscript, since we will not be concerned with
the geometrical details of the interferometers we describe, only how
the spatial modes are connected. Also, to avoid notational clutter we
will use operator hats only for non-unitary and non-Hermitian
operators, as well as cases where omission of the hat would lead to
confusion. 

An important optical component is the single-mode {\em phase
shift}. It changes the phase of the electromagnetic field in a given
mode:
\begin{equation}\label{ps2}
 \hat{a}^{\dagger}_{\rm out} = e^{i\phi\hat{a}_{\rm
 in}^{\dagger}\hat{a}_{\rm in}}\, \hat{a}_{\rm in}^{\dagger}\,
 e^{-i\phi\hat{a}_{\rm in}^{\dagger} \hat{a}_{\rm in}} =
 e^{i\phi} \hat{a}_{\rm in}^{\dagger}\; , 
\end{equation}
with the interaction Hamiltonian $H_{\varphi}=\phi\,\hat{a}_{\rm
in}^{\dagger}\hat{a}_{\rm in}$ (here, and throughout this review, we use the
convention that $\hbar=1$, and the time dependence is absorbed in
$\phi$). This Hamiltonian is proportional to the number 
operator, which means that the photon number is conserved. Physically,
a phase shifter is a slab of transparent material with an index of
refraction that is different from that of free space.

Another important component is the {\em beam splitter} (see
Fig.~\ref{fig:bs}).  Physically, it consists of a semi-reflective
mirror: when light falls on this mirror, part will be reflected and
part will be transmitted. The theory of the lossless beam splitter is central
to LOQC, and was developed by Zeilinger (1981) \nocite{zeilinger81} and Fearn
and Loudon (1987) \nocite{fearn87}. Lossy beam splitters were studied by
Barnett et al.\ (1989) \nocite{barnett89}. The transmission and
reflection properties of general dielectric media were studied by Dowling
(1998). \nocite{dowling98} Let the two incoming modes on either side of
the beam splitter be denoted by $\hat{a}_{\rm in}$ and $\hat{b}_{\rm
  in}$, and the outgoing modes by $\hat{a}_{\rm out}$ and
$\hat{b}_{\rm out}$. When we parameterise the probability amplitudes of
these possibilities as $\cos\theta$ and $\sin\theta$, and the relative phase
as $\varphi$, then the beam splitter yields an evolution in operator form
\begin{eqnarray}\label{2bs}
 \hat{a}_{\rm out}^{\dagger} &=& \cos\theta\,\hat{a}_{\rm in}^{\dagger} + 
 i e^{-i\varphi} \sin\theta\,\hat{b}_{\rm in}^{\dagger} \; , \cr
 \hat{b}_{\rm out}^{\dagger} &=& i e^{i\varphi}\sin\theta\,\hat{a}_{\rm
 in}^{\dagger} +  \cos\theta\,\hat{b}_{\rm in}^{\dagger} \; , 
\end{eqnarray}
The reflection
and transmission coefficients $R$ and $T$ of the beam splitter  are
$R=\sin^2\theta$ and $T=1-R=\cos^2\theta$. The relative phase shift $i e^{\pm
  i\varphi}$ ensures that the transformation is unitary. Typically, we choose
either $\varphi=0$ or $\varphi=\pi/2$. Mathematically, the two parameters
$\theta$ and $\varphi$ represent the angles of a rotation about two
orthogonal axes in the Poincar\'e sphere. The physical beam splitter
can be described by any choice of $\theta$ and $\varphi$, provided the
correct phase shifts are applied to the outgoing modes. 

In general the Hamiltonian $H_{\rm BS}$ of the beam splitter
evolution in Eq.~(\ref{2bs}) is given by
\begin{equation}\label{su2plus}
 H_{\rm BS} = \theta e^{i\varphi} \hat{a}_{\rm  in}^{\dagger}\hat{b}_{\rm in}
 + \theta e^{-i\varphi} \hat{a}_{\rm in}\hat{b}_{\rm in}^{\dagger}\; .
\end{equation}
Since the operator $H_{\rm BS}$ commutes with the total number operator,
$[H_{\rm BS},\hat{n}]=0$, the photon number is conserved in the lossless beam 
splitter, as one would expect. 

The same mathematical description applies to the evolution due to a 
{\em polarisation rotation}, physically implemented by quarter- and
half-wave plates. Instead of having two different spatial modes 
$a_{\rm in}$ and $b_{\rm in}$, the two incoming modes have different 
polarisations. We write $\hat{a}_{\rm in}\rightarrow\hat{a}_x$ and 
$\hat{b}_{\rm in}\rightarrow\hat{a}_y$ for some orthogonal set of 
coordinates $x$ and $y$ (i.e., $\langle x|y\rangle=0$). The parameters
$\theta$ and $\varphi$ are now angles of rotation:
\begin{eqnarray}\label{2polrot}
 \hat{a}_{x'}^{\dagger} &=& \cos\theta\,\hat{a}_x^{\dagger} + 
 i e^{-i\varphi} \sin\theta\,\hat{a}_y^{\dagger} \; , \cr
 \hat{a}_{y'}^{\dagger} &=& i e^{i\varphi} \sin\theta\,\hat{a}_x^{\dagger} + 
 \cos\theta\,\hat{a}_y^{\dagger} \; .
\end{eqnarray}
This evolution has the same Hamiltonian as the beam splitter, and it
formalises the equivalence between the so-called polarisation and
dual-rail logic. These transformations are sufficient to implement any photonic
single-qubit operation \cite{simon90}. 

\begin{figure}[t]
  \begin{center}
  \begin{psfrags}
       \epsfig{file=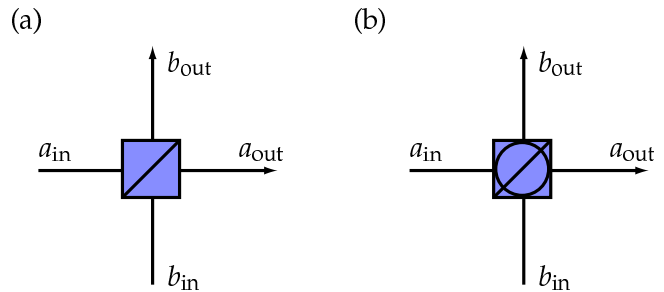}
  \end{psfrags}
  \end{center}
  \caption{The polarising beam splitter cut to different polarisation
   bases. (a) The horizontal-vertical basis. (b) The diagonal basis.}
  \label{fig:pbs}
\end{figure}

The last linear optical element that we highlight here is the
polarising beam splitter (PBS). In circuit diagrams, it is usually
drawn as a box around a regular beam splitter (see
Fig.~\ref{fig:pbs}a). If the PBS is cut to  separate horizontal and
vertical polarisation, the transformation of the incoming modes
($a_{\rm in}$ and $b_{\rm in}$) yields the following outgoing modes
(($a_{\rm out}$ and $b_{\rm out}$):
\begin{eqnarray}\label{eq:pbs}
  \hat{a}_{{\rm in},H} \rightarrow \hat{a}_{{\rm out},H} &
  ~\text{and}~ & \hat{a}_{{\rm in},V} \rightarrow \hat{b}_{{\rm out},V}
  \cr 
  \hat{b}_{{\rm in},H} \rightarrow \hat{b}_{{\rm out},H} &
  ~\text{and}~ & \hat{b}_{{\rm in},V} \rightarrow \hat{a}_{{\rm out},V} 
\end{eqnarray}
We can also cut the PBS to different polarisation directions (e.g., $L$ and
$R$), in which case we make the substitution $H \leftrightarrow L$, $V
\leftrightarrow R$. Diagrammatically a PBS cut with a different
polarisation usually has a circle drawn inside the box (Fig.~\ref{fig:pbs}b). 

At this point, we should devote a few words to the term ``linear
optics''. Typically this denotes the set of optical elements whose
interaction Hamiltonian is bilinear in the creation and annihilation
operators: 
\begin{equation}\label{eq:bilinear}
 H = \sum_{jk} A_{jk} \hat{a}_j^{\dagger} \hat{a}_k\; .
\end{equation}
An operator of this form commutes with the total number operator, and
has the property that a simple mode transformation of creation
operators into a linear combination of other creation operators
affects only the matrix $A$, but does not introduce terms that are
quadratic (or higher) in the creation or annihilation
operators. However, from a field-theoretic point of view, the most
general linear Bogoliubov transformation of creation and annihilation
operators is given by  
\begin{equation}
 \hat{a}_j \rightarrow \sum_k u_{jk} \hat{a}_k + v_{jk}
 \hat{a}_k^{\dagger}\; .
\end{equation}
Clearly, when such a transformation is substituted in
Eq.~(\ref{eq:bilinear}) this will give rise to terms such as
$\hat{a}_j \hat{a}_k$ and $\hat{a}_j^{\dagger} \hat{a}_k^{\dagger}$,
i.e., squeezing. The number of photons is not conserved in such a
process. For the purpose of this review, we exclude squeezing as a
resource other than as a method for generating single photons.

With the linear optical elements introduced in this section we can
build large optical networks. In particular, we can make computational
circuits by using known states as the input and measuring the output
states. Next we will study these optical circuits in more detail.

\subsection{$N$ port interferometers and optical circuits}

\noindent
An optical circuit can be thought of as a black box with incoming
and outgoing modes of the electromagnetic field. The black box
transforms a state of the incoming modes into a different state of
the outgoing modes. The modes might be mixed by beam splitters, or
they may pick up a relative phase shift or polarisation
rotation. These operations all belong to a class of optical components
that preserve the photon number, as described in the previous
section. In addition, the box may include measurement devices, the
outcomes of which may modify optical components on the remaining
modes. This is called {\em feed-forward}
detection, and it is an important technique that can increase the
efficiency of a device \cite{lapaire03,clausen03}. 

Optical circuits can also be thought of as a general unitary
transformation on $N$ modes, followed by the detection of a subset of
these modes (followed by unitary transformation on the remaining modes,
detection, and so on). The interferometric part of this circuit is
also called an $N$ port interferometer. $N$ ports yield a unitary
transformation $U$ of the spatial field modes $a_k$, with
$j,k\in\{1,\ldots,N\}$: 
\begin{equation}\label{trans1}
 \hat{b}_k \rightarrow \sum_{j=1}^N U_{jk} \hat{a}_j
 \qquad\mbox{and}\qquad
 \hat{b}^{\dagger}_k \rightarrow \sum_{j=1}^N U_{jk}^*\hat{a}^{\dagger}_j\; ,
\end{equation} 
where the incoming modes of the $N$ port are denoted by $a_j$ and the
outgoing  modes by $b_j$. The explicit form of $U$ is given by the
repeated application of transformations such as given in Eqs.~(\ref{ps2}),
(\ref{2bs}), and (\ref{2polrot}).

\begin{figure}[t]
  \begin{center}
  \begin{psfrags}
       \epsfig{file=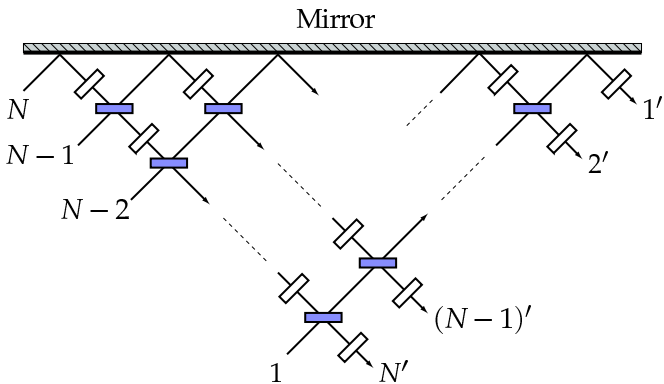}
  \end{psfrags}
  \end{center}
 \caption{Decomposing an $N$ port unitary $U(N)$ into $SU(2)$ group
  elements, i.e., beam splitters and phase shifters. Moreover, this is
  an efficient process: the maximum number of beam splitters needed is
  $N(N-1)/2$.} 
 \label{fig:reck}
\end{figure}

The two-mode operators $\hat{L}_+ = \hat{a}^{\dagger}\hat{b}$,
$\hat{L}_- =  \hat{a}\hat{b}^{\dagger}$, and $\hat{L}_0 =
(\hat{a}^{\dagger}\hat{a} - \hat{b}^{\dagger}\hat{b})/2$ form an
$su(2)$ Lie algebra: 
\begin{equation}
 [\hat{L}_0,\hat{L}_{\pm}] = \pm \hat{L}_{\pm} \quad\mbox{and}\quad
 [\hat{L}_+,\hat{L}_-] = 2 \hat{L}_0\; .
\end{equation}
This means that any two-mode
interferometer exhibits $U(2)$ symmetry\footnote{Two remarks: Algebras
  are typically denoted in lower-case, while the group itself is
  denoted in upper-case. Secondly, single-mode phase shifts break the
  ``special'' symmetry ($\det U = 1$), which is why an interferometer
  is described by $U(N)$, rather than $SU(N)$.}. In general, an $N$
port interferometer can be described by a transformation from the group
$U(N)$. Reck et al.\ (1994) demonstrated that the converse is also
true, i.e., that any unitary transformation of $N$ optical modes can
be implemented efficiently with an $N$ port interferometer
\nocite{reck94}. They showed how a general $U(N)$ element can be broken
down into $SU(2)$ elements, for which we have a complete physical
representation in terms of beam splitters and phase shifters (see
Fig.~\ref{fig:reck}). The primitive element is a matrix $T_{pq}$
defined on the modes $p$ and $q$, which corresponds to a beam splitter
and phase shifts. Implicit in this notation is the identity operator
on the rest of the optical modes, such that $T_{pq} \equiv T_{pq}
\otimes \unity_{\rm rest}$. We then have
\begin{equation}
 U(N) \cdot T_{N,N-1}\cdots T_{N,1} = U(N-1)\oplus e^{i\phi}\; ,
\end{equation}
where $\phi$ is a single-mode phase. Concatenating
this procedure leads to a full decomposition of $U(N)$ into $T$
elements, which in turn are part of $SU(2)$. The maximum number of
beam splitter elements $T$ that are needed is $N(N-1)/2$. This
procedure is thus manifestly scalable. 

Subsequently, it was shown by T{\"o}rm{\"a} et al.\ (1995, 1996) and
Jex et al.\ (1995) how one can construct multi-mode
Hamiltonians that generate these unitary mode-transformations,
\nocite{torma95,jex95,torma96b} and a three-path Mach-Zehnder 
interferometer was demonstrated experimentally by Weihs et al.\
(1996). \nocite{weihs96b} A good introduction to linear optical
networks is given by Leonhardt (2003) \nocite{leonhardt03} and a
determination of effective Hamiltonians is given by Leonhardt and  
Neumaier (2004). \nocite{leonhardt04} For a treatment of optical
networks in terms of their {\em permanents}, see Scheel
(2004). \nocite{scheel04c}  Optical circuits in a (general)
relativistic setting are described by Kok and Braunstein
(2006). \nocite{kok06}

\subsection{Qubits in linear optics}\label{qulo}

\noindent
Formally, a qubit is a quantum system with an $SU(2)$ symmetry. We saw
above that two optical modes form a natural implementation of this
symmetry. In general, two modes with 
fixed total photon number $n$ furnish natural irreducible representations of
this group with the dimension of the representation given by $n+1$
\cite{biedenharn81}. It is at this point not specified whether we should use
spatial or polarisation modes. In linear optical quantum computing, the qubit 
of choice is usually taken to be a single photon that has the choice of two
different modes $|0\rangle_L = |1\rangle \otimes |0\rangle \equiv |1,0\rangle$
and $|1\rangle_L = |0\rangle \otimes |1\rangle \equiv |0,1\rangle$. This is
called a dual-rail qubit. When the two modes represent the internal
polarisation degree of freedom of the photon ($|0\rangle_L = |H\rangle$ and
$|1\rangle_L = |V\rangle$), we speak of a {\em polarisation qubit}. In this
review we will reserve the term ``dual rail'' for a qubit with two spatial
modes. As we showed earlier, these two representations are mathematically
equivalent, and we can physically switch between them using polarisation beam
splitters. In addition, some practical applications (typically involving a
dephasing channel such as a fibre) may call for so-called {\em time-bin}
qubits, in which the two computational qubit values are ``early'' and ``late''
arrival times in a detector. However, this degree of freedom does not exhibit
a natural internal $SU(2)$ symmetry: Arbitrary single-qubit operations are
very difficult to implement. In this review we will be concerned mainly with
polarisation and dual-rail qubits.

In order to build a quantum computer, we need both single-qubit
operations as well as two-qubit operations. Single-qubit operations
are generated by the Pauli operators $\sigma_x$, $\sigma_y$, and
$\sigma_z$, in the sense that the operator $\exp(i\theta\sigma_j)$ is
a rotation about the $j$-axis in the Bloch sphere with angle $\theta$.
As we have seen, these operations can be implemented with phase shifters, beam
splitters, and polarisation rotations on polarisation and dual-rail
qubits. In this review, we will use the convention that $\sigma_x$,
$\sigma_y$, and $\sigma_z$ denote physical processes, while we use $X$,
$Y$, and $Z$ for the corresponding logical operations on the
qubit. These two representations become inequivalent when we deal with
logical qubits that are encoded in multiple physical qubits.

Whereas single-qubit operations are straightforward in the polarisation
and dual-rail representation, the two-qubit gates are more
problematic. Consider, for example, the transformation from a state in
the computational basis to a maximally entangled Bell state: 
\begin{equation}
 |H,H\rangle_{ab} ~\rightarrow~ \frac{1}{\sqrt{2}} \left(
   |H,V\rangle_{cd} + |V,H\rangle_{cd} \right) \; . 
\end{equation}
This is the type of transformation that requires a two-qubit gate. In
terms of the creation operators (and ignoring normalisation), the
linear optical circuit that is supposed to create Bell states out of
computational basis states is described by a Bogoliubov transformation of
both creation operators
\begin{eqnarray}\label{limitations}
 \hat{a}_H^{\dagger} \hat{b}_H^{\dagger} &\rightarrow& \left(
 \sum_{k=H,V} \alpha_k \hat{c}_k^{\dagger} + \beta_k
 \hat{d}_k^{\dagger} \right) \left( \sum_{k=H,V} \gamma_k
 \hat{c}_k^{\dagger} + \delta_k \hat{d}_k^{\dagger} \right) \cr
 &\neq& \hat{c}_H^{\dagger} \hat{d}_V^{\dagger} + \hat{c}_V^{\dagger}
 \hat{d}_H^{\dagger} \; .
\end{eqnarray}
It is immediately clear that the right-hand sides in both lines cannot
be made the same for any choice of $\alpha_k$, $\beta_k$, $\gamma_k$,
and $\delta_k$: The top line is a separable expression in the creation
operators, while the bottom line is an entangled expression in the
creation operators. Therefore, linear optics alone cannot create
maximal polarisation entanglement from single polarised photons in a
deterministic manner \cite{kok00b}. Entanglement that is generated
by changing the definition of our subsystems in terms of the global field
modes is inequivalent to the entanglement that is generated by applying true
two-qubit gates to single-photon polarisation or dual-rail qubits.

Note also that if we choose our representation of the qubit differently, we
{\em can} implement a two-qubit transformation. Consider the {\em single-rail}
qubit encoding $|0\rangle_L = |0\rangle$ and $|1\rangle_L = |1\rangle$. That
is, the qubit is given by the vacuum and the single-photon state. We can then
implement the following (unnormalised) transformation deterministically: 
\begin{equation}
 |1,0\rangle ~\rightarrow~ |1,0\rangle + |0,1\rangle\; .
\end{equation}
This is a 50:50 beam splitter transformation. However, in this
representation the single-qubit operations cannot be implemented
deterministically with linear optical elements, since these
transformations do not preserve the photon number \cite{paris00}. This 
implies that we cannot implement single-qubit and two-qubit gates
deterministically for the same physical representation. For linear
optical quantum computing, we typically need the ability to (dis-)
entangle field modes. We therefore have to add a non-linear component
to our scheme. Two possible approaches are the use of Kerr
nonlinearities, which we briefly review in the next section, and
the use of {\em projective measurements}. In the rest of this review, we
concentrate mainly on linear optical quantum computing with projective 
measurements, based on the work by Knill, Laflamme, and Milburn.

Finally, in order to make a quantum computer with light that can
outperform any classical computer, we need to understand more about
the criteria that make quantum computers ``quantum''. For example,
some simple schemes in quantum communication require only
superpositions of quantum states to distinguish them from their
corresponding classical ones. However, we know that this is not
sufficient for general computational tasks. 

First, we give two definitions. The {\em Pauli group} $P$ is the set of Pauli
operators with coefficients $\{ \pm 1, \pm i\}$. For instance, the Pauli
group for one qubit is $\{ \unity, \pm X, \pm Y, \pm Z, \pm i \unity \pm i X,
\pm i Y, \pm i Z, \}$ where $\unity$ is the identity matrix. The Pauli group
for $n$ qubits consists of elements that are products of $n$ Pauli operators,
including the identity. In addition, we define the {\em Clifford group} $C$ of
transformations that leave the Pauli group invariant. In other words, for any
element of the Clifford group $c$ and any element of the Pauli group $p$, we
have 
\begin{equation}
 c p c^{\dagger} = p' \quad\text{with}\quad p'\in P\; .
\end{equation}
Prominent members of the Clifford group are the Hadamard transformation,
phase transformations, and the controlled-not (CNOT)\footnote{See
  Eq.~(\ref{eq:cnotdef}) for a definition of the CNOT operation.} Note that
the Pauli group is a subgroup of the Clifford group. 

The Gottesman-Knill theorem (1999) \nocite{gottesman99b} states that
any quantum algorithm that initiates in the computational basis and
employs only transformations (gates) from the the Clifford
group, along with projective measurements in the computational
basis, can be efficiently simulated on a classical computer. This
means there is no computational advantage in restricting the quantum
computer to such circuits. A classical machine could simulate them
efficiently.  

In discrete-variable quantum information
processing, the Gottesman-Knill theorem provides a valuable tool for
assessing the classical  complexity of a given process.  (For a
precise formulation and proof of this theorem, see
Nielsen and Chuang, \nocite{nielsen00} page 464). Although the set of
gates in the Pauli and Clifford groups does not satisfy the universality
requirements, the addition of a single-qubit $\pi/8$ gate will render
the set universal. In single-photon quantum information processing we
have easy access to such single-qubit operations.

\subsection{Early optical quantum computers and nonlinearities}\label{sec:kerr}

\begin{figure}[t]
  \begin{center}
  \begin{psfrags}
       \epsfig{file=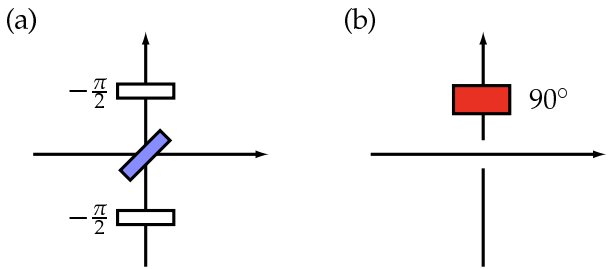}
  \end{psfrags}
  \end{center}
 \caption{Linear optical quantum computing simulation according to
  Cerf, Adami, and Kwiat. (a) Hadamard gate. (b) CNOT gate. The four
  two-qubit degrees of freedom are carried by which-path and
  polarization information. The broken line indicates that there is no
  interaction between the crossing modes.} 
 \label{fig:cerf}
\end{figure}

\noindent
Before the work of Knill, Laflamme, and Milburn (KLM), quantum information
processing with linear optics was studied (among other things) in
non-scalable architectures by Cerf, Adami, and Kwiat
(1998, 1999). \nocite{cerf98,adami99}  Their linear optical protocol
can be considered a {\em simulation} of a quantum computer: $n$ qubits are
represented by a single photon in $2^n$ different paths. In such an
encoding, both single- and two-qubit gates are easily implemented
using (polarisation) beam splitters and phase shifters. For example,
let a single qubit be given by a single photon in two optical modes:
$|0\rangle_L = |1,0\rangle$ and $|1\rangle_L = |0,1\rangle$. The
Hadamard gate acting on this qubit can then be implemented with a
50:50 beam splitter given by Eq.~(\ref{2bs}) with $\varphi=0$, and two
$-\pi/2$ phase shifters (see Fig.~\ref{fig:cerf}a):
\begin{eqnarray}\nonumber
 |1,0\rangle & \rightarrow &  |1,0\rangle + i |0,1\rangle
 \rightarrow |1,0\rangle + |0,1\rangle \cr
 |0,1\rangle & \rightarrow & -i \left( i |1,0\rangle +
  |0,1\rangle \right) \rightarrow |1,0\rangle -
  |0,1\rangle_{\rm out}\; ,
\end{eqnarray}
where we suppressed the normalisation. 

The CNOT gate in the Cerf, Adami, and Kwiat protocol is even  simpler:
suppose that the two optical modes in Fig.~\ref{fig:cerf}b carry
polarisation. The spatial degree of freedom carries the control qubit,
and the polarisation carries the target. If the photon is in the
vertical spatial mode, it will undergo a polarisation rotation; thus
implementing a CNOT. The control and target qubits can be interchanged
trivially using a polarisation beam splitter.

Since this protocol requires an exponential number of optical modes,
this is a simulation rather than a fully scalable quantum
computer. Other proposals in the same spirit include work by Clauser
and Dowling (1996), Summhammer (1997), Spreeuw (1998), and Ekert
(1998). \nocite{clauser96,summhammer97,spreeuw98,ekert98} Using this
simulation, a classical version of Grover's search algorithm can be
implemented \cite{kwiat00}. General quantum logic using polarised
photons  was studied by T\"orm\"a and Stenholm (1996), Stenholm
(1996), and Franson and Pittman
(1999). \nocite{torma96,stenholm96,franson99} 

Prior to KLM, it was widely believed that scalable all-optical quantum
computing needed a nonlinear component, such as a {\em Kerr
medium}. These media are typically characterised by a refractive index
$n_{\rm Kerr}$ that has a nonlinear component:  
\begin{equation} n_{\rm Kerr} = n_0 + \chi^{(3)} E^2\; .
\end{equation} Here, $n_0$ is the ordinary refractive index, and $E^2$
is the optical intensity of a probe beam with proportionality constant
$\chi^{(3)}$. A beam traversing through a Kerr medium will then
experience a phase shift that is proportional to its intensity. 

\begin{figure}[t]
  \begin{center}
  \begin{psfrags} 
\psfrag{i}{$a$} 
\psfrag{n}{$b'$} 
\psfrag{b}{$a_s$} 
\psfrag{k}{PBS} 
 \epsfig{file=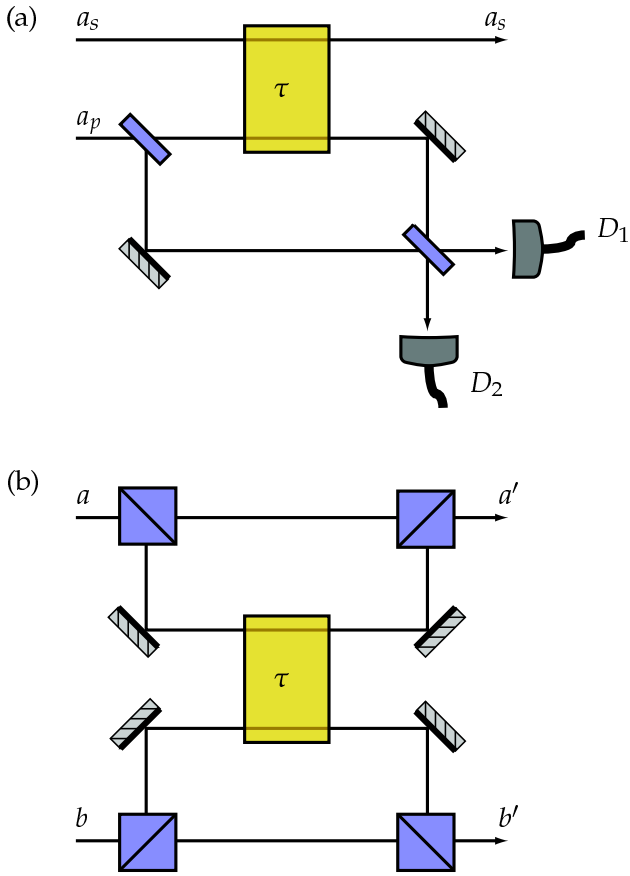}
  \end{psfrags}
  \end{center}
 \caption{Using cross-Kerr nonlinearities ($\tau$) in optical
information processing. (a) Single-photon quantum non-demolition
measurement. The Mach-Zehnder interferometer is balanced, such that
the presence of a photon in the signal mode directs the probe field to
the dark output port. (b) Single-photon CZ gate. When both photons in
modes $a$ and $b$ are vertically polarised, the two-photon state
acquires a relative phase. This results in an entangling gate that,
together with single-photon rotations, is sufficient for universal
quantum computing.} 
 \label{fig:kerr}
\end{figure}

A variation on this is the {\em cross-Kerr} medium, in which the phase
shift of a signal beam is proportional to the intensity of a second
probe beam. In the language of quantum optics, we describe the
cross-Kerr medium by the Hamiltonian 
\begin{equation} H_{\rm Kerr} = \kappa\, \hat{n}_s \hat{n}_p\; ,
\end{equation}  where $\hat{n}_s$ and $\hat{n}_p$ are the number
operators for the signal and probe mode, respectively. Compare $H_{\rm
Kerr}$ with the argument of the exponential in Eq.~(\ref{ps2}):
Transforming the probe (signal) mode using this Hamiltonian induces a
phase shift that depends on the number of photons in the signal
(probe) mode. Indeed, the mode transformations of the signal and probe
beams are 
\begin{equation} \hat{a}_s \rightarrow \hat{a}_s e^{-i\tau \hat{n}_p}
\quad\text{and}\quad \hat{a}_p \rightarrow \hat{a}_p e^{-i\tau
\hat{n}_s}\; ,
\end{equation}  with $\tau \equiv \kappa t$. When the cross-Kerr
medium is placed in one arm of a balanced Mach-Zehnder interferometer,
a sufficiently strong phase shift $\tau$ can switch the field from one
output mode to another (see Fig.~\ref{fig:kerr}a). For example, if the
probe beam is a (weak) optical field, and the signal mode may or may
not be populated with a single photon, then the detection of the
output ports of the Mach-Zehnder interferometer reveals whether there
was a photon in the signal beam. Moreover, we obtain this information
{\em without destroying the signal photon}. This is called a quantum
non-demolition measurement \cite{imoto85}.

It is not hard to see that we can use this mechanism to create an
all-optical CZ gate for photonic qubits [for the definition of a CZ
gate, see Eq.~(\ref{eq:czdef})]. Such a gate would give us the
capability to build an all-optical quantum computer. Let's assume that
our qubit states are single photons with horizontal or vertical
polarisation. In Fig.~\ref{fig:kerr}b, we show how the cross-Kerr
medium should be placed. The mode transformations are\footnote{Note
that the phase factors in these operator transformations are evaluated
for the vacuum state of modes $a$ and $b$.} 
\begin{eqnarray} \hat{a}_H \hat{b}_H \rightarrow \hat{a}_H' \hat{b}_H'
&\qquad& \hat{a}_V \hat{b}_H \rightarrow \hat{a}_V' \hat{b}_H' \cr
\hat{a}_H \hat{b}_V \rightarrow \hat{a}_H' \hat{b}_V' &\qquad&
\hat{a}_V \hat{b}_V \rightarrow \hat{a}_V' \hat{b}_V'\,  e^{i\tau}\; ,
\end{eqnarray} which means that the strength of the Kerr nonlinearity
should be $\tau = \pi$ in order to implement a CZ gate. It is trivial
to transform this gate into a CNOT gate. A Kerr-based Fredkin gate was
developed by Yamamoto  al.\ (1988) \nocite{yamamoto88} and  Milburn
(1989). \nocite{milburn89} Architectures based on these or similar
nonlinear optical gates were studied by Chuang and Yamamoto (1995),
\nocite{chuang95} Howell and Yeazell (2000b, 2000c),
\nocite{howell00b,howell00c} and d'Ariano et al.\
(2000). \nocite{dariano00a} Nonlinear interferometers are treated in
Sanders and Rice (2000), \nocite{sanders00} while state transformation
using Kerr media is the subject of Clausen et al.\
(2002). \nocite{clausen02} Recently, Hutchinson and Milburn (2004)
\nocite{hutchinson04} proposed cross-Kerr nonlinearities to create
cluster states for quantum computing. We will discuss cluster state
quantum computing in some detail in section \ref{sec:cluster}. 

Unfortunately, even the largest natural cross-Kerr nonlinearities are
extremely weak ($\chi^{(3)} \approx
10^{-22}$~m$^2$V$^{-2}$). Operating in the optical single-photon
regime with a mode volume of approximately 0.1~cm$^3$, the Kerr phase
shift is only $\tau \approx 10^{-18}$ \cite{kok02a}. This makes
Kerr-based optical quantum computing extremely challenging, if not
impossible. Much larger cross-Kerr nonlinearities of $\tau \approx
10^{-5}$ can be obtained with electromagnetically-induced transparent
materials \cite{schmidt96}. However, this value of $\tau$ is still
much too small to  implement the gates we discussed above. Towards the
end of this review  we will  indicate how such small-but-not-tiny
cross-Kerr nonlinearities may be used for quantum computing.

Turchette et al.\ (1995) \nocite{turchette95} proposed a different
method of inducing a phase shift when a signal mode $s$ and probe mode
$p$ of different frequency are both populated by a single polarised
photon. By sending both modes through a cavity containing caesium
atoms, they obtain a phase shift that is dependent on the
polarisations of the input modes:
\begin{eqnarray} |L,L\rangle_{sp} &\rightarrow& |L,L\rangle_{sp} \cr
|R,L\rangle_{sp} &\rightarrow& e^{i\phi_s}|R,L\rangle_{sp} \cr
|L,R\rangle_{sp} &\rightarrow& e^{i\phi_p}|L,R\rangle_{sp} \cr
|R,R\rangle_{sp} &\rightarrow&
e^{i(\phi_s+\phi_p+\delta)}|R,R\rangle_{sp},
\end{eqnarray} where $|L\rangle = |H\rangle + i|V\rangle$ and
$|R\rangle = |H\rangle - i|V\rangle$. Using weak coherent pulses,
Turchette et al.\ found $\phi_s = 17.5^{\circ} \pm 1$, $\phi_p =
12.5^{\circ} \pm 1$, and $\delta = 16^{\circ} \pm 3$.  An improvement
of this system was proposed by Hofmann et
al. (2003). \nocite{hofmann03} These authors showed how a phase shift
of  $\pi$ can be achieved with a single two-level atom in a one-sided
cavity. The cavity effectively enhances the tiny nonlinearity of the
atom. The losses in this system are negligible.  

In section \ref{sec:outlook} we will return to systems in which
(small) phase shifts can be generated using nonlinear optical
interactions, but the principal subject of this review is how {\em
projective  measurements} can induce enough of a nonlinearity to make
linear optical quantum computing possible.

\section{A new paradigm for optical quantum computing}\label{klm}

\noindent In 2000, Knill, Laflamme, and Milburn proved that it is
indeed possible to create universal quantum computers with linear
optics, single photons, and photon detection \cite{knill01}. They
constructed an explicit protocol, involving off-line resources,
quantum teleportation, and error correction. In this section, we will
describe this new paradigm, which has become known as the {\em KLM
scheme}, starting from the description of linear optics that we
developed in the previous section. In sections \ref{elga},
\ref{sec:parity} and \ref{expo}, we introduce some elementary
probabilistic gates and their experimental realizations, followed by a
characterisation of gates in section \ref{sec:char}, and a general
discussion on nonlinear unitary gates with projective measurements in
section \ref{genga}. We then describe how to teleport these gates into
an optical computational circuit in sections \ref{tele} and
\ref{sec:klm}, and the necessary error correction is outlined in
section \ref{sec:er}. Recently, Myers and Laflamme (2005)
\nocite{myers05} published a tutorial on the original ``KLM theory.'' 

\subsection{Elementary gates}\label{elga}

\noindent Physically, the reason why we cannot construct deterministic
two-qubit gates in the polarisation and dual-rail representation, is
that photons do not interact with each other. The only way in which
photons can directly influence each other is via the bosonic symmetry
relation. Indeed, linear optical quantum computing exploits exactly
this property, i.e., the bosonic commutation relation
$[\hat{a},\hat{a}^{\dagger}]=1$. To see what we mean by this
statement, consider two photons in separate spatial modes interacting
on a 50:50 beam splitter. The transformation will be 
\begin{eqnarray} |1,1\rangle_{ab} &=& \hat{a}^{\dagger}
\hat{b}^\dagger |0\rangle \cr &\rightarrow& \frac{1}{2}\left(
\hat{c}^{\dagger} + \hat{d}^{\dagger} \right) \left( \hat{c}^{\dagger}
- \hat{d}^{\dagger} \right) |0\rangle_{cd} \cr  &=& \frac{1}{2}\left(
\hat{c}^{\dagger 2} - \hat{d}^{\dagger 2} \right) |0\rangle_{cd} \cr
&=& \frac{1}{\sqrt{2}}\left( |2,0\rangle_{cd} - |0,2\rangle_{cd}
\right) \; .
\end{eqnarray} It is clear (from the second and third line) that the
bosonic nature of the electromagnetic field gives  rise to {\em photon
bunching}: the incoming photons pair off together. This is a strictly
quantum-mechanical effect, since classically, the two photons could
equally well end up in different output modes. In terms of quantum
interference, there are two paths leading from the input state
$|1,1\rangle_{\rm in}$ to the output state $|1,1\rangle_{\rm out}$:
Either both photons are transmitted, or both photons are
reflected. The relative phases of these paths are determined by the
beam splitter equation (\ref{2bs}):
\begin{eqnarray} |1,1\rangle_{\rm in} &\rightarrow_{\rm trans.}&
\cos^2\theta\, |1,1\rangle_{\rm out} , \cr   |1,1\rangle_{\rm in}
&\rightarrow_{\rm refl.}& -\sin^2\theta\, e^{i\varphi} e^{-i\varphi}
|1,1\rangle_{\rm out} .
\end{eqnarray} For a 50:50 beam splitter, we have $\cos^2\theta =
\sin^2\theta = 1/2$, and the two paths cancel exactly, irrespective of
the value of $\varphi$.

\begin{figure}[t]
  \begin{center}
  \begin{psfrags} 
\epsfig{file=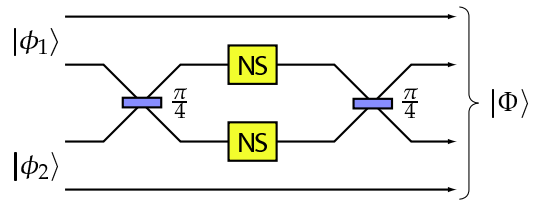}
  \end{psfrags}
  \end{center}
 \caption{The conditional phase gate (CZ). This gate uses two NS gates
to change the relative phase of the two qubits: when both qubits are
in the $|1\rangle$ state, the two photons interfere on the 50:50 beam
splitter ($\cos^2(\pi/4) = 1/2$). The Hong-Ou-Mandel effect then
ensures that both photons exit the same output mode, and the NS gates
induce a relative phase $\pi$. Upon recombination on the second beam
splitter, this phase shows up only in the states where both qubits
were in the $|1\rangle$ state.} 
 \label{fig:CZ_KLM}
\end{figure}

The absence of the $|1,1\rangle_{cd}$ term is called the
Hong-Ou-Mandel effect \cite{hong87}, and it lies at the heart of
linear optical quantum computing. However, as we have argued in
section \ref{qulo}, this is not enough to make deterministic linear
optical quantum computing possible, and we have to turn our attention
instead to {\em probabilistic gates}. 

As was shown by Lloyd (1995), \nocite{lloyd95} almost any two-qubit
gate is universal for quantum computing (in  addition to single-qubit
gates), but in linear optics we usually consider the controlled-phase
gate (CZ, also sometimes known as CPHASE or CSign) and the
controlled-not gate (CNOT). In terms of a truth table, they induce the
following transformations 
\begin{equation}
\begin{tabular}{c|c|c|c} ~Control~ & ~Target~ & ~CZ~ & ~CNOT~ \\
\hline $|0\rangle$ & $|0\rangle$ & $~\phantom{-}|0,0\rangle~$ &
$|0,0\rangle$ \\ $|0\rangle$ & $|1\rangle$ &
$~\phantom{-}|0,1\rangle~$ & $|0,1\rangle$ \\ $|1\rangle$ &
$|0\rangle$ & $~\phantom{-}|1,0\rangle~$ & $|1,1\rangle$ \\
$|1\rangle$ & $|1\rangle$ & $~-|1,1\rangle~$ & $|1,0\rangle$ \\
\end{tabular}
\end{equation} which is identical to
\begin{subequations}
\begin{eqnarray}\label{eq:czdef} |q_1,q_2\rangle & ~\rightarrow_{\rm
CZ}~ & (-1)^{q_1 q_2} |q_1,q_2\rangle \\
 \label{eq:cnotdef} |q_1,q_2\rangle & ~\rightarrow_{\rm CNOT}~ &
|q_1,q_2 \oplus q_1\rangle \; .
\end{eqnarray}
\end{subequations} Here $q_k$ takes the qubit values 0 and 1, while
$q_2 \oplus q_1$ is taken modulo 2.

A CZ gate can be constructed in linear optics using two {\em nonlinear
sign} (NS) gates. The NS gate acts on the three lowest Fock states in
the following manner: 
\begin{equation} \alpha |0\rangle + \beta |1\rangle + \gamma |2\rangle
~\rightarrow~ \alpha |0\rangle + \beta |1\rangle - \gamma |2\rangle \;
. 
\end{equation}   Its action on higher number states is irrelevant, as
long as it does not change the amplitudes of $|0\rangle$, $|1\rangle$,
or $|2\rangle$.  Consider the optical circuit drawn in
Fig.~\ref{fig:CZ_KLM}, and suppose the (separable) input state is
given by $|\phi_1\rangle \otimes |\phi_2\rangle = (\alpha |0,1\rangle
+ \beta |1,0\rangle)(\gamma |0,1\rangle + \delta
|1,0\rangle)$. Subsequently, we apply the beam splitter transformation
to the first and third mode, and find the Hong-Ou-Mandel effect only
when both modes are populated by one photon. The NS gates will then
induce a phase shift of $\pi$. Applying a second beam splitter
operation yields
\begin{eqnarray} |\Phi\rangle &=& \alpha\gamma |0,1,0,1\rangle +
\alpha\delta |0,1,1,0\rangle \cr && + \beta\gamma |1,0,0,1\rangle -
\beta\delta |1,0,1,0\rangle\; . 
\end{eqnarray} This is no longer separable in general. In fact, when
we choose $\alpha = \beta = \gamma = \delta = 1/\sqrt{2}$, then the
output state is a maximally entangled state. The overall probability
of this CZ gate $p_{CZ} = p_{NS}^2$. 

\begin{figure}[t]
  \begin{center}
  \begin{psfrags} 
\psfrag{b}{$|0\rangle$} 
\psfrag{d}{$\eta_1$} 
\psfrag{i}{``0''} \epsfig{file=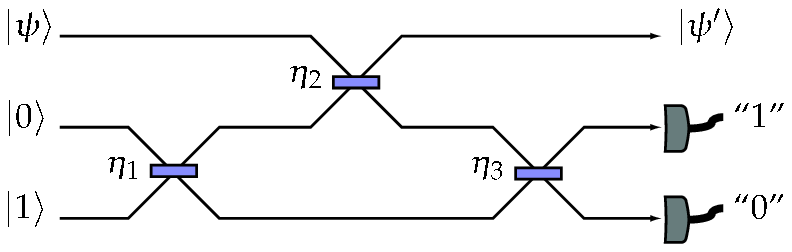}
  \end{psfrags}
  \end{center}
 \caption{The nonlinear sign (NS) gate according to Knill, Laflamme
and Milburn. The beam splitter transmission amplitudes are $\eta_1 =
\eta_3 = 1/(4-2\sqrt{2})$ and $\eta_2 = 3-2\sqrt{2}$.}
 \label{fig:NS_KLM}
%
  \begin{center}
  \begin{psfrags} 
\psfrag{g}{$\eta_1$} 
\epsfig{file=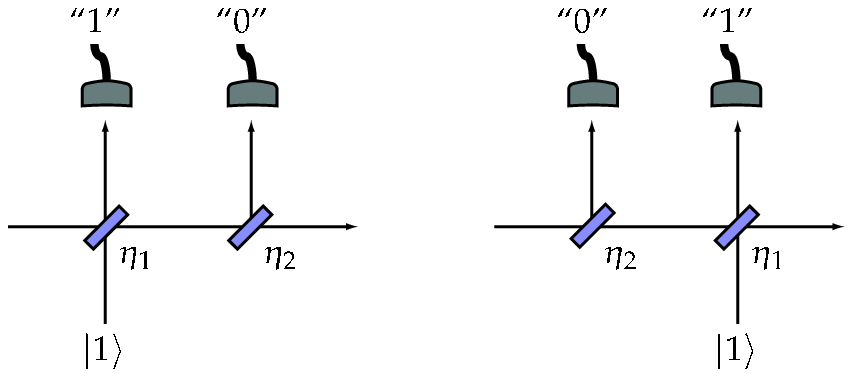}
  \end{psfrags}
  \end{center}
 \caption{The two equivalent versions of the NS gate by Ralph et al.\
(2002b). Only two beam splitters are used, while the other resources
are identical to the NS gate by Knill. Laflamme and Milburn. The
success probability of this gate is $(3-\sqrt{2})/7$.} 
 \label{fig:NS_RWMM}
\end{figure}

It is immediately clear that we cannot make the NS gate with a regular
phase shifter, because only the state $|2\rangle$ picks up a phase. A
linear optical phase shifter would also induce a factor $i$ (or $-i$)
in the state $|1\rangle$. However, it is possible to perform the
NS-gate {\em probabilistically} using projective measurements. The
fact that two NS gates can be used to create a CZ gate was first
realized by Knill, Laflamme, and Milburn \nocite{knill01}
(2001). Their probabilistic NS gate is a 3-port device, including two
ancillary modes the output of which is measured with perfect
photon-number discriminating detectors (see
Fig.~\ref{fig:NS_KLM}). The input states for the ancill\ae\ are the
vacuum and a single-photon, and the gate succeeds when the detectors
$D_1$ and $D_2$ measure zero and one photons, respectively. For an
arbitrary input state $\alpha |0\rangle + \beta |1\rangle + \gamma
|2\rangle$, this occurs with probability $p_{NS} = 1/4$. The general
upper bound for such gates was found to be 1/2 \cite{knill03}. Without
any feed-forward mechanism, the success probability of the NS gate
cannot exceed 1/4. It was shown numerically by Scheel and L\"utkenhaus
(2004) \nocite{scheel04} and proved analytically by Eisert (2005)
\nocite{eisert04} that, in general, the NS$_N$ gate defined by
\begin{equation} \sum_{k=0}^N c_k |k\rangle
~\rightarrow_{\mathrm{NS}_N}~ \sum_{k=0}^{N-1} c_k |k\rangle - c_N
|N\rangle 
\end{equation} can be implemented with probability $1/N^2$ [see also
Scheel and Audenaert (2005)]. \nocite{scheel05}

Several simplifications of the NS gate were reported shortly after the
original KLM proposal. First, a 3-port NS gate with only marginally
lower success probability $p_{NS}' = (3-\sqrt{2})/7$ was proposed by
Ralph et al.\ \nocite{ralph02b} (2002b). This gate uses only two beam
splitters (see  Fig.~\ref{fig:NS_RWMM}). Secondly, similar schemes
using two ancillary photons have been proposed
\cite{zou02c,scheel04b}. These protocols have success probabilities of
20\% and 25\%, respectively. 

\begin{figure}[b]
  \begin{center}
  \begin{psfrags}  
\psfrag{b}{$a$}  
\psfrag{e}{$|V\rangle$}  
\psfrag{h}{$\theta$}  
\psfrag{k}{$|\psi_{\rm out}\rangle$} \epsfig{file=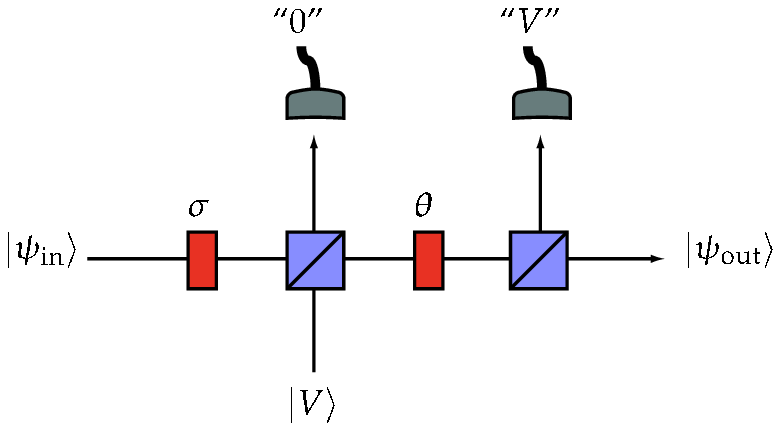}
  \end{psfrags}
  \end{center}
 \caption{The NS gate by Rudolph and Pan. Based on a vacuum detection
of the first output port, and a single vertically polarised photon on
the second output port, the interferometer applies an NS gate to the
input state. The success probability is also $(3-\sqrt{2})/7$, which
is close to the optimal value of 1/4.}  
 \label{fig:NS_rudolph}
%
  \begin{center}
  \begin{psfrags}  
  \epsfig{file=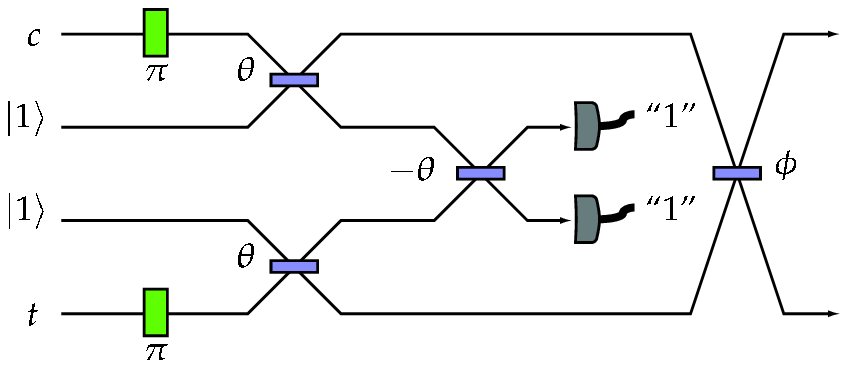}
  \end{psfrags}
  \end{center}
 \caption{The Knill CZ gate based on two ancilla photons and two
detected photons. The beam splitter angles are $\theta=54.74$ and
$\phi=17.63$, such that the transmission amplitudes are given by
$\cos\theta$ and $\cos\phi$, respectively.}  
 \label{fig:knill}
\end{figure}

Finally, a scheme equivalent to the one by Ralph et al.\ was
proposed by Rudolph and Pan (2001), in which the variable beam
splitters are replaced with polarisation rotations
\nocite{rudolph01}. These might be more convenient to implement
experimentally, since the irrational transmission and reflection
coefficients of the beam splitters are translated into polarisation
rotation angles (see Fig.~\ref{fig:NS_rudolph}). For pedagogical
purposes, we treat this gate in a little more detail. Assume that the
input mode is horizontally polarised. The polarisation rotation then
gives $\hat{a}_H \rightarrow \cos\sigma \hat{a}_H + \sin\sigma
\hat{a}_V$ and the input state transforms according to 

\begin{widetext}
\begin{equation}\nonumber \left( \alpha + \beta \hat{a}_H^{\dagger} +
\frac{\gamma}{\sqrt{2}}\hat{a}_H^{\dagger 2} \right)
\hat{b}_V^{\dagger} |0\rangle  \rightarrow \left[ \alpha + \beta
\cos\sigma\, \hat{a}_H + \beta \sin\sigma\, \hat{a}_V +
\frac{\gamma}{\sqrt{2}} \left( \cos^2\sigma\, \hat{a}_H^{\dagger 2} +
\sin 2\sigma\, \hat{a}_H^{\dagger} \hat{a}_V^{\dagger} + \sin^2\sigma\,
\hat{a}_V^{\dagger 2} \right)\right] \hat{b}_V^{\dagger} |0\rangle \; .
\end{equation} 
Detecting no photons in the first output port yields
\begin{equation}\nonumber \left( \alpha + \beta\cos\sigma\,
\hat{a}_H^{\dagger} + \frac{\gamma}{\sqrt{2}} \cos^2\sigma\,
\hat{a}_H^{\dagger 2} \right) \hat{b}_V^{\dagger} |0\rangle \; ,
\end{equation} 
after which we apply the second polarisation rotation:
$\hat{a}_H \rightarrow \cos\theta\, \hat{a}_H + \sin\theta\, \hat{a}_V$
and $\hat{a}_V \rightarrow -\sin\theta\, \hat{a}_H + \cos\theta\,
\hat{a}_V$. This gives the output state 
\begin{equation}\nonumber \left[ \alpha + \beta\cos\sigma \left(
\cos\theta\, \hat{a}_H^{\dagger} + \sin\theta\, \hat{a}_V^{\dagger}
\right) + \frac{\gamma}{\sqrt{2}} \cos^2\sigma \left( \cos\theta\,
\hat{a}_H^{\dagger} + \sin\theta\, \hat{a}_V^{\dagger} \right)^2 \right]
\left( -\sin\theta\, \hat{a}_H^{\dagger} + \cos\theta\,
\hat{a}_V^{\dagger} \right) |0\rangle\; .
\end{equation} 
After detecting a single vertically polarised photon in the second
output port, we have  
\begin{equation}\nonumber |\psi_{\rm out}\rangle = \alpha\cos\theta
|0\rangle + \beta\cos\sigma\cos 2\theta |1\rangle +
\gamma\cos^2\sigma\cos\theta (1-\sin^2 3\theta) |2\rangle. 
\end{equation}
\end{widetext} 
When we choose $\sigma \simeq 150.5^{\circ}$ and
$\theta \simeq 61.5^{\circ}$, this yields the NS gate with the same
probability $\cos^2\theta = (3-\sqrt{2})/7$.  Finally, in
Fig.~\ref{fig:knill}, the circuit of the CZ gate by Knill
\nocite{knill02} (2002) is shown. The success probability is
2/27. This is the most efficient CZ gate known to date.

Sometimes it might be sufficient to apply destructive two-photon
gates. For example, a Bell measurement in teleportation does not need
to be non-destructive in order to sucessfully teleport a photon. In
this case, we can increase the probability of success of the gate
considerably.  A CNOT gate that needs post-selection to make sure
there is one polarised photon in each output mode was proposed by
Ralph et al.\ \nocite{ralph02a} (2002a). It makes use only of beam
splitters with reflection coefficient of 1/3, and polarising beam
splitters. The success probability is 1/9. An identical gate was
proposed independently by Hofmann and Takeuchi \nocite{hofmann02a}
(2002). It was also shown that the success probability of an array of
$n$ CZ gates of this type can be made to operate with a probability of
$p=(1/3)^{n+1}$, rather than $p=(1/9)^n$ \cite{ralph04}.

\subsection{Parity gates and entangled ancill\ae}\label{sec:parity}

\noindent A special optical gate that will become important in section
\ref{sec:imp} is the so-called {\em parity check}. It consists of a
single polarizing beam splitter, followed by photon detection in the
complementary basis of one output mode. If the input modes are denoted
by $a$ and $b$, and the output modes are $c$ and $d$, then the
Bogoliubov transformation is given by Eq.~(\ref{eq:pbs}). For two
input qubits in the computational basis $\{ |H\rangle,|V\rangle\}$
this gate induces the following transformation:
\begin{eqnarray} |H,H\rangle_{ab} &\rightarrow& |H,H\rangle_{cd}, \cr
|H,V\rangle_{ab} &\rightarrow& |HV,0\rangle_{cd}, \cr |V,H\rangle_{ab}
&\rightarrow& |0,HV\rangle_{cd}, \cr |V,V\rangle_{ab} &\rightarrow&
|V,V\rangle_{cd} ,
\end{eqnarray} where $|HV,0\rangle_{cd}$ denotes a vertically and
horizontally polarised photon in mode $c$, and nothing in mode $d$.
Making a projective measurement in mode $c$ onto the complementary
basis $(|H\rangle \pm |V\rangle)/\sqrt{2}$ then yields a parity check:
If we detect a single photon in mode $c$, we know that the input
qubits had the same logical value. This value is transmitted into the
output qubit in mode $d$ (up to a $\sigma_z$ transformation depending
on the measurement result). On the other hand, if we detect zero or
two photons in mode $c$, the input qubits were not identical. In this
case, the state of the output mode is no longer in the single-qubit
subspace.  

\begin{figure}[t]
  \begin{center}
  \begin{psfrags} 
\psfrag{b}{out} 
\epsfig{file=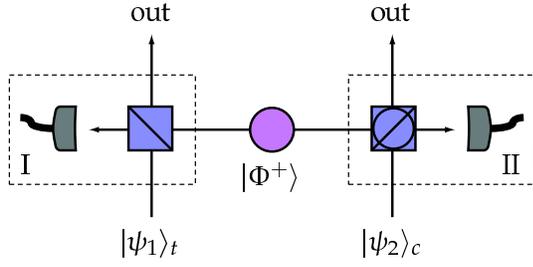}
  \end{psfrags}
  \end{center}
 \caption{The CNOT gate by Pittman et al.\ (2001). The two boxes I and
II are parity gates in two complementary bases, where the detector
measures in a complementary basis with respect to the polarising beam
splitter. The gate makes use of  a maximally entangled ancillary state
$|\Phi^+\rangle$, which boosts the success probability up to one
quarter. The target $|\psi_1\rangle_t$ and control $|\psi_2\rangle_c$
input states will evolve into an entangled output state conditioned on
the required detector signature.} 
 \label{fig:CNOT-franson}
\end{figure}

This gate was used by Cerf, Adami, and Kwiat to construct small
optical quantum circuits \nocite{cerf98} (1998). As we have seen in
section \ref{sec:kerr}, however, their approach is not scalable since
$n$-qubit circuits involve $2^n$ distinct paths. When two parity gates
in complementary bases are combined with a maximally entangled ancilla
state $|\Phi^+\rangle = (|H,H\rangle + |V,V\rangle)/\sqrt{2}$, a CNOT
gate with success probability 1/4 is obtained
\cite{pittman01,koashi01}. The setup is shown in
Fig.~\ref{fig:CNOT-franson}.   

For a detailed analysis of several probabilistic gates, see Lund and
Ralph (2002), Gilchrist et al.\ (2003), and Lund et al.\ (2003).
\nocite{lund02,gilchrist03,lund03} General two-qubit gates based on
Mach-Zehnder interferometry were proposed by Englert et al.\ (2001).
\nocite{englert01} For a general discussion of entanglement in quantum
information processing see Paris et al.\ (2003). \nocite{paris03}  

\bigskip

\noindent All the gates that we have discussed so far are
probabilistic, and indeed all two-qubit gates based on projective
measurements {\em must} be probabilistic. However, it is in principle
still possible that feed-forward protocols can increase the
probability to unity. As mentioned before, Knill (2003)
\nocite{knill03} demonstrated that this is not the case, and that
instead the highest possible success probability for the NS gate
(using feed-forward) is one half. He did not show that this bound is
tight. Indeed, numerical evidence strongly suggests an upper bound of
one third for infinite feed-forward without entangled ancill\ae\
\cite{scheel05b}. This indicates that the benefit of feed-forward
might not outweigh its cost.

\subsection{Experimental demonstrations of gates}\label{expo}

\noindent A number of experimental groups have already demonstrated
all-optical probabilistic quantum gates. Early experiments involved a
parity check of two polarisation qubits on a polarising beam splitter
\cite{pittman02b}, and a two-photon conditional phase switch
\cite{resch02}. A destructive CNOT gate was demonstrated by
\nocite{franson03} Franson et al.\ (2003) and O'Brien et al.\
\nocite{obrien03} (2003). In this section we will describe the
experimental demonstration of three CNOT gates. 

\begin{figure}[t]
  \begin{center}
  \begin{psfrags} 
 \epsfig{file=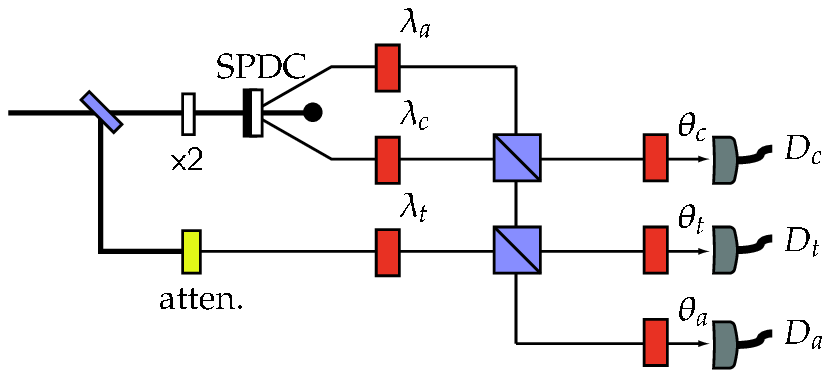}
  \end{psfrags}
  \end{center}
 \caption{Schematic diagram of the experimental setup of  the
three-photon CNOT gate of Pittman {et al}.\ (2003). The gate starts by
preparing the qubits with polarization rotations $\lambda_i$, followed
by mixing  the ancilla and control qubits on a polarising beam
splitter.  The ancilla qubit then is mixing with the target qubit on
the second polarising beam splitter. The gate is implemented upon a
three-fold detector coincidence in  the control, target, and ancilla
modes. The polarization rotations $\theta_i$ are used to select
different polarization bases.}   
 \label{fig:experiment}
%
  \begin{center}
  \begin{psfrags} 
 \epsfig{file=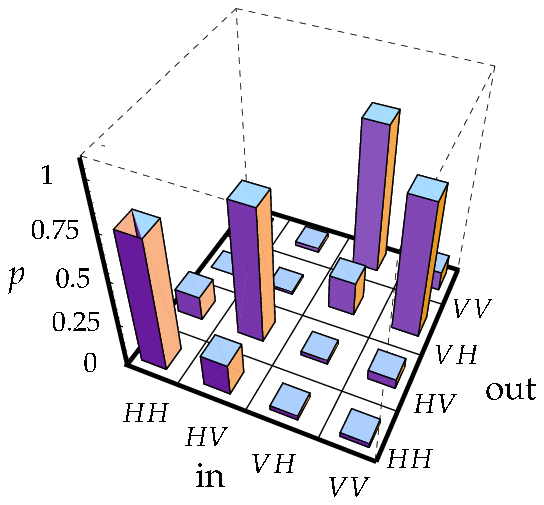}
  \end{psfrags}
  \end{center}
 \caption{Experimental demonstration of the CNOT gate by Pittman et
al.\ (2003).  The figure shows the probability of 3-fold coincidences
as a function of  the output qubit analysers for all four
computational basis states  $HH,HV,VH,VV$ in the input registers. The
error in the gate is approximately 21\%. } 
 \label{fig:basisstates}
\end{figure}

First, we consider the three-photon CNOT gate performed by Pittman,
Fitch, Jacobs, and Franson \nocite{pittman03} (2003). The gate is
shown in Fig.~\ref{fig:experiment} and consists of three
polarisation-encoded  single photon qubits and two polarising beam
splitters. Two of the polarisation qubits represent the control and
target qubits and are initially in an arbitrary two qubit state
$|\psi\rangle_{\rm in} = \alpha_{1}| H H \rangle_{ct} + \alpha_{2}|H
V\rangle_{ct}+\alpha_{3}|V H\rangle_{ct} +\alpha_{4}|V V\rangle_{ct}
$. The third photon is used as an ancilla qubit and is initially
prepared in the state $(|H\rangle + |V\rangle)/\sqrt{2}$. In the
Pittman et al.\ experiment the control qubit and the ancillary qubit
are created using pulsed parametric down conversion. The target qubit
is generated by an attenuated laser pulse where the pulse is branched
off the pump laser. The pulse is converted by a frequency doubler to
generate entangled photon pairs at the same frequency as the photon
constituting the target qubit. The CNOT gate is then implemented as
follows: The action of the polarising beam  splitters on the control,
target and ancilla qubits transforms them according to 
\begin{equation}\label{eq:cnottransform} |\psi\rangle_{\rm out}
\propto  |H\rangle_{a} U_{\rm C} |\psi\rangle_{\rm in} + |V\rangle_{a}
\widetilde{U}_{\rm C} |\psi\rangle_{\rm in} + \sqrt{6}
|\xi\rangle_{act},
\end{equation} where $U_{\rm C}$ is the CNOT operator between the
control and target modes $c$ and $t$ and $\widetilde{U}_{\rm C} =
(\unity\otimes\sigma_x)\, U_{\rm C}\, (\unity\otimes\sigma_x)$. The
state $|\xi\rangle_{act}$ represent terms with zero, two, or three
photons present in the modes $a$, $c$, and $t$.  Depending on the
polarisation of the measured ancillary  photon in mode $a$ (and one
photon in the control and  target modes) a CNOT gate up to a local
transformation is applied to the control and target qubits. For a
horizontally measured $|H\rangle_{a}$ photon the CNOT gate is exactly
implemented, while for a vertically measured $|V\rangle_{a}$ photon
the control and target qubits undergo a CNOT gate up to a bit flip on
the target qubit. In Fig.~\ref{fig:basisstates} the truth table is
shown as a function of the output qubit analysers for all four
computational basis states $HH,HV,VH,VV$ in the input. The success
probability for this gate is $p=1/4$ with an error of approximately
21\%.

\bigskip

\begin{figure}[t]
  \begin{center}
  \begin{psfrags} 
\epsfig{file=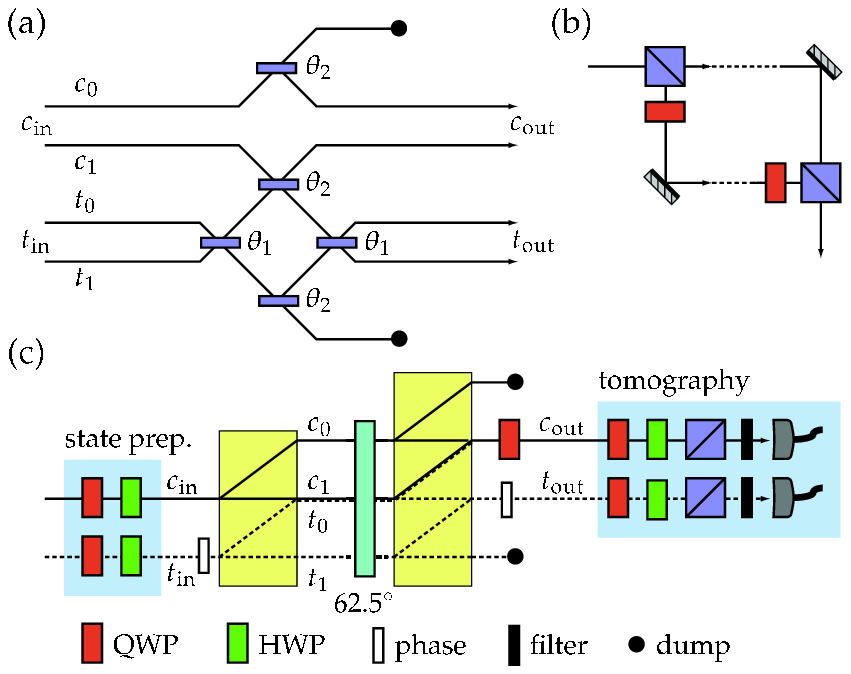, width=8.5cm}
  \end{psfrags}
  \end{center}
 \caption{Schematic diagram of the CNOT gate demonstrated by O'Brien
et al.\ (2003). (a) Concept of the two-qubit gate: The beam splitter
coefficients are $\theta_1 = \pi/4$ and $\theta_2 = \arccos
(1/\sqrt{3})$. (b) Translation circuit for converting polarisation and
dual rail qubits. (c) Schematic  of the experimental
setup. Simultaneous  detection of a single photon at each of the
detectors heralds the successful operation of the gate.}
 \label{fig:white}
  \begin{center}
  \begin{psfrags} 
\epsfig{file=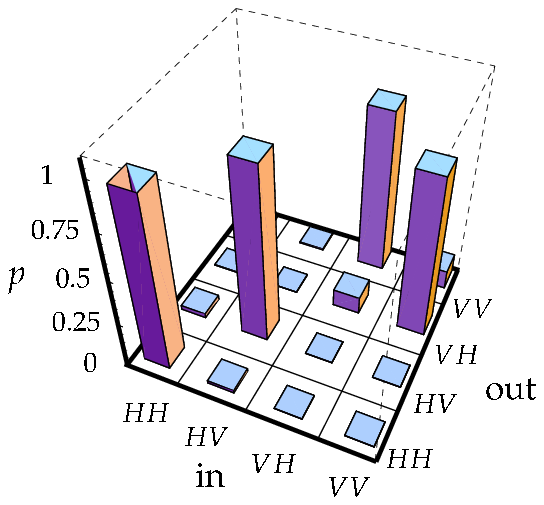}
  \end{psfrags}
  \end{center}
 \caption{Experimental demonstration of the CNOT gate by O'Brien et
al.\ (2003) in the logical qubit basis. The data is obtained by full
state tomography of the output states.} 
 \label{fig:truth}
\end{figure}

\noindent The second experiment we consider is the CNOT gate by
O'Brien, Pryde, White, Ralph, and Branning, depicted in
Fig.~\ref{fig:white} \cite{obrien03}, which is an implementation of
the gate proposed  by Ralph et al.\ (2002a). This is a post-selected
two-photon gate where the  two polarised qubits are created in a
parametric down conversion event. The polarisation qubits can be
converted into which-path qubits via a translation  stage depicted in
Fig.~(\ref{fig:white}b). Both the control and target  qubits can be
prepared in an arbitrary pure superposition of the computational basis
states.

The gate is most easily understood in terms of dual spatial rails,
Fig.~(\ref{fig:white}a). 
The two spatial modes that support the target qubit are mixed on a
50:50 beam splitter ($\theta_1 = \pi/4$). One of these output modes is
mixed with a spatial mode of the  control qubit on  a beam splitter
with $\cos\theta_2 = 1/\sqrt{3}$ (that is, a beam splitter with a
reflectivity of 33$\frac{1}{3}$\%).  To balance the probability
distribution of the CNOT gate, two ``dump ports'' consisting of
another beam splitter with $\cos\theta_2 = 1/\sqrt{3}$ are introduced
in one of the control and  target modes. The gate works as follows: If
the control qubit is in the state where the photon occupies the top
mode $c_0$ there is no interaction between the control and the target
qubit. On the other hand, when the control photon is in the lower
mode, the control and target photons interfere non-classically at the
central beam splitter with $\cos\theta_2 = 1/\sqrt{3}$. This
two-photon quantum interference causes a $\pi$ phase shift in the
upper arm of the target interferometer $t_0$, and as a result, the
target photon is switched from one output mode to the other. In other
words, the target state experiences a bit flip. The control qubit
remains unaffected, hence the interpretation of this experiment as a
CNOT gate. We do not always observe a single photon in each of the
control and target outputs. However, when a control and target photon
are detected we know that the CNOT operation has been correctly
realized. The success probability of such an event is $1/9$. The
detection of the control and target qubits could in principle be
achieved by a quantum non-demolition measurement (see section
\ref{sec:detectors}) and would not destroy the information encoded on
the qubits. Experimentally, beam displacers are used to spatially
separate the polarisation modes, and waveplates are used for the
beam-mixing. 

In Fig.~\ref{fig:truth}, we show the truth table for the CNOT
operation in the coincidence basis. The fidelity of the gate is
approximately 84\% with conditional fringe visibilities exceeding 90\%
in non-or\-thogonal bases. This indicates that entanglement has been
generated in the  experiment: The gate can create entangled output
states from separable input states.

\bigskip

\noindent The last experiment we consider in some detail is the
realization of an optical CNOT by Gasparoni, Pan, Walther, Rudolph,
and Zeilinger \nocite{gasparoni04} (2004). The experiment is based on
the  four-photon logic gate by Pittman et al.\ (2001) depicted in
Fig.~\ref{fig:CNOT-franson}.

\begin{figure}[t]
  \begin{center}
  \begin{psfrags} 
 \epsfig{file=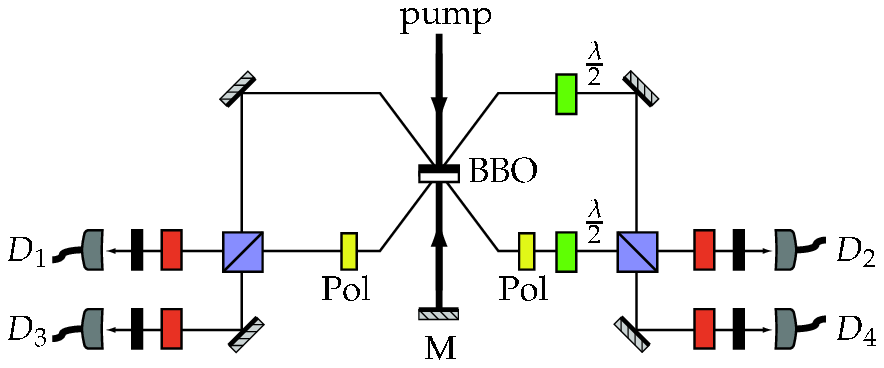, width=8.5cm}
  \end{psfrags}
  \end{center}
 \caption{Schematic diagram of the four-photon CNOT gate by Gasparoni
et al.\ (2004). A parametric down conversion source is used to create
the control and target input qubits in the spatial modes $a_1$ and
$a_2$, as well as a maximally entangled ancilla pair in the spatial
modes $a_3$ and $a_4$. Polarising filters (Pol) can be used to destroy
the initial entanglement in $a_1$ and $a_2$ if necessary.}   
 \label{fig:setupnew}
%
  \begin{center}
  \begin{psfrags} 
\epsfig{file=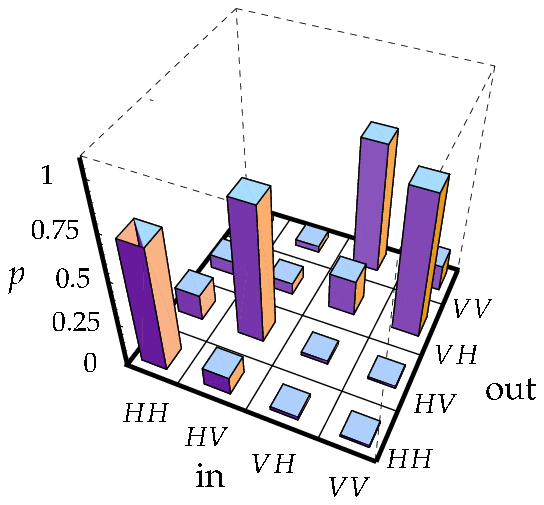}
  \end{psfrags}
  \end{center}
 \caption{Experimental demonstration of the CNOT gate by Gasparoni et
al.\ (2004). Four-fold coincidences for all combinations of inputs and
outputs are shown.}   
 \label{fig:cnot3}
\end{figure}

The Gasparoni et al.\ experiment employs a type-II parametric down
conversion source operated in a ``double-pass'' arrangement.  The
down-conversion process naturally produces close to maximally
entangled photon pairs. This means that, depending on the input state
for the control and target qubits, we may have to destroy or decrease
any initial polarisation entanglement. This is achieved by letting the
photons pass through appropriate polarisation filters. After this, any
two-qubit input state can be prepared. The gate depicted in
Fig.~\ref{fig:setupnew} works as follows: The control qubit and one
half of the Bell state are sent into a polarising beam splitter, while
the target qubit with the second half of the Bell state are sent
through a second polarising beam splitter. The detection of the
ancilla photons heralds the operation of the CNOT gate (up to a known
bit or sign flip on the control and/or target qubit). The probability
of success of this gate is  $1/4$. Due to a lack of detectors that can
resolve the  difference between one and two photons and the rather low
source and detector efficiencies, four-fold
coincidence detection was employed to confirm the presence of photons
in the output control and target ports. In principle, this
post-selection can be circumvented by using deterministic Bell pair
sources and detectors that differentiate between one and two incoming
photons.

The CNOT truth table for this experiment, based on four-fold
coincidences, is shown in Fig.~\ref{fig:cnot3}. This shows the
operation of the CNOT gate. In addition,   Gasparoni et al.\ showed
that an equal superposition of $H$ and $V$  for the control qubit and
$H$ for the target qubit generated the maximally entangled state
$|HH\rangle+|VV\rangle$ with a fidelity  of $81\% $. This clearly
shows that the gate is creating entanglement.

As experiments become more sophisticated, more demonstrations of
optical gates are reported. We cannot describe all of them here, but
other recent experiments include the nonlinear sign shift
\cite{sanaka04}, a non-destructive CNOT \cite{zhao05}, another CNOT
gate \cite{fiorentino04}, and three-qubit optical quantum circuits
\cite{takeuchi00a,takeuchi01}. Four-qubit {\em cluster states}, which
we will encounter later in this review, were demonstrated by Walther
et al.\ \nocite{walther05} (2005).

\subsection{Characterisation of linear optics gates}\label{sec:char}

\begin{figure}[t]
  \begin{center}
  \begin{psfrags} 
\epsfig{file=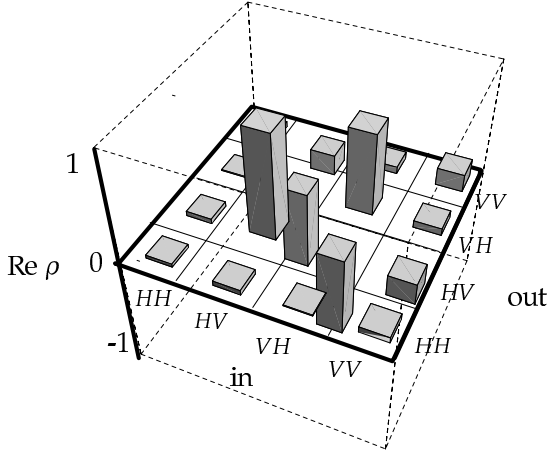}
  \end{psfrags}
  \end{center}
  \caption{Plot of the real part of the density matrix reconstructed
from  quantum process tomography for the input state $(|H\rangle_c -
|V\rangle_c) |V\rangle_t$. This shows the highly entangled singlet
state of the form $|H\rangle_c |V\rangle_t-|V\rangle_c |H\rangle_t$}
  \label{fig:tomoreal} 
\end{figure}

\noindent In section \ref{expo}, we showed the experimentally realised
CNOT truth table for three different experimentally realised
gates. However, the construction of the truth table is in itself not
sufficient to show that a CNOT  operation has been performed. It is
essential to show the quantum coherence of the gate. One of the
simplest ways to show coherence is to apply the gate to an initial
separate state and show that the gate creates an entangled state (or
vice versa). For instance, the operation of a CNOT gate on the initial
state $(|H\rangle_c-|V\rangle_c) |V\rangle_t$ creates the maximally
entangled singlet state $|H\rangle_c |V\rangle_t-|V\rangle_c
|H\rangle_t$. This is sufficient  to show the coherence properties of
the gate. However, showing such coherences does not fully characterise
the gate. To this end, we can perform state tomography. We show an
example of this for the CNOT gate demonstrated by O'Brien et al.\
(2003) in Fig.~\ref{fig:tomoreal}. The reconstructed density matrix
clearly indicates that a high-fidelity singlet state has been
produced.

To fully understand the operation of a gate we need to create a
complete map $\hat{\mathcal{E}}$ of all the input states to the output
states. 
\begin{eqnarray} \hat{\mathcal{E}}(\rho)=\sum_{m,n=0}^{d-1} \chi_{mn}
{\hat A}_m \rho {\hat A}^\dag_n,  
\end{eqnarray} This map represents the process acting on an arbitrary
input state $\rho$, where the operators ${\hat A_m}$ form a basis for
the operators acting on $\rho$. The matrix  $\chi$ describes
completely the process $\hat{\mathcal{E}}$. Once this map has been
constructed, we know everything  about the process, including the
purity of the operation and the  entangling power of the gate. This
information can then  be used to fine-tune the gate
operation. Experimentally, the map is obtained by  performing {\it
quantum process tomography} \cite{chuang97,poyatos97}. A set of
measurements is made on the output of the $n$-qubit quantum gate,
given a complete set of input states. The input states and measurement
projectors each form a basis for the set of $n$-qubit density
matrices. For the two-qubit CNOT gate ($d=16$), we require  256
different settings of input states and measurement projectors.

\begin{figure}[t]
  \begin{center}
  \begin{psfrags} 
\epsfig{file=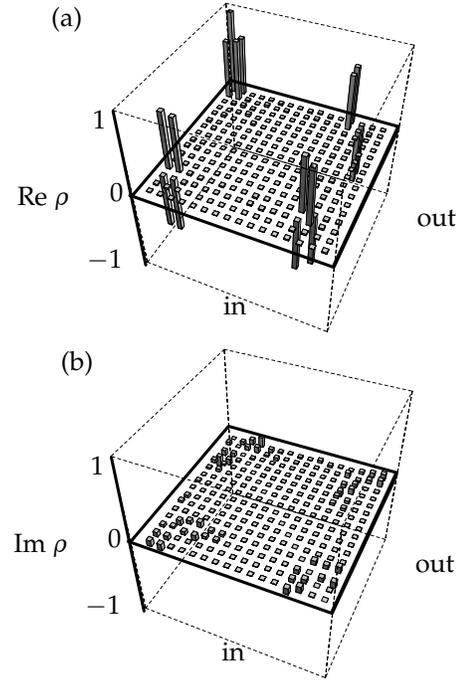}
  \end{psfrags}
  \end{center}
  \caption{Plot of the real (a) and imaginary (b) parts of the
reconstructed process matrix of the CNOT gate by O'Brien et al.\
(2003). The ideal CNOT can be written as a coherent sum: ${\hat
U}_{\rm CNOT}=\tfrac{1}{2}(\unity \otimes \unity + \unity \otimes X +
Z \otimes \unity - Z \otimes X);$ of the tensor products of Pauli
operators $\{\unity, X, Y, Z \}$ acting on control and target qubits
respectively.}   
  \label{fig:tomouq} 
\end{figure}

In Fig.~\ref{fig:tomouq}, we reproduce the reconstructed process
matrix $\chi$ for the CNOT gate performed by O'Brien et al.\
(2003). The ideal CNOT can be written as  a coherent sum: ${\hat
U}_{\rm CNOT}=\tfrac{1}{2}(\unity\otimes\unity + \unity \otimes
X+Z\otimes\unity -Z\otimes X)$ of tensor products of Pauli operators
$\{\unity, X, Y, Z\}$ acting on control and target qubits
respectively. The process matrix shows the populations/coherences
between the basis operators making up the gate. The process fidelity
for this gate exceeds 90\% (see also \nocite{obrien04} O'Brien et al.\
2004).  For a general review of quantum state tomography with an
emphasis on quantum information processing, see Lvovsky and Raymer
(2005). \nocite{lvovsky05}

\subsection{General probabilistic nonlinear gates}\label{genga}

\noindent The two-qubit gates described above are special cases of $N$
ports acting on a set of input states, followed by a projective
measurement. For quantum computing applications, however, we usually
want the resulting nonlinear transformation $\mathsf{M}$ to be {\em
unitary}. This is because non-unitary operations will reveal
information about the qubits in the projective measurement, and hence
corrupts the computation. We can derive a simple criterion that the
$N$ ports and the projective measurements must satisfy
\cite{lapaire03}.

Suppose the qubits undergoing $\mathsf{M}$ span a Hilbert space
${\mathcal{H}}_Q$, and the auxiliary qubits span
${\mathcal{H}}_A$. Furthermore, let $U$ be the unitary transformation
of the $N$ port in Eq.~(\ref{trans1}) and $P_k$ the projector on the
auxiliary states denoting the measurement outcome labelled by
$k$. $P_k$ must be a projector on the Hilbert space with dimension
$\dim{\mathcal{H}}_A$ for $\mathsf{M}$ to be unitary. Given an
arbitrary input state $\rho$ of the qubits and a state $\sigma$ of the
auxiliary systems, the output state can be written as
\begin{equation}\label{testout} \rho_{\mathrm{out}}^{(k)} =
\frac{\mathrm{Tr}_A \left[ U (\rho\otimes\sigma) U^{\dagger} P_k
\right]}{\mathrm{Tr}_{QA} \left[ U (\rho\otimes\sigma) U^{\dagger} P_k
\right]}\; .
\end{equation} When we define $d(\rho) \equiv \mathrm{Tr}_{QA} \left[
U (\rho\otimes\sigma) U^{\dagger} P_k \right]$, we find that
$\mathsf{M}$ is unitary if and only if $d(\rho)$ is independent of
$\rho$. We can then construct a {\em test operator} $\hat{T} = \mathrm{Tr}_A
\left(\sigma U^{\dagger} P_k U \right)$. The induced operation on the
qubits in ${\mathcal{H}}_Q$ is then unitary if and only if $\hat{T}$ is
proportional to the identity, or
\begin{equation} 
 \hat{T} = \mathrm{Tr}_A \left(\sigma U^{\dagger} P_k U
 \right) \propto\unity ~\Leftrightarrow~ d(\rho) = d\; .
\end{equation} Given the auxiliary input state $\sigma$, the $N$ port
transformation $U$ and the projective measurement $P_k$, it is
straightforward to check whether this condition holds. The success
probability of the gate is given by $d$.

In Eq.~(\ref{testout}), the projective measurement was in fact a
projection operator $(P_k^2 = P_k)$. However, in general, we might
want to include generalised measurements, commonly known as Positive
Operator-Valued Measures, or POVMs. These are particularly useful when
we need to distinguish between nonorthogonal states, and they can be
implemented with $N$ ports as well \cite{myers97}. Other optical
realizations of non-unitary transformations were studied by Bergou et
al.\ (2000). \nocite{bergou00}

The inability to perform a deterministic two-qubit gate such as the
CNOT with linear optics alone is intimately related to the
impossibility of complete Bell measurements with linear optics
\cite{lutkenhaus99,vaidman99,calsamiglia02}. Since quantum computing
can be cast into the shape of single-qubit operations and two-qubit
projections \cite{nielsen03,leung01}, we can approach the problem of
making nonlinear gates via complete discrimination of multi-qubit
bases.

Van Loock and L{\" u}tkenhaus gave straightforward criteria for the
implementation of complete projective measurements with linear optics
\cite{loock04}. Suppose the basis states we want to identify without
ambiguity are given by $\{ |s_k\rangle \}$, and the auxiliary state is
given by $|\psi_{\mathrm{aux}}\rangle$. Applying the unitary $N$ port
transformation yields the state $|\chi_k\rangle$. If the outgoing
optical modes are denoted by $a_j$, with corresponding annihilation
operators $\hat{a}_j$, then the set of conditions that have to be
fulfilled for $\{ |\chi_k\rangle \}$ to be completely distinguishable
are
\begin{eqnarray} \langle\chi_k| \hat{a}_j^{\dagger} \hat{a}_j
|\chi_l\rangle = 0 && \forall j\cr \langle\chi_k| \hat{a}_j^{\dagger}
\hat{a}_j \hat{a}_{j'}^{\dagger} \hat{a}_{j'} |\chi_l\rangle = 0 &&
\forall j,j'\cr \langle\chi_k| \hat{a}_j^{\dagger} \hat{a}_j
\hat{a}_{j'}^{\dagger} \hat{a}_{j'} \hat{a}_{j''}^{\dagger}
\hat{a}_{j''}|\chi_l\rangle = 0 && \forall j,j',j'' \cr \vdots\qquad
&& \vdots
\end{eqnarray} Furthermore, when we keep the specific optical
implementation in mind, we can use intuitive physical principles such
as photon number conservation and group-theoretical techniques such as
the decomposition of $U(N)$ into smaller groups. This gives us an
insight into how the auxiliary states and the photon detection affects
the (undetected) signal state \cite{scheel03}.

So far we have generally focused on the means necessary to perform
single-qubit rotations and CNOT gates. It is well known that such
gates are sufficient for universal computation. However, it is not
necessary to restrict ourselves to such a limited set of
operations. Instead, it is possible to extend our operations to
general circuits that can be constructed from linear elements, single
photon sources, and detectors. This is analogous to the shift in
classical computing from a RISC (Reduced Instruction Set Computer)
architecture to the CISC (Complex Instruction Set Computer)
architecture. The RISC-based architecture in quantum computing terms
could be thought of as a device built only from the minimum set of
gates, while a CISC-based machine would be built from a much larger
set; a natural set of gates allowed by the fundamental resources.  The
quantum SWAP operation illustrates this point. From fundamental {\em
gates}, three CNOTs are required to build such an operation. However,
from fundamental optical {\em resources} only two beam splitters and a
phase shifter are necessary. Scheel et al.\ (2003) focused their
attention primarily on one-mode and two-mode situations, though the
approach is easily extended to multi-mode situations. They
differentiated between operations that are easy and that are
potentially difficult. For example, operations that cause a change in
the Fock layers (for instance the Hadamard operator) are generally
difficult but not impossible.

\subsection{Scalable optical circuits and quantum
teleportation}\label{tele}

\begin{figure}[t]
  \begin{center}
  \begin{psfrags} 
 \epsfig{file=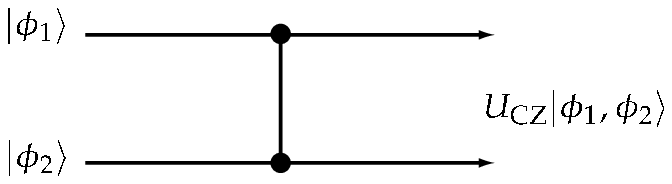}
  \end{psfrags}
  \end{center}
 \caption{The CZ applied to two qubits inside a quantum circuit. If it
fails, then the two qubit states are lost.} 
 \label{fig:cnot-inline}
%
  \begin{center}
  \begin{psfrags} 
 \epsfig{file=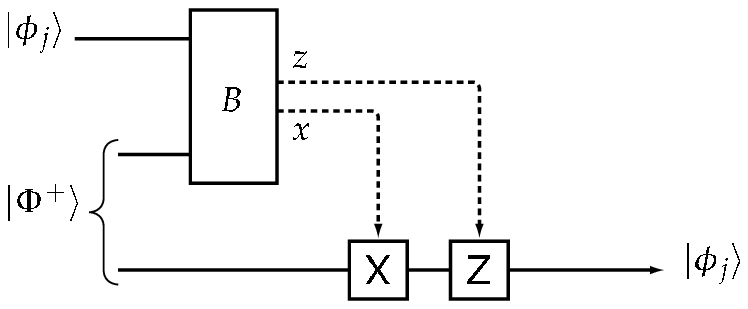}
  \end{psfrags}
  \end{center}
 \caption{The teleportation circuit. The state $|\phi_j\rangle$ is
teleported via an entangled quantum channel $|\Phi^+\rangle$ and a
Bell measurement $B$. The binary variables $x$ and $z$ parameterise
the outcome of the Bell measurement and determine which Pauli operator
is applied to the output mode.} 
 \label{fig:teleportation}
\end{figure}

\noindent When the gates in a computational circuit succeed only with
a certain probability $p$, then the entire calculation that uses $N$
such gates succeeds with probability $p^{N}$. For large $N$ and small
$p$, this probability is minuscule. As a consequence, we have to
repeat the calculation on the order of $p^{-N}$ times, or run $p^{-N}$
such systems in parallel. Either way, the resources (time or circuits)
scale exponentially with the number of gates. Any advantage that
quantum algorithms might have over classical protocols is thus
squandered on retrials or on the amount of hardware we need. In order
to do useful quantum computing with probabilistic gates, we have to
take the probabilistic elements out of the running calculation.

In 1999, Gottesman and Chuang proposed a trick that removes the
probabilistic gate from the quantum circuit, and places it in the
resources that can be prepared off-line \cite{gottesman99}. It is
commonly referred to as the {\em teleportation trick}, since it
``teleports the gate into the quantum circuit.''

Suppose we need to apply a probabilistic CZ gate to two qubits with
quantum states $|\phi_1\rangle$ and $|\phi_2\rangle$ respectively. If
we apply the gate directly to the qubits, we are very likely to
destroy the qubits (see Fig.~\ref{fig:cnot-inline}). However, suppose
that we teleport both qubits from their initial mode to a different
mode. For one qubit, this is shown in
Fig.~\ref{fig:teleportation}. Here, $x$ and $z$ are binary variables,
denoting the outcome of the Bell measurement, which determine the
unitary transformation that we need to apply to the output mode. If
$x=1$, we need to apply the $\sigma_x$ Pauli operator (denoted by
$X$), and if $z=1$, we need to apply $\sigma_z$ (denoted by $Z$). If
$x,z=0$ we do not apply the respective operator. For teleportation to
work, we also need the entangled resource $|\Phi^+\rangle$, which can
be prepared off-line. If we have a suitable storage device, we do not
have to make $|\Phi^+\rangle$ on demand: we can create it with a
probabilistic protocol using several trials, and store the output of a
successful event in the storage device.

\begin{figure}[t]
  \begin{center}
  \begin{psfrags}  
  \epsfig{file=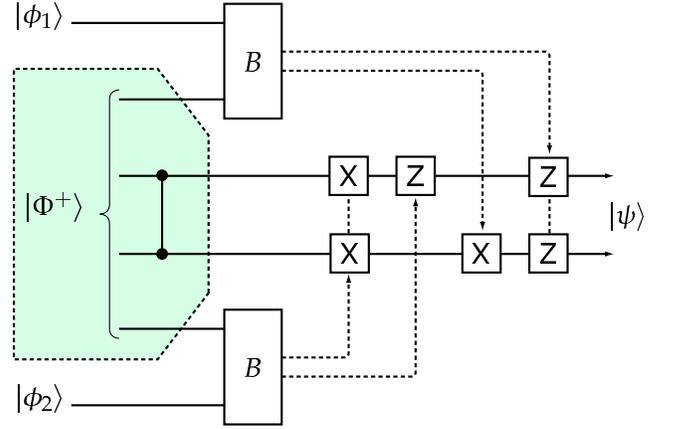}
  \end{psfrags}
  \end{center}
 \caption{The CZ gate using teleportation: here, $|\psi\rangle =
U_{\rm CZ} |\phi_1\phi_2\rangle$. By commuting the CZ gate through the
Pauli gates from the computational circuit to the teleportation
resources, we have taken the probabilistic part off-line. We can
prepare the teleportation channel (the shaded area, including the  CZ)
in many trials, without disrupting the quantum computation.}  
 \label{fig:cnot-gc}
\end{figure}

When we apply the probabilistic CZ gate to the output of the two
teleportation circuits, we effectively have again the situation
depicted in Fig.~\ref{fig:cnot-inline}, except that now our circuit is
much more complicated. Since the CZ gate is part of the Clifford
group, we can commute it through the Pauli operators $X$ and $Z$ at
the cost of more Pauli operators. This is good news, because that
means we can move the CZ gate from the right to the left, and only
incur the optically available single-qubit Pauli gates. Instead of
preparing two entangled quantum channels $|\Phi^+\rangle$, we now have
to prepare the resource $\unity \otimes U_{\rm CZ} \otimes \unity
|\Phi^+\rangle \otimes |\Phi^+\rangle$ (see
Fig.~\ref{fig:cnot-gc}). Again, with a suitable storage device, this
can be done off-line with a probabilistic protocol. There are now no
longer any probabilistic elements in the computational circuit.

\subsection{The Knill-Laflamme-Milburn protocol}\label{sec:klm}

\noindent  Unfortunately, there is a problem with the teleportation
trick when applied to linear optics: In our qubit representation the
Bell measurement (which is essential to quantum teleportation) is not
complete, and works at best only half of the time
\cite{lutkenhaus99,vaidman99}. It seems that we are back where we
started. This is one of the problems of linear optical quantum
computing that was solved by Knill, Laflamme, and Milburn (2001).
\nocite{knill01}

In the KLM scheme, the qubits are chosen from the dual-rail
representation. However, in the KLM protocol the teleportation trick
applies to the single-rail state $\alpha |0\rangle + \beta |1\rangle$,
where $|0\rangle$ and $|1\rangle$ denote the vacuum and the
single-photon Fock state respectively, and $\alpha$ and $\beta$ are
complex coefficients (this is because the CZ gate involves only one
optical mode of each qubit). Linearity of quantum mechanics ensures
that if we can  teleport this state, we can also teleport any coherent
or incoherent superposition of such a state.

\begin{figure}[t]
  \begin{center}
  \begin{psfrags} 
 \epsfig{file=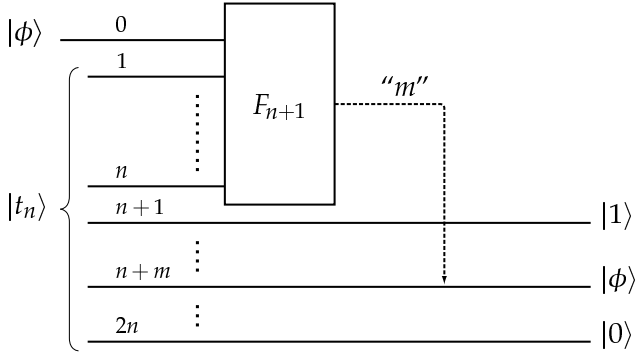}
  \end{psfrags}
  \end{center}
 \caption{Near-deterministic teleportation according to Knill,
Laflamme and Milburn. The input state $|\phi\rangle = \alpha |0\rangle
+ \beta |1\rangle$ is teleported to the $m^{\rm th}$ outgoing mode,
where $m$ is the number of detected photons in the measurement of the
$(n+1)$-point quantum Fourier transform. Note that $|\psi\rangle$ is a
single-rail state; $0$ and $1$ denote photon numbers here.} 
 \label{fig:tp-klm}
\end{figure}

Choose the quantum channel to be the $2n$-mode state
\begin{equation}\label{telaux} |t_n\rangle = \frac{1}{\sqrt{n+1}}
\sum_{j=0}^n |1\rangle^j |0\rangle^{n-j} \; |0\rangle^j
|1\rangle^{n-j} ,
\end{equation} where $|k\rangle^j \equiv |k\rangle_1 \otimes \ldots
\otimes |k\rangle_j$. We can then teleport the state $\alpha |0\rangle
+ \beta |1\rangle$ by applying an $n+1$-point discrete quantum Fourier
transform (QFT) to the input mode and the first $n$ modes of
$|t_n\rangle$, and count the number of photons $m$ in the output
mode. The input state will then be teleported to mode $n+m$ of the
quantum channel (see Fig.~\ref{fig:tp-klm}).

The discrete quantum Fourier transform $F_n$ can be written in matrix
notation as:
\begin{equation} (F_n)_{jk} = \frac{1}{\sqrt{n}} \exp\left[ 2\pi i\;
\frac{(j-1)(k-1)}{n} \right] .
\end{equation} It erases all path information of the incoming modes,
and can be interpreted as the $n$-mode generalisation of the 50:50
beam splitter. To see how this functions as a teleportation protocol,
it is easiest to consider an example.

\begin{figure}[t]
  \begin{center}
  \begin{psfrags} 
  \epsfig{file=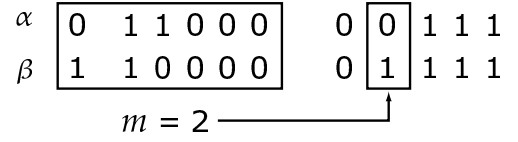}
  \end{psfrags}
  \end{center}
 \caption{The 5-photon ancillary scheme for near-deterministic
teleportation.} 
 \label{fig:tp2-klm}
\end{figure}

Suppose, we choose $n=5$, such that the state $|t_n\rangle$ describes
ten optical modes, and assume further that we count two photons
($m=2$). This setup is given in Fig.~\ref{fig:tp2-klm}. The two rows
of zeros and ones denote two terms in the superposition
$|t_5\rangle$. The five numbers on the left are the negative of the
five numbers on the right (from which we will choose the outgoing
qubit mode). It is then clear from this diagram that when we find two
photons, there are only two ways this could come about: either the
input mode did not have a photon (associated with amplitude $\alpha$),
in which case the two photons originated from $|t_5\rangle$, or the
input mode did have a photon, in which case the state $|t_5\rangle$
provided the second photon. However, by construction of $|t_5\rangle$,
the second mode of the five remaining modes must have the same number
of photons as the input mode. And because we erased the which-path
information of the measured photons using the $F_6$ transformation,
the two possibilities are added coherently. This means that we
teleported the input mode to mode $5+2=7$. In order to keep the
amplitudes of the output state equal to those of the input state, the
relative amplitudes of the terms in $|t_n\rangle$ must be equal. 

Sometimes, this procedure fails, however. When we count either zero or
$n+1$ photons in the output of the QFT, we collapsed the input state
onto zero or one photons respectively. In those cases we know that the
teleportation failed. The success rate of this protocol is $n/(n+1)$
(where we used that $|\alpha|^2 + |\beta|^2 = 1$). We can make the
success probability of this protocol as large as we like by increasing
the number of modes $n$. The success probability for teleporting a
two-qubit gate is then the {\em square} of this probability,
$n^2/(n+1)^2$, because we need to teleport {\em two} qubits
successfully. The quantum teleportation of a superposition state of a
single photon with the vacuum was realized by Lombardi et al.\
(2002) \nocite{lombardi02} using spontaneous parametric
down-conversion. 

Now that we have a (near-) deterministic teleportation protocol, we
have to apply the probabilistic gates to the auxiliary states
$|t_n\rangle$. For the CZ gate, we need the auxiliary state
\begin{eqnarray} |cz_n\rangle &=& \frac{1}{n+1} \sum_{i,j=0}^n
(-1)^{(n-i)(n-j)} |1\rangle^i |0\rangle^{n-i} \cr && \; \times
|0\rangle^i |1\rangle^{n-i}\; |1\rangle^j |0\rangle^{n-j} \;
|0\rangle^j |1\rangle^{n-j}\; .
\end{eqnarray} The cost of creating this state is quite high. In the
next section we will see how the addition of error correcting codes
can alleviate this resource count somewhat.

At this point, we should resolve a paradox: Earlier results have shown
that it is impossible to perform a deterministic Bell measurement with
linear optics. However, teleportation relies critically on a Bell
measurement of some sort, and we have just shown that we can perform
near-deterministic teleportation with only linear optics and photon
counting. The resolution in the paradox lies in the fact that the
impossibility proofs are concerned with {\em exact} deterministic Bell
measurements. The KLM variant of the Bell measurement always has an
arbitrarily small error probability $\epsilon$. We can achieve
scalable quantum computing by making $\epsilon$ smaller than the
fault-tolerant threshold. 

One way to boost the probability of success of the teleportation
protocol is to minimise the amplitudes of the $j=0$ and $j=n$ terms in
the superposition $|t_n\rangle$ of Eq.~(\ref{telaux}). At the cost of
changing the relative amplitudes (and therefore introducing a small
error in the teleported output state), the success probability of
teleporting a single qubit can then be boosted to $1-1/n^2$
\cite{franson02}. The downside of this proposal is that the errors
become less well-behaved: Instead of perfect teleportation of the
state $\alpha |0\rangle + \beta |1\rangle$ with an occasional
$\sigma_z$ measurement of the qubit, the Franson variation will yield
an output state $c_j \alpha |0\rangle + c_{j-1} \beta |1\rangle$,
where $j$ is known and the $c_j$ are the amplitudes of the modified
$|t_n\rangle$. There is no simple two-mode unitary operator that
transforms this output state into the original input state without
knowledge about $\alpha$ and $\beta$. This makes error correction much
harder. 

Another variation on the KLM scheme due to Spedalieri et al.\ (2005)
\nocite{spedalieri05} redefines the teleported qubit $\alpha|0\rangle
+ \beta|1\rangle$ and Eq.~(\ref{telaux}). The vacuum state is replaced
with a single horizontally polarised photon, $|0\rangle \rightarrow
|H\rangle$, and the one-photon state is replaced with a vertically
polarised photon, $|1\rangle \rightarrow |V\rangle$. There are now
$2n$ rather than $n$ photons in the state $|t_n\rangle$. The
teleportation procedure remains the same, except that we now count the
total number of vertically polarised photons. The advantage of this
approach is that we know that we should detect exactly $n$ photons. If
we detect $m\neq n$ photons, we know that something went wrong, and
this therefore provides us with a level of {\em error detection} (see
also section \ref{sec:error}). 

Of course, having a near-deterministic two-qubit gate is all very
well, but if we want to do arbitrarily long quantum computations, the
success probability of the gates must be close to one. Instead of
making larger teleportation networks, it might be more cost effective
or easier to use a form of error correction to make the gates
deterministic. This is the subject of the next section.

\subsection{Error correction of the probabilistic gates}\label{sec:er}

\noindent As we saw in the previous section the probability of success
of teleportation gates can be increased arbitrarily by preparing
larger entangled states. However the asymptotic behaviour to unit
probability is quite slow as a function of $n$. A more efficient
procedure is to encode against gate failure. This is possible because
of the well-defined failure mode of the teleporters. We noted in the
previous section that the teleporters fail if zero or $n+1$ photons
are detected because we can then infer the logical state of the input
qubit. In other words the failure mode of the teleporters is to
measure the logical value of the input qubit. If we can encode against
accidental measurements of this type then our qubit will be able to
survive gate failures and the probability of eventually succeeding in
applying the gate will be increased. 

KLM introduced the following logical encoding over two polarisation
qubits: 
\begin{eqnarray} |0\rangle_L & = & |HH\rangle + |VV\rangle \nonumber\\
|1\rangle_L & = & |HV\rangle + |VH\rangle
\end{eqnarray} This is referred to as {\it parity encoding} as the
logical zero state is an equal superposition of the even parity states
and the logical one state is an equal superposition of the odd parity
states. Consider an arbitrary logical qubit: $\alpha |0\rangle_L +
\beta |1\rangle_L$. Suppose a measurement is made on one of the
physical  qubits returning the result $H$. The effect on the logical
qubit is the projection: 
\begin{equation} \alpha |0\rangle_L + \beta |1\rangle_L \to \alpha
|H\rangle + \beta |V\rangle 
\label{eq:ec1}
\end{equation} That is, the qubit is not lost, the encoding is just
reduced from parity to polarisation. Similarly if the measurement
result is $V$ we have: 
\begin{equation} \alpha |0\rangle_L + \beta |1\rangle_L \to \alpha
|V\rangle + \beta |H\rangle
\label{eq:ec2}
\end{equation} Again the superposition is preserved, but this time a
bit-flip occurs. However, the bit-flip is heralded by the measurement
result and can therefore be corrected. 

Suppose we wish to teleport the logical value of a parity qubit with
the $t_1$ teleporter. We attempt to teleport one of the polarisation
qubits. If we succeed we measure the value of the remaining
polarisation qubit and apply any necessary correction to the
teleported qubit. If we fail we can use the result of the teleporter
failure (did  we find zero photons or two photons?) to correct the
remaining polarisation qubit. We are then able to try again. In this
way the probability of success of teleportation is increased from
$1/2$ to $3/4$. At this point we have lost our encoding in the process
of teleporting. However, this can be fixed by introducing the
following entanglement resource: 
\begin{equation} |H\rangle |0\rangle_L + |V\rangle |1\rangle_L 
\label{eq:ec3}
\end{equation} If teleportation is successful, the output state
remains encoded. The main observation is that the resources required
to construct the entangled state of Eq.~(\ref{eq:ec3}) are much less
than those required to construct $|t_3\rangle$. As a result, error
encoding turns out to be a more efficient way to scale up
teleportation and hence gate success. 

Parity encoding of an arbitrary polarisation qubit can be achieved by
performing a CNOT gate between the arbitrary qubit and an ancilla
qubit prepared in the diagonal state, where the arbitrary qubit is the
target and the ancilla qubit is the control. This operation has been
demonstrated experimentally \cite{obrien05}. In this experiment the
projections given by Eqs.~(\ref{eq:ec1}) and (\ref{eq:ec2}) were
confirmed up to fidelities of 96\%. In a subsequent experiment by
Pittman et al., the parity encoding was prepared in a somewhat
different manner and, in order to correct the bit-flip errors, a
feed-forward mechanism was implemented \cite{pittman05}.

To boost the probability of success further, we need to increase the
size of the code. The approach adopted by Knill, Laflamme and Milburn
\nocite{knill01} (2001) was to concatenate the code. At the first
level of concatenation the parity code states become:  
\begin{eqnarray} |0\rangle_{L}^{(4)} & = & |00\rangle_L + |11\rangle_L
\nonumber\\ |1\rangle_{L}^{(4)} & = & |01\rangle_L + |10\rangle_L 
\end{eqnarray} This is now a four-photon encoded state. At the second
level of concatenation we would obtain an eight-photon state etc. At
each higher level of concatenation, corresponding encoded
teleportation circuits can be constructed that operate with higher and
higher probabilities of success.  

If we are to use encoded qubits we must consider a universal set of
gates on the logical qubits. An arbitrary rotation about the $x$-axis,
defined by the operation $X_{\theta} = \cos{(\theta/2)} I - i
\sin{(\theta/2)} X$, is implemented on a logical qubit by simply
implementing it on one of the constituent polarisation
qubits. However, to achieve arbitrary single qubit rotations we also
require a $\pi/2$ rotation about the $z$-axis,
i.e. $Z_{\pi/2}=1/\sqrt{2}(I -i Z)$. This can be implemented on the
logical qubit by applying  $Z_{\pi/2}$ to each constituent qubit and
then applying a CZ gate between the constituent qubits. The CZ gate is
of course non-deterministic and so the $Z_{\pi/2}$ gate becomes
non-deterministic for the logical qubit. Thus both the $Z_{\pi/2}$ and
the logical CZ gate must be implemented with the teleportation gates
in order to form a universal gate set for the logical qubits. In
Ref.~\cite{knill00} it is reported that the probability of
successfully implementing a $Z_{\pi/2}$ gate on a parity qubit in this
way is $P_Z = 1-F_Z$ where 
\begin{equation} F_Z = \frac{f^2(2-f)}{1-f(1-f)}
\label{eq:ec4}
\end{equation} and $f$ is the probability of failure of the
teleporters acting on the constituent polarisation qubits. One can
obtain the probability of success after concatenation iteratively. For
example the probability of success after one concatenation is $P_{Z1}
= 1-F_{Z1}$ where $F_{Z1} = F_Z^2(2-F_Z)/(1-F_Z(1-F_Z))$. The
probability of success for a CZ gate between two logical qubits is
$P_{CZ} = (1-F_Z)^2$. Notice that, for this construction, an overall
improvement in gate success is not achieved unless $f<1/2$. Using
these results one finds that first level concatenation and $t_3$
($f=1/4$) teleporters are required to achieve a CZ gate with better
than 95\% probability of success. It can be estimated that of order
$10^4$ operations would be required in order to implement such a gate
\cite{hayes04}. 

So the physical resources for the original KLM protocol, albeit
scalable, are daunting. For linear optical quantum computing to become
a viable technology, we need more efficient quantum gates. This is the
subject of the next section.

\section{Improvements on the KLM protocol}\label{sec:imp}

\noindent We have seen that the KLM protocol explicitly tells us how
to build scalable quantum computers with single-photon sources, linear
optics, and photon counting. However, showing scalability and
providing a practical architecture are two different things. The
overhead cost of a two-qubit gate in the KLM proposal, albeit
scalable, is prohibitively large.  

If linear optical quantum computing is to become a practical
technology, we need less resource-intensive protocols. Consequently,
there have been a number of proposals that improve the scalability of
the KLM scheme. In this section we review these proposals. Several
improvements are based on cluster-state techniques
\cite{yoran03,nielsen04,browne04}, and recently a circuit-based model
of optical quantum computing was proposed that circumvents the need
for the very costly KLM-type teleportation \cite{gilchrist05}. After a
brief introduction to cluster state quantum computing, we will
describe these different proposals. 

\begin{figure}[t]
  \begin{center} \begin{psfrags} \epsfig{file=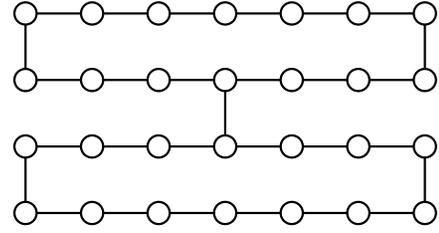, height=3cm}
  \end{psfrags} \end{center} \caption{A typical cluster state. Every
circle represents a logical qubit, and the vertices represent CZ
operations. A quantum computation proceeds by performing single-qubit
measurements on the left column of qubits, thus removing them from the
cluster and teleporting the quantum information through the cluster
state. The vertical links induce two-qubit
operations.}    \label{fig:cluster}
\end{figure}

\begin{figure}[b]
  \begin{center}
  \begin{psfrags} 
 \epsfig{file=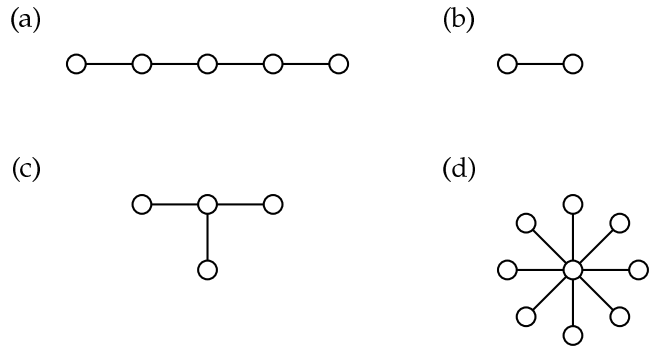}
  \end{psfrags}
  \end{center}
  \caption{Different cluster and graph states. (a) A linear cluster of
five qubits. (b) A cluster representing the Bell states. (c) A
four-qubit GHZ state. This state can be obtained by an $X$-measurement
of the central qubit in (a). (d) A general GHZ state.}
  \label{fig:graphs} 
\end{figure}

\subsection{Cluster states in optical quantum
computing}\label{sec:cluster}

\noindent In the traditional circuit-based approach to quantum
computing, quantum information is encoded in qubits, which
subsequently undergo single- and two-qubit operations. There is,
however, an alternative model, called the {\em cluster-state model} of
quantum computing \cite{raussendorf01}. In this model, the quantum
information encoded in a set of qubits is teleported to a new set of
qubits via entanglement and single-qubit measurements. It uses a
so-called cluster state in which physical qubits are represented by
nodes and entanglement between the qubits is represented by connecting
lines (see Fig.~\ref{fig:cluster}). Suppose that the qubits in the
cluster state are arranged in a lattice. The quantum computation then
consists of performing single-qubit measurements on a ``column'' of
qubits, the outcomes of which determine the basis for the measurements
on the next column. Single qubit gates are implemented by  choosing a
suitable basis for the single-qubit measurement, while two-qubit gates
are induced by local measurements of two qubits exhibiting a vertical
link in the cluster state. 

Two-dimensional cluster states, i.e., states with vertical as well as
horizontal links, are essential for quantum computing, as linear
cluster-state computing can be efficiently simulated on classical
computers \cite{nielsen05}. Since single-qubit measurements are
relatively easy to perform when the qubits are photons, this approach
is potentially suitable for linear optical quantum computing: Given
the right cluster state, we need to perform only the photon detection
and the feed-forward post-processing. Verstraete and Cirac (2004)
\nocite{verstraete04} demonstrated how the teleportation-based
computing scheme of Gottesman and Chuang could be related to
clusters. They derived their results for generic implementations and
did not address the special demands of optics.  

Before we discuss the various proposals for efficient cluster-state
generation, we present a few more properties of cluster states. Most
importantly, a cluster such as the one depicted in
Fig.~\ref{fig:cluster} does not correspond to a unique quantum state:
It represents a family of states that are equivalent up to local
unitary transformations of the qubits. More precisely, a cluster state
$|C\rangle$  is an eigenstate of a set of commuting operators $S_i$
called the {\em stabiliser generators} \cite{raussendorf03}:
\begin{equation} S_i |C\rangle = \pm |C\rangle \quad \forall i\; .
\end{equation} Typically, we consider the cluster state that is a +1
eigenstate for all $S_i$. Given a graphical representation of a
cluster state, we can write down the stabiliser generators by
following a simple recipe: Every qubit $i$ (node in the graph)
generates an operator $S_i$. Suppose that a qubit labelled $q$ is
connected to $k$  neighbours labelled $1$ to $k$. The stabiliser
generator $S_q$ for qubit $q$ is then given by 
\begin{equation} S_q = X_q \,\overset{k}{\underset{j=1}{\mbox{\Large
$\otimes$}}}\, Z_{j}\; ,
\end{equation} For example, a (simply connected) linear cluster chain
of five qubits labelled $a$, $b$, $c$, $d$, and $e$
(Fig.~\ref{fig:graphs}a) is uniquely determined by the following five
stabiliser generators: $S_a = X_a Z_b$, $S_b = Z_a X_b Z_c$, $S_c =
Z_b X_c Z_d$, $S_d = Z_c X_d Z_e$, $S_e = Z_d X_e$. It is easily
verified that these operators commute. Note that this recipe applies
to general graph states, where every node (i.e., a qubit) can have an
arbitrary number of links with other nodes. The rectangular shaped
cluster states are a subset of the set of graph states. 

Consider the following important examples of cluster and graph states:
The connected two-qubit cluster state is locally equivalent to the
Bell states (Fig.~\ref{fig:graphs}b), and a linear three-qubit cluster
state is locally equivalent to a three-qubit GHZ
(Greenberger-Horne-Zeilinger) state. These are states that are locally
equivalent to $|0,\ldots,0\rangle + |1,\ldots,1\rangle$. In general,
GHZ states can be represented by a star-shaped graph such as shown in
Figs.~\ref{fig:graphs}c and \ref{fig:graphs}d.  

To build the cluster state that is needed for a quantum computation,
we can transform one graph state into another using entangling
operations, single-qubit operations and single-qubit measurements.  A
$Z$ measurement removes a qubit from a cluster and severs all the
bonds that it had with the cluster \cite{raussendorf03,hein04}.  An
$X$ measurement on a qubit in a cluster removes that qubit from the
cluster, and it will transfer all the bonds of the original qubit to a
neighbour. All the other neighbours become single connected qubits to
the neighbour that inherited the bonds \cite{raussendorf03,hein04}.

There is a well-defined physical recipe for creating cluster or graph
states, such as the one shown in Fig.~\ref{fig:cluster}. First of all,
we prepare all qubits in the state $(|0\rangle + |1\rangle)/
\sqrt{2}$. Secondly, we apply a CZ-gate to all qubits that are to be
linked with a horizontal or vertical line, the order of which does not
matter. 

To make a quantum computer using the one-way quantum computer, we need
two-dimensional cluster states \cite{nielsen05}. Computation on linear
cluster chains can be simulated efficiently on a classical
computer. Furthermore, two-dimensional cluster states can be created
with Clifford group gates. The Gottesman-Knill theorem then implies
that the single-qubit measurements implementing the quantum
computation must include non-Pauli measurements.

It is the entangling operation that is problematic in optics, since a
linear optical CZ gate in our qubit representation is inherently
probabilistic. There have been, however, several proposals for making
cluster or graph states with linear optics and photon detection, and
we will discuss them in chronological order. 

\subsection{The Yoran-Reznik protocol}

\begin{figure}[t]
  \begin{center}
  \begin{psfrags} 
 \epsfig{file=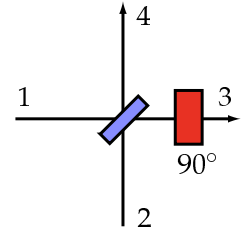}
  \end{psfrags}
  \end{center}
 \caption{Using the ``hyper-entanglement'' of the polarisation and
which-path observables, a single photon spans a four-dimensional
Hilbert space $\{ |H,1\rangle, |H,2\rangle, |V,1\rangle, |V,2\rangle
\}$. A simple 50:50 beam-splitter and polarisation rotation then
furnishes a deterministic transformation from the computational basis
to the Bell basis.}  
 \label{fig:bm-popescu}
\end{figure}

\noindent The first proposal for linear optical quantum computing
along these lines by Yoran and Reznik (2003) \nocite{yoran03} is not
strictly based on the cluster-state model, but it has many attributes
in common. Most notably, it uses ``entanglement chains'' of photons in
order to pass the quantum information through the circuit via
teleportation.  

\begin{figure}[t]
  \begin{center}
  \begin{psfrags} 
\epsfig{file=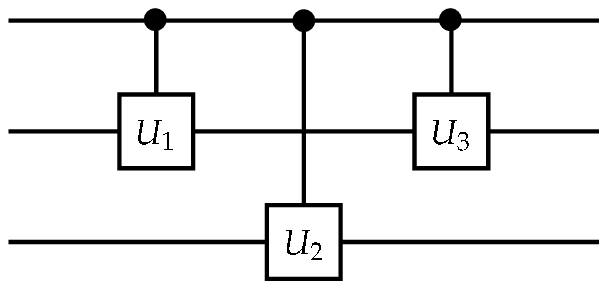}
  \end{psfrags}
  \end{center}
 \caption{A typical three-qubit quantum computational circuit.}
 \label{fig:circuit}
%
  \begin{center}
  \begin{psfrags} 
\epsfig{file=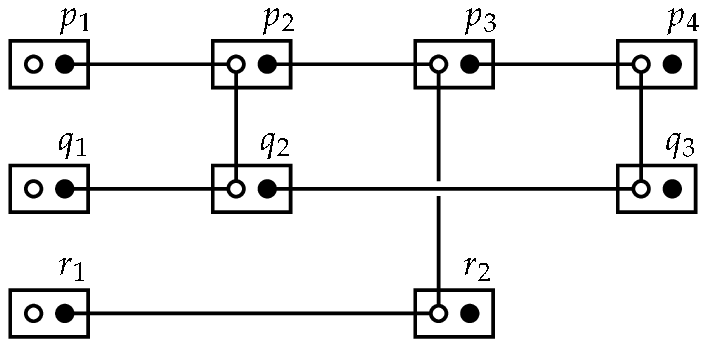}
  \end{psfrags}
  \end{center}
 \caption{The computational circuit of Fig.~\ref{fig:circuit} in terms
of the physical implementation by \cite{yoran03}. This is reminiscent
of the cluster-state model of quantum computing. The open and closed
dots represent the polarisation and which-path degrees of freedom,
respectively.} 
 \label{fig:circuit-yr}
\end{figure}

First of all, for this protocol to work, the nondeterministic nature
of optical teleportation must be circumvented. We have already
remarked several times that complete (deterministic) Bell measurements
cannot be performed in the dual-rail and polarisation qubit
representations of linear optical quantum computing. However, in a
different representation this is no longer the case. Instead of the
traditional dual-rail implementation of qubits, we can encode the
information of two qubits in a single photon when we include both the
polarisation and the spatial degree of freedom. Consider the device
depicted in Fig.~\ref{fig:bm-popescu}. A single photon carrying
specific polarisation and path information is then transformed as
\cite{popescu95}:   
\begin{eqnarray}\label{popbellmeas} |H,1\rangle & ~\rightarrow~ &
\frac{1}{\sqrt{2}} \left( |V,3\rangle + |H,4\rangle \right) \cr
|V,1\rangle & ~\rightarrow~ & \frac{1}{\sqrt{2}} \left( |V,4\rangle -
|H,3\rangle \right) \cr |H,2\rangle & ~\rightarrow~ &
\frac{1}{\sqrt{2}} \left( |H,4\rangle - |V,3\rangle \right) \cr
|V,2\rangle & ~\rightarrow~ & \frac{1}{\sqrt{2}} \left( |V,4\rangle +
|H,3\rangle \right)\; . 
\end{eqnarray} These transformations look tantalisingly similar to the
transformation from the computational basis to the Bell
basis. However, there is only one photon in this system. The second
``qubit'' is given by the which-path information of the input
modes. By performing a polarisation measurement of the output modes
$3$ and $4$, we can project the input modes onto a ``Bell
state''. This type of entanglement is sometimes called {\em
hyper-entanglement}, since it involves more than one observable of a
single system \cite{kwiat98,barreiro05,cinelli05}. A teleportation
experiment based on this mechanism was performed by Boschi { et al}.\
(1998). \nocite{boschi98}

It was shown by Yoran and Reznik how these transformations can be used
to cut down on the number of resources: Suppose we want to implement
the computational circuit given in Fig.~\ref{fig:circuit}. We will
then create (highly entangled) chain states of the form 
\begin{eqnarray}\label{eq:chain} && (\alpha |H\rangle_{p_1} + \beta
|V\rangle_{p_1})(|1\rangle_{p_1} |H\rangle_{p_2} + |2\rangle_{p_1}
|V\rangle_{p_2}) \times \ldots \cr && \times (|2n-1\rangle_{p_n}
|H\rangle_{p_{n+1}} + |2n\rangle_{p_{n}} |V\rangle_{p_{n+1}})
|2n+1\rangle_{p_{n+1}}, \cr &&
\end{eqnarray} where the individual photons are labelled by $p_j$.
This state has the property that a Bell measurement of the form of
Eq.~(\ref{popbellmeas}) on the first photon $p_1$ will teleport the
input qubit $\alpha |H\rangle + \beta |V\rangle$ to the next photon
$p_2$. 

\begin{figure}[t]
  \begin{center}
  \begin{psfrags} 
\epsfig{file=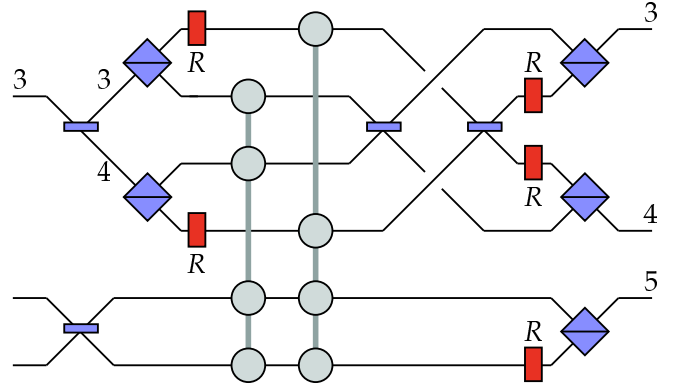}
  \end{psfrags}
  \end{center}
 \caption{How to add a link to the YR chain. This will create the
state given in Eq.~(\ref{eq:chain}) with $n+1 = 5$. The four
vertically connected grey circles represent the probabilistic CZ
gate. Note that we need two of them.} 
 \label{fig:add-yr}
\end{figure}

Let's assume that we have several of these chains running in parallel,
and that furthermore, there are vertical ``cross links'' of
entanglement between different chains, just where we want to apply the
two-qubit gates $U_1$, $U_2$, and $U_3$. This situation is sketched in
Fig.~\ref{fig:circuit}. The translation into optical chain states is
given in Fig.~\ref{fig:circuit-yr}. The open circles represent
polarisation, and the dots represent the path degree of freedom. In
Fig.~\ref{fig:add-yr}, the circuit that adds a link to the chain is
shown. The unitary operators $U_1$, $U_2$, and $U_3$ are applied to
the polarisation degree of freedom of the photons. 

Note that we still need to apply two probabilistic CZ gates in order
to add a qubit to a chain. However, whereas the KLM scheme needs the
teleportation protocol to succeed with very high probability (scaling
as $n^2/(n+1)^2$) in the protocol proposed by Yoran and Reznik the
success probability of creating a link in the chain must be larger
than one half. This way, the entanglement chains grow {\em on
average}. This is a very important observation and plays a key role in
the protocols discussed in this section. Similarly, a vertical link
between the entanglement chains  can be established with a two-qubit
unitary operation on the polarisation degree of  freedom of both
photons (c.f. the vertical lines between the open dots in
Fig.~\ref{fig:circuit-yr}). If the gate fails, we can grow longer
chains and try again until the gate succeeds. 

\subsection{The Nielsen protocol}\label{sec:nielsen}

\noindent
A more explicit use of cluster-state quantum computing was made by
Nielsen (2004). As Yoran and Reznik, Nielsen recognised that in order to
build cluster states, the probability of adding a link to the cluster
must be larger than one half, rather than arbitrarily close to one. Otherwise
the cluster will shrink on 
average.  The KLM teleportation protocol allows us to apply a
two-qubit gate with probability $n^2/(n+1)^2$, depending on  the
number $n$ of ancillary photons. Let us denote a CZ gate with this success
probability by $CZ_{n^2/(n+1)^2}$. This gate can be
used to add qubits to a cluster chain. When the gate fails, it removes
a qubit from the cluster. This means that, instead of using very large
$n$ to make the CZ gate near-deterministic, links can be added on
average with a modest $CZ_{9/16}$-gate, or $n=3$. This leads to
similarly reduced resource requirements as the Yoran-Reznik protocol,
while still keeping (in principle) error-free quantum
computing. However, there is an extra gain in resources available when
we try to add a qubit to a chain \cite{nielsen04}.

Suppose that we wish to add a single qubit to a cluster chain via the
teleportation-based CZ gate. Instead of teleporting the two qubits
simultaneously, we {\em first} teleport the disconnected qubit, and
{\em secondly} teleport the qubit at the end of the cluster. We know
that a teleportation failure will remove the qubit from the cluster,
so we attempt the second teleportation protocol {\em only after the
first has succeeded}. The first teleportation protocol then becomes
part of the off-line resource preparation, and the CZ gate effectively
changes from $CZ_{n^2/(n+1)^2}$ to $CZ_{n/(n+1)}$. The growth
requirement of the cluster state then becomes $n/(n+1) > 1/2$, or
$n=2$, and we make another substantial saving in resources.

Apart from linear cluster states, we also need the ability to make
the two-dimensional clusters depicted in Fig.~\ref{fig:cluster}. This
is equivalent to linking a qubit to two cluster chains, and hence
needs two successful CZ gates. Arguing along the same lines as before,
it is easily shown that the success probability is 4/9 for this procedure
using two ancill\ae\ per teleportation gate. Since this is
smaller than one half, this procedure on average {\em removes} qubits
from the cluster. However, we can first {\em add} extra qubits with
the previous procedure, such that there is a buffer of qubits in the
cluster state. This way, the average shrinkage of the cluster due to
vertical links is absorbed by the buffer region. 

Finally, Nielsen introduces so-called {\em micro-clusters} consisting
of multiple qubits connected to the end point of a cluster chain. Such
a micro-cluster is depicted in Fig.~\ref{fig:graphs}d, where the
central qubit is an endpoint of a cluster chain. Having such a fan of
qubits at the end of a chain, we can retry the entangling gate as many 
times as there are ``dangling'' qubits. This removes the lower limit
on the success probability of the CZ gate at the cost of making large
GHZ states \cite{nielsen04b}. Therefore, {\em any} optical two-qubit
gate with arbitrary success probability $p$ can be used to make
cluster states efficiently. 

\subsection{The Browne-Rudolph protocol}\label{sec:br}

\noindent
There is still a cheaper way to grow cluster states. In order for a
cluster chain to grow on average without using expensive micro-clusters, the
success probability of adding a 
single qubit to the chain must be larger than one half. However, if we
can add small chains of qubits to the cluster, this requirement may be
relaxed. Suppose that the success probability of creating a link
between two cluster chains is $p$, and that in each successful linking
of two chains we lose $d_s$ qubits from the chain. This might happen
when the entangling operation joining the two clusters involves the
detection of qubits in the cluster. Similarly, in an unsuccessful
attempt, we may lose $d_f$ qubits from the existing cluster chain (we
do not count the loss of qubits in the small chain that is to be
added). If our existing cluster chain has  length $N$, and the chain
we wish to add has length $m$, then we can formulate the following
{\em growth requirement} \cite{barrett05,browne04}:  
\begin{eqnarray}
 && p (N+m-d_s) + (1-p) (N-d_f) > N  \cr
 &\Leftrightarrow& m > \frac{p\, d_s + (1-p) d_f}{p}\; . 
\end{eqnarray}
Given a specific strategy $(d_s,d_f)$ and success probability $p$, we
need to create chains of length $m$ off-line in order to make large
cluster chains efficiently. Note that, again, there is no lower limit
to the success probability $p$ of the entangling operation in
principle. This allows us to choose the optical gates with the most
desirable physical properties (other than high success probability),
and it means that we do not have to use the expensive and error-prone
$CZ_{n/(n+1)}$ gates.  

\begin{figure}[t]
  \begin{center}
  \begin{psfrags}
       \epsfig{file=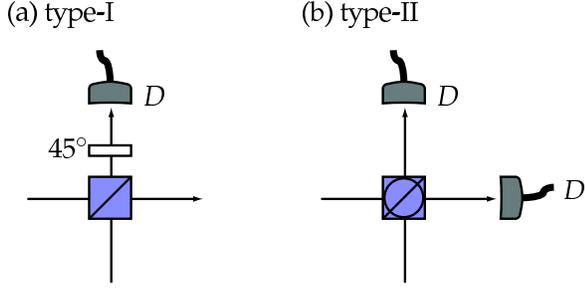}
  \end{psfrags}
  \end{center}
 \caption{Two types of fusion operators. (a) the type-I fusion operator
  employs a polarisation beam splitter (PBS1) followed by the detection
  $D$ of a single output mode in the 45$^{\circ}$ rotated polarisation
  basis. This operation determines the parity of the input mode with
  probability 1/2. (b) the type-II fusion operator uses a diagonal
  polarisation beam splitter (PBS2), detects both output modes, and
  projects the input state onto a maximally entangled Bell state with
  probability 1/2.}  
 \label{fig:fusion}
\end{figure}

Indeed, Browne and Rudolph introduced a protocol for generating cluster
states using the probabilistic parity gates of section \ref{sec:parity}
\cite{browne04,cerf98,pittman01}. The notable advantage of this gate
is that it is relatively easy to implement in practice
\cite{pittman02b}, and that it can be made robust against common
experimental errors. Initially these gates were called {\em parity gates}, but
following Browne and Rudolph, we call these the
type-I and type-II {\em fusion} gates (see Fig.~\ref{fig:fusion}). 

Let us first consider the operation of the type-I fusion gate in
Fig.~\ref{fig:fusion}a. Given the detection of one and only one photon
with polarisation $H$ or $V$ in the detector $D$, the gate induces the
following projection on the input state:
\begin{eqnarray}
 \text{``H''}: && \frac{1}{\sqrt{2}} \left( |H\rangle\langle H,H| -
 |V\rangle\langle V,V| \right) \cr
 \text{``V''}: && \frac{1}{\sqrt{2}} \left( |H\rangle\langle H,H| +
 |V\rangle\langle V,V| \right) .
\end{eqnarray}
It is easily verified that the probability of success for this gate is
$p=1/2$. When the type-I fusion gate is applied to two photons
belonging to two different cluster states containing $n_1$ and $n_2$
photons, respectively, a successful operation will generate a cluster
chain of $n_1 + n_2 -1$ photons. However, when the gate fails, it
effectively performs a $\sigma_z$ measurement on both photonic qubits,
and the two cluster states both lose the qubit that was detected. The
type-I fusion gate is therefore a $(1,1)$-strategy, i.e.,
$d_s=d_f=1$ (recall that we count only the loss of qubits on one
cluster to determine $d_f$). The ideal growth requirement is
$m>1/p=2$.   

Browne and Rudolph also introduced the type-II fusion operator (see
Fig.~\ref{fig:fusion}b). This operation involves the photon detection
of both output modes of a polarisation beam splitter, and a successful
event is heralded by a detector coincidence (i.e., one photon with a
specific polarisation in each detector). When successful, this gate
projects the two incoming qubits onto one of two polarisation Bell
states, depending on the detection outcome:  
\begin{eqnarray}
 \text{``H,V'' or ``V,H''}: && \frac{1}{\sqrt{2}} \left( |H,H\rangle +
  |V,V\rangle \right) \cr  
 \text{``H,H'' or ``V,V''}: && \frac{1}{\sqrt{2}} \left( |H,V\rangle +
  |V,H\rangle \right) .
\end{eqnarray} 
The success probability of this gate is $p=1/2$, and it is a
$(2,1)$-strategy (i.e., $d_s=2$ and $d_f=1$). The ideal growth
requirement is thus $m>(1+p)/p = 3$. The type-II fusion gate is
essentially a version of the incomplete optical Bell measurement 
\cite{braunstein96,weinfurter94}. 

Note that in order to grow long chains, we must be able to create
chains of three qubits. Given a plentiful supply of Bell pairs as our
fundamental resource, we can make three-qubit chains only with the
type-I fusion gate, since the type-II gate necessarily destroys two
qubits. This also indicates a significant difference between this
protocol and the previous ones: Using only single-photon sources, the
fusion gates alone cannot create cluster states. We can, however, use
any method to create the necessary Bell pairs (such as the CZ gate in 
section \ref{elga}), as they constitute an off-line resource.

Upon successful operation, both the type-I and the type-II fusion
gates project two qubits that are part of a cluster onto a polarisation
Bell state. When we apply a Hadamard operation to one of the qubits
adjacent to the detected qubit(s), the result will again be a cluster
state. However, upon failure the characteristics of the fusion gates
are quite different from each other. When the type-I gate fails, it
performs a $Z$ measurement on the input qubits. When the type-II gate
fails, it performs an $X$ measurement on the input qubits. Recall that
there is a fundamental difference between a $Z$ and an $X$ measurement
on qubits in cluster states: A $Z$ measurement will break all bonds
with the qubits neighbours and removes it from the cluster. An $X$
measurement will also remove the qubit from the cluster, but it will
join its neighbours into a {\em redundantly encoded} qubit. In terms of
the graphs, this corresponds to a qubit with dangling bonds called {\em
leaves} or {\em cherries} (see also Fig.~\ref{fig:graphs}c).

When the measured qubits are both end points of cluster states (i.e.,
they have only one link to the rest of the cluster), failing type-I
and type-II fusion gates have  similar effects on the cluster states:
They remove the qubits from the cluster. However, when the fusion
gates are applied to two qubits inside a cluster (i.e., the qubits
have two or more links to other qubits in the cluster), then the
failure modes of the two fusion gates differ dramatically: In particular
when we apply the fusion gate to a qubit in a chain, a failed type-I
gate will break the chain, while a failed type-II gate will only
shorten the chain and create one redundantly encoded qubit next to the
measured qubit. Since it is costly to re-attach a broken chain, it is
best to avoid the type-I gate for this purpose. The redundancy induced
by a failed type-II fusion gate is closely related to the error
correction model in section \ref{sec:error}. We will explore this
behaviour further in the next section. 

Again, we need at least two-dimensional cluster states in order to
achieve the level of quantum computing that cannot be simulated
efficiently on a classical computer. Using the failure behaviour of
the type-II fusion gate, we can construct an efficient way of creating
the vertical links between linear cluster chains. We attempt a type-II
fusion between two qubits that are part of different chains. If the
gate succeeds, we have created a vertical link on the neighbouring
qubits. If the gate fails, one neighbouring qubit to each detected
qubit becomes redundantly encoded. The type-I fusion can now be
attempted once more on the redundantly encoded qubits. If the gate
succeeds, we established a vertical link. If the gate fails, we end up
with two disconnected chains that are both two qubits shorter. Given
sufficiently long linear cluster chains, we can repeat this protocol
until we have succeeded in creating a vertical link. A
proof-of-principle experiment demonstrating optical cluster-state
quantum computing with four photons was performed in Vienna
\cite{walther05}.

\subsection{Circuit-based optical quantum computing
revisited}\label{sec:circuit} 

\begin{figure}[t]
 \begin{center}
  \begin{psfrags}
     \epsfig{file=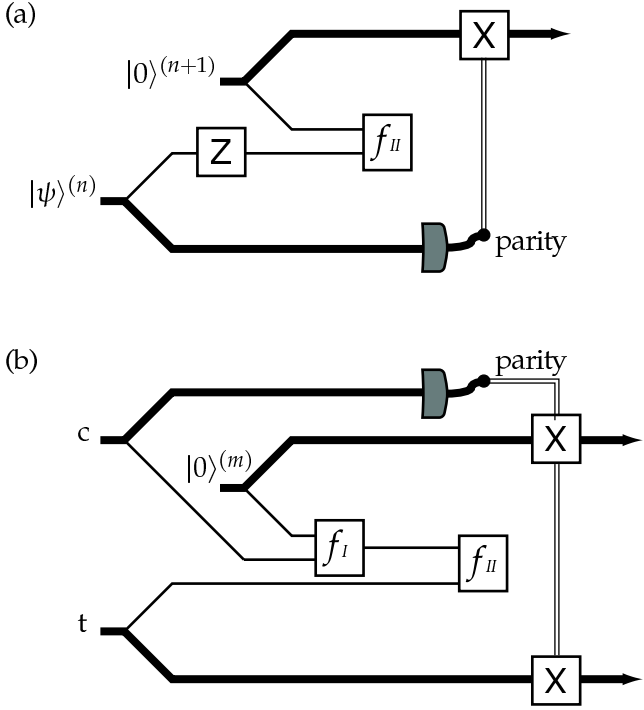}
  \end{psfrags}
  \end{center}
  \caption{The two probabilistic gates that complete a universal
  set. (a) the $Z_{\pi/2}$ gate uses a deterministic single-photon
  rotation and a single type-II fusion gate. (b) the CNOT gate uses one
  type-I and one type-II fusion gate. Both gates also need a parity
  measurement, which is implemented by $\sigma_z$ measurements on the
  individual physical qubits. }
  \label{fig:gates-ghr}
\end{figure}

\noindent
After all this, one might conclude that the cluster-state approach to
linear optical quantum information processing has completely replaced
the circuit-based model. However, such a conclusion would be
premature. In fact, in a slightly altered form, the redundancy that we
encountered in the Browne-Rudolph protocol can be used to make a
scalable circuit-based optical quantum computer
\cite{hayes04,gilchrist05}. We will now show how this is done. 

We can encode a logical qubit in $n$ physical qubits using the
{\em parity encoding} we encountered earlier in section \ref{sec:er}:
\begin{eqnarray}\label{eq:rhg}
 |0\rangle^{(n)} &\equiv& \frac{1}{\sqrt{2}} \left( |+\rangle^{\otimes
  n} + |-\rangle^{\otimes n} \right) \cr 
 |1\rangle^{(n)} &\equiv& \frac{1}{\sqrt{2}} \left( |+\rangle^{\otimes
  n} - |-\rangle^{\otimes n} \right) ,
\end{eqnarray} 
where $|\pm\rangle = (|H\rangle \pm |V\rangle)/\sqrt{2}$. The
superscript $(n)$ denotes the level of encoding. This encoding has the
attractive property that a computational-basis measurement of one of
the physical qubits comprising the logical qubit $|\psi\rangle^{(n)}$
will yield $|\psi\rangle^{(n-1)}$, up to a local unitary on a single
(arbitrary) physical qubit. In other words, no quantum information has
been lost \cite{gilchrist05}. 

One way to generate this encoding is to use the type-II fusion operator
without the two polarisation rotations (half-wave plates) in the input
ports of the polarising beam splitter. This yields
\begin{equation}
  f_{II} |\psi\rangle^{(n)}|0\rangle^{(m)}\rightarrow\left\{ 
  \begin{array}{cl}
    |\psi\rangle^{(n+m-2)} & \mbox{(success)}\\
    |\psi\rangle^{(n-1)} |0\rangle^{(m-1)} & \mbox{(failure),}
  \end{array}\right.
\end{equation}
with $|\psi\rangle^{(n)} \equiv \alpha |0\rangle^{(n)} + \beta
|1\rangle^{(n)}$. From this, we can immediately deduce that, given a
success probability $p$, the growth requirement for the redundancy
encoding is $m>(1+p)/p$. This is exactly the same scaling behaviour as
the Browne-Rudolph protocol when clusters are grown with the type-II
fusion gate.

In order to build circuits that are universal for quantum computing,
we need a set of single-qubit operations, and at least one two-qubit
entangling gate at the level of the parity encoded quits. As we have
seen in section \ref{sec:er}, we can perform 
the operations $X_{\theta}$ and $Z$ deterministically: The operator
$X_{\theta} \equiv \cos(\theta/2) \unity + i\sin(\theta/2) \sigma_x$
can be implemented by applying this single-qubit operator to {\em only
one} physical qubit of the encoding. The $Z$ gate for an encoded qubit
corresponds to a $\sigma_z$ operation on {\em all} the physical qubits.

To complete the universal set of gates, we also need the single-qubit
gate $Z_{\pi/2}$ and the CNOT. These gates cannot be implemented
deterministically on parity-encoded qubits. In
Fig.~\ref{fig:gates-ghr} we show how to implement these gates using
the fusion operators. The thickness of the lines denotes the level of
encoding. In addition to the fusion operators, we need to perform a
parity measurement on one of the qubits by doing $\sigma_z$
measurements on all the physical qubits in a circuit line. Since this
measurement is performed on a subset of the physical qubits comprising
a logical qubit, no quantum information is lost in this procedure. It
should also be noted that we attempt the probabilistic fusion gates
before the destructive parity measurement, so that in the case of a
failed fusion operation, we still have sufficient redundancy to try
the fusion again. It has been estimated that universal gate operations
can be implemented with $>99\%$ probability of success with about
$10^2$ operations \cite{gilchrist05}. 

This circuit-based protocol for linear optical quantum computation has
many features in common with the Browne-Rudolph protocol. Although the
cluster-state model is conceptually quite different from the
circuit-based model, they have similar resource requirements. Another
point that the reader might be wondering about, is whether these
schemes are tolerant to photon loss and other practical noise. The
errors that we discussed so far originate from the probabilistic
nature of linear optical photon manipulation, but can we also correct
errors that arise from, e.g., detection inefficiencies? We will
discuss the realistic errors of linear optical component in the next
section, and the possible fault tolerance of LOQC in the presence of
these errors in section \ref{sec:error}.

\section{Realistic optical components and their errors}\label{sec:real}

\noindent
In order to build a real quantum computer based on linear optics,
single-photon sources, and photon detection, our design must be able to
deal with errors: The unavoidable errors in practical implementations
should not erase the quantum information that is present in the
computation. We have already seen that the teleportation trick in the
KLM scheme employs error correction to turn the non-deterministic
gates into near-deterministic gates. However, this assumes that the
photon sources, the mode matching of the optical circuits, and the
photon counting are all perfect. In the real world, this is far from
true. 

What are the types of errors that can occur in the different stages of
the quantum computation? We can group them according to the optical
components: detection errors, source errors, and circuit errors
\cite{takeuchi00b}. In this section we will address these errors. In
addition, we will address an assumption that has received little
attention thus far: the need for quantum memories. 

\subsection{Photon detectors}\label{sec:detectors}

\noindent
In linear quantum optics, the main method for gaining information
about the quantum states is via photon detection. Theoretically, we can
make a distinction between at least two types of detectors: ones that
tell us exactly how many photons there are in an input state, and ones
that give a binary output ``nothing'' or ``many''. There are many more
possible distinctions between detectors, but these two are the most
important. The first type is called a {\em number-resolving} detector
or a detector with {\em single-photon resolution}, while the second
type is often called a {\em bucket} or {\em vacuum} detector. The
original KLM proposal relies critically on the availability of
number-resolving detectors. On the other hand, typical photon detectors
in LOQC experiments are bucket detectors. In recent years there has
been a great effort to bridge the gap between the requirements of LOQC
and the available photon detectors, leading to the development of
number-resolving detectors and LOQC protocols that rely less on high
photon-number counting. In this section, we state the common errors
that arise in realistic photon detection and review some of the
progress in the development of number-resolving detectors. 

Real photon detectors of any kind give rise to two different types of
errors: 
\begin{enumerate}
 \item The detector counts fewer photons than were actually present in
   the input state. This is commonly known as {\em photon loss};
 \item the detector counts more photons than were actually present in
   the input state. These are commonly known as {\em dark counts}.
\end{enumerate}
Observe that it is problematic to talk about the number of photons
``that were actually present'' in the input state: When the input
state is a superposition of different photon number states, the photon
number in the state prior to detection is ill-defined. However, we
{\em can} give a general meaning to the concepts of photon loss and
dark counts for arbitrary input states when we define the loss or dark
counts as a property of the detector (i.e., independent of the input
state). The {\em detector efficiency} $\eta \in [0,1]$ can be defined
operationally as the probability that a single photon input state will
result in a detector count, while the dark counts can be defined as
the probability that a vacuum input state will result in a detector
count. Subsequently, these definitions can be modified to take into
account non-poissonian errors.  

Whereas perfect number-resolving detectors can be modelled using the
projection operators onto the Fock states $|n\rangle\langle n|$,
realistic detectors give rise to Positive Operator Valued Measures, or
POVMs. A standard photon loss model is to have a perfect detector be
preceded by a beam splitter with transmission coefficient $\eta$ and
reflection coefficient $1-\eta$. The reflected mode is considered lost
(mathematically, this mode is traced over), so only a fraction $\eta$
of the input reaches the detector. In this model, every incoming
photon has the same probability of being lost, leading to Poissonian
statistics. The POVM for a number-resolving photon detector
corresponding to this model is \cite{scully69} 
\begin{equation}
  \hat{E}_n = \sum_{k=n}^{\infty} \left({k}\atop{n}\right) \eta^{n}
  (1-\eta)^{k-n}\; |k\rangle\langle k|\; . 
\end{equation}
Using the same loss model, the POVM describing the effect of a bucket
detector is \cite{kok00}  
\begin{eqnarray}
  \hat{E}_0 &=& \sum_{n=0}^{\infty}\; (1-\eta)^n\; |n\rangle\langle n| \cr
  \hat{E}_1 &=& \sum_{n=0}^{\infty}\; [1-(1-\eta)^n]\; |n\rangle\langle
  n|\; ,
\end{eqnarray}
where 1 and 0 denote a detector click and no detector click,
respectively. For an analysis including dark counts, see Lee et al.\
\nocite{lee04} (2004a).

Currently, the most common detectors in experiments on LOQC are {\em
Avalanche Photo-Diodes} (APDs). When a photon hits the active
semi-conductor region of an APD, it will induce the emission of an
electron into the conductance band. This electron is subsequently accelerated
in an electric potential, causing an 
avalanche of secondary electrons. The resulting current tells us that
a photon was detected. The avalanche must be stopped by reversing the
potential, which leads to a dead time of a few nanoseconds in
the detector. Any subsequent photon in the input mode can therefore
not be detected, and this means that we have a bucket detector. A
typical (unfiltered) detector efficiency for such a detector is 85\%
at a wavelength of 660~nm. Dark counts can be made as low as $6\cdot 10^3$~Hz
at room temperature and around $25~$Hz at cryogenic temperatures.

Several attempts have been made to create a number resolving detector
using only bucket detectors and linear optics, but no amount of linear
optics and bucket detection can lead to perfect, albeit inefficient,
single-photon resolution \cite{kok03b}. On the other hand, we can create
approximate number-resolving detectors using only bucket detectors
via {\em detector cascading}. In this setup, the incoming optical mode 
is distributed equally over $N$ output modes, followed by bucket
detection. When the number of modes in the cascade is large compared
to the average photon number in the input state, and the detector
efficiencies of the bucket detectors are relatively high, then good
fidelities for the photon number measurement can be obtained
\cite{kok01b,rohde05a}. Detector cascading in the time-domain using 
increasingly long fibre delays is called {\em time multiplexing}
\cite{fitch03,achilles03,banaszek03}. However, the fibre length (and
hence the detection time) must increase exponentially for this
technique to work. An alternative way to create number-resolving
detectors is to use photon-number assisted homodyne detection
\cite{nemoto02a,branczyk03}. When an (imperfect) quantum copier is available,
extra information can be extracted from the qubits (Deuar and Munro, 2000a,
2000b). \nocite{deuar00a} \nocite{deuar00b}

Fully-fledged number-resolving photon detectors are also being
developed, such as the Visible Light Photon Counter (VLPC)
\cite{takeuchi99,kim99}. An excellent recent introduction to this
technology is given by Waks et al.\ (2003). \nocite{waks03} The VLPCs
operate at a temperature of a few Kelvin in order to minimise dark
counts. They consist of an active area that is divided into many
separate active regions. When a photon  triggers such a region, it is
detected while leaving the other regions fully operational. Once a
region has detected a photon, it experiences a dead time in which no
photon detection can take place. Multiple photon detections in different
regions then generate a current that is proportional to the number of
photons. The VLPC is thus effectively a large detector cascade ($N\approx
10^4$) with high detection efficiency ($\approx 88\%$ at 694~nm). The dark
count rate of $2\cdot 10^4$~Hz is about an order of magnitude higher
than the dark count rate for off-the-shelf APDs.

An alternative technique uses a superconducting transition-edge sensor
that acts as a calorimeter. It measures the rise in temperature of an
absorber, which is quickly heated by incoming photons in the visible
light and near infra-red \cite{rosenberg05}. This device operates at
temperatures well below 100~mK, and has a measured detection
efficiency greater than $88\%$. The dark counts are negligible, but the
repetition rate is rather slow (of order 10~kHz) due to the cooling
mechanism after a photon has been detected. In addition to these
experimental schemes, there are theoretical proposals for
number-resolving detectors involving atomic vapours \cite{james02},
electromagnetically induced transparency \cite{imamoglu02},
and resonant nonlinear optics \cite{johnsson03}.  

Finally, we briefly mention quantum non-demolition (QND)
measurements. In the photon detectors that we described so far, the
state of the electromagnetic field is invariably destroyed by the
detector. However, in a QND measurement there is a freely propagating
field mode after the measurement. In particular, the outcome of the
QND measurement faithfully represents the state of the field after
detection \cite{grangier98}. Several schemes for single-photon QND
measurements have been proposed, either with linear optics
\cite{howell00e,kok02a}, optical quantum relays \cite{jacobs02}, or other
implementations \cite{brune90,brune92,roch97,munro05}. The experimental
demonstration of a single-photon QND was reported by Nogues et al.\ (1999)
using a cavity QED system, \nocite{nogues99} and a linear-optical QND
measurement was performed by Pryde et al.\ \nocite{pryde04}
(2004). However, this last experiment has led to a controversy about
the nature of the fidelity measure that was used (see
Kok and Munro, 2005, and Pryde et al.\ 2005). \nocite{kok05}
\nocite{pryde05} 

So far, we have considered only photon-number detection. However, in many
implementations of LOQC the qubit is encoded in a single polarised
photon. A qubit detector must therefore extract the polarization of the
photon, which may have had unwanted interactions with the
environment. A change in the polarization of the photon will then
induce an error in the computational circuit.

One mechanism that leads to errors in the polarization is inherent in
{\em any} photon detector, and deserves a special mention here. In the
Coulomb gauge, the polarization is perpendicular to the direction of
propagation, and the plane of detection must therefore be perpendicular
to the Poynting vector $\vec{k}$. A complication
arises when we consider beams that are not perfectly collimated. We
can write the $\vec{k}$-vector of the beam as  
\begin{equation}
 \vec{k}(\theta,\phi) = (\sin\theta\cos\phi,\sin\theta\sin\phi,\cos\theta) .
\end{equation} 
A realistic, reasonably well-collimated beam will have a narrow distribution
of $\theta$ and $\phi$ around $\theta_0$ and $\phi_0$. If we model the
active area of 
a detector as a flat surface perpendicular to $\vec{k}(\theta_0,\phi_0)$,
some modes in the beam will hit the detector at an angle. Fixing the 
gauge of the field in the detection plane then causes a mixing of left- and
right-handed polarization. This introduces a detection error that is {\em
fundamental}, since the uncertainty principle prevents the  transverse
momentum in a beam from being exactly zero \cite{peres03}. At first sight
this effect might seem negligible, but later we will see that concatenation of
error correcting codes will amplify small errors. It is therefore important to
identify all possible sources of errors.

\subsection{Photon sources}

\noindent
The LOQC protocols described in this review all make critical use of perfect
single-photon sources. In this section we wish to make more precise what is
meant by a single-photon source. We have thus far considered interferometric
properties of monochromatic plane waves with exactly one field
excitation. Such states, while a useful heuristic, are not physical.  Our
first objective is to give a general description of a single-photon state
followed by a description of current experimental realisations.  

The notion of a single photon conjures up an image of a single particle-like
object localised in space and time. However it was conclusively demonstrated
long ago by Newton and Wigner \nocite{newton49} (1949), and also Wightman
\nocite{wightman62} (1962), that a single photon cannot be localised in the
same sense that a single massive particle can be localised. Here we
are only concerned with temporal localisation, which is ultimately due to the
fact that the energy spectrum of the field is bounded from below. In this
section we take a simpler operational view. A photon refers to a single
detection event in a counting time window $T$. A single photon source leads
to a periodic sequence of single detection events with one and only one,
photon detected in each counting window. Further refinement of this
definition, via the output counting statistics of interferometers,  is needed
to specify the kind of single-photon sources necessary for LOQC.
  
Consider a one-dimensional cavity of length $L$. The allowed wave vectors for
plane wave modes form a denumerable set given by $k_n= n\pi/L$, 
with corresponding frequencies $\omega_n=ck_n$. If we measure time in units of
$\pi L/c$, the allowed frequencies may simply be denoted by an integer
$\omega_n = n \in \Bbb{N}$. Similarly if we measure length in units
of $\pi/L$, 
the allowed wave vectors are also integers.  We are primarily interested in
multi mode fields with an optical carrier frequency, $\Omega\gg 1$.  We define
the positive-frequency field component as, 
\begin{equation}
 \hat{a}(t)=\sum_{n=1}^{\infty} \hat{a}_n e^{-int}.
\label{dynamics}
\end{equation}
The bosonic annihilation and creation operators are given by
Eq.~(\ref{eq:commutation}).  From this point on we assume the detector is
located at $x=0$ and thus 
evaluate all fields at the spatial origin. Following the standard theory of
photo-detection, the probability per unit time for detecting a single photon is
given by 
\begin{equation}
 p_1(t)=\eta n(t),
\end{equation}
where
\begin{equation}
 n(t)=\langle \hat{a}^\dagger (t) \hat{a}(t)\rangle,
\end{equation}
and the parameter $\eta$ characterises the detector.   

A single-photon state may be defined as 
\begin{equation}
|1;f\rangle=\sum_{m=1}^\infty f_m \hat{a}_m^\dagger |0\rangle,
\end{equation}
where $|0\rangle=\prod_m|0\rangle_m$ is the multi-mode global vacuum state, and
we require that the {single-photon amplitude} $f_m$ satisfies 
\begin{equation}
\sum_{m=0}^\infty |f_m|^2=1.
\end{equation}
The counting probability is then determined by
\begin{equation}
n(t)=\left |\sum_{k=1}^\infty f_k e^{-ikt}\right |^2.
\end{equation}
This function is clearly periodic with a period $2\pi$.
As the spectrum is bounded from below by $n=1$, it is not possible to choose
the amplitudes $f_n$ so that the functions $n(t)$ have arbitrarily narrow
support on $t\in [0,2\pi)$.  

\begin{figure}[t]
  \begin{center}
  \begin{psfrags}
       \epsfig{file=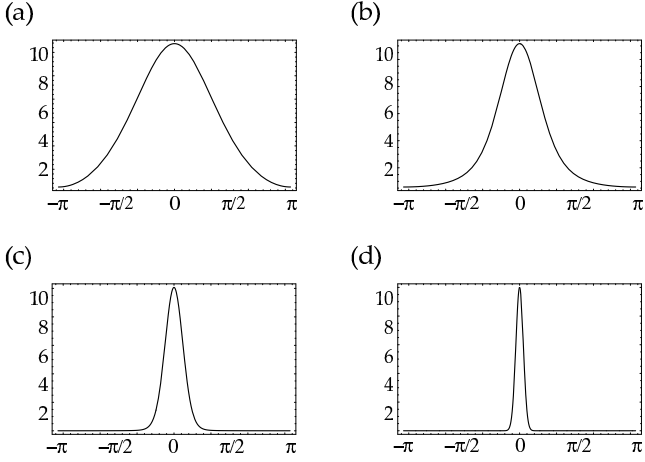}
  \end{psfrags}
  \end{center}
 \caption{The function $n(t)$ in arbitrary units in the domain $-\pi \leq t
   \leq \pi$ for 
   different values of $\mu$ and $N=100$. (a) $\mu = 0.001$; (a) $\mu =
   0.001$; (b) $\mu = 0.01$; (c) $\mu = 0.05$; (d) $\mu = 0.49$.} 
 \label{fig:mu}
\end{figure}

As an example we take
\begin{equation}
f^N_m=\frac{1}{\sqrt{1-(1-\mu)^N}}{N\choose m}^{1/2}\mu^{m/2}(1-\mu)^{(N-m)/2},
\label{binomial}
\end{equation}
where we have introduced a cut-off frequency, $N$, making infinite sums
finite, and $0\leq \mu\leq 0.5$. For $N \gg 1$ the normalisation is very close
to 
unity, so we will drop it in the following.  The dominant frequency in this
distribution is $\Omega=\mu N$, which we call the carrier frequency. In this
case  
\begin{equation}
n(t)=\left |\sum_{k=1}^N e^{-ikt}{N\choose
    k}^{1/2}\mu^{k/2}(1-\mu)^{(N-k)/2}\right |^2 .
\end{equation}
This function is shown in Fig.~\ref{fig:mu} for various values of $\mu$. The
probability per unit time is thus a periodic function of time, with period
$2\pi$ and pulse width determined by $\mu$ when $N$ is fixed. If we fix the
carrier frequency $\Omega=\mu N$ and let $N$ become large we must let $\mu$
become small. In the limit $N\rightarrow \infty,\ \mu\rightarrow 0$ with
$\Omega$ fixed we obtain a Poisson distribution for the single photon
amplitude.  

A second example is the Lorentzian,
\begin{equation}
f_n^N=\frac{1}{A}\frac{\sqrt{\mu}}{\mu+in}.
\label{lorentzian}
\end{equation}
In the limit $N\rightarrow  \infty$, the normalisation constant is 
\begin{equation}
A=\frac{\pi e^{\mu\pi}}{2\sinh(\mu\pi)}-\frac{1}{2\mu}.
\end{equation}

While a field for which exactly one photon is counted in one counting
interval, and zero in all others, is no doubt possible, it does not correspond
to a more physical situation in which a source is {\em periodically} producing
pulses with exactly one photons per pulse.  To define such a field state we
now introduce time-bin operators. For simplicity we assume that only field
modes $n\leq N$ are excited and all others are in the vacuum state. It would
be more physical to assume only field modes are excited in some band,
$\Omega-B\leq n\leq \Omega+B$. Here $\omega$ is the carrier frequency and $2B$
is the bandwidth. However, this adds very little to the discussion.

Define the operators
\begin{equation}
 \hat{b}_\nu=\frac{1}{\sqrt{N}}\sum_{m=1}^N e^{-i\tau m\nu}\hat{a}_m,
\end{equation}
where $\tau=2\pi/N$. This can be inverted to give
\begin{equation}
 \hat{a}_m=\frac{1}{\sqrt{N}}\sum_{\nu=1}^N e^{i\tau m\nu}\hat{b}_\nu.
\end{equation}
The temporal evolution of the positive frequency components of the field modes
then follows from Eq.~(\ref{dynamics}) 
\begin{equation}
 \hat{a}(t)  =  \sum_{\mu=1}^N g_\mu(t)\hat{b}_\mu,
\label{time-bin-series}
\end{equation}
where
\begin{equation}
g_\mu(t)=\frac{1}{\sqrt{N}}\left [1-e^{i(\mu\tau-t)}\right ]^{-1}.
\end{equation}
The time-bin expansion functions $g_\mu(t)$ are a function of $\mu\tau-t$
alone and thus are simple translations of the functions at $t=0$. The form
of Eq.~(\ref{time-bin-series}) is a special case of a more sophisticated way to
define time-bin modes. If we were to regard $\hat{a}(t)$ as a classical signal,
then 
the decomposition in Eq.~(\ref{time-bin-series}) could be generalised as a
wavelet transform where the integer $\mu$ labels the  translation index for
the wavelet functions. In that case the functions $g_\mu(t)$ could be made
rather less singular.   In an experimental context, however, the form of the
functions $g_\mu(t)$ depends upon the details of the generation process.  

The linear relationship between the plane wave modes $a_m$ and the time bin
modes $b_\nu$ is realised by a unitary transformation that does not
change particle number, so the vacuum state for the time-bin modes is the same
as the vacuum state for the global plane wave modes. We can then define a
one-photon time-bin state as 
\begin{equation}
 \tilde{|1\rangle}_\mu=\hat{b}_\mu^\dagger|0\rangle.
\end{equation}
The mean photon number for this state is, 
\begin{equation}
 n(t)=|g_\mu(t)|^2.
\end{equation}
This function is periodic on $t\in[0,2\pi)$ and corresponds to a pulse
localised in time at $t=\mu\tau$. Thus the integer $\mu$ labels the temporal
coordinate of the single-photon pulse.  

\begin{figure}[t]
  \begin{center}
  \begin{psfrags}
       \epsfig{file=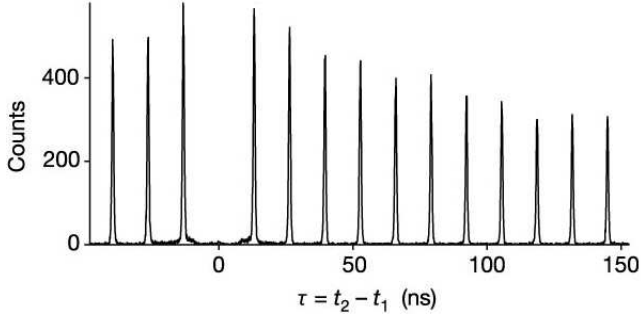, width=8.5cm}
  \end{psfrags}
  \end{center}
 \caption{The $G^{(2)}(\tau)$ for the InAs quantum dot single-photon source of
   Santori et al.\ (2002b). Note that the variable $T$ in the text is here
   replaced with $\tau$. Reprinted with permission from Nature
   Publishing Group.} 
 \label{fig:g2}
\end{figure}

We are now in a position to define an $N$-photon state with one photon per
pulse. In addition to the mean photon number, $n(t)$ we can now compute
two-time correlation functions such as the second order correlation function,
$G^{(2)}(\tau)$ defined by 
\begin{equation}
 G^{(2)}(T)=\langle \hat{}a^\dagger(t) \hat{a}^\dagger(t+T) \hat{a}(t+T)
 \hat{a}(t)\rangle .
\end{equation}
The simplest example  for $N=2$ is 
\begin{equation}
 |1_\mu,1_\nu\rangle=\hat{b}_\mu^\dagger\hat{b}_\nu^\dagger|0\rangle\ \ \
  \ \ \mu\neq\nu .
\end{equation} 
The corresponding mean photon number is 
\begin{equation}
 n(t)=|g_\mu(t)|^2+|g_\nu(t)|^2,
\end{equation}
as would be expected. The two-time correlation function is,
\begin{equation}
 G^{(2)}(T)=|g_\mu(t)g_\nu(t+T)+g_\nu(t)g_\mu(t+T)|^2.
\end{equation} 
Clearly this has a zero at $T=0$ reflecting the fact that the probability to
detect a single photon immediately after a single photon detection is zero, as
the two pulses are separated in time by $|\mu-\nu|$. This is known as {\em
  anti-bunching} and is the first essential diagnostic for a sequence of
single photon pulses with one and only one photon per pulse.  When
$T=|\mu-\nu|\tau$, however, there is a peak in the two-time correlation
function 
as expected. In Fig.~\ref{fig:g2} we have reproduced the experimental results
for the $G^{(2)}(T)$ from Santori et al.\ (2002b).

\begin{figure}[t]
  \begin{center}
  \begin{psfrags}
       \epsfig{file=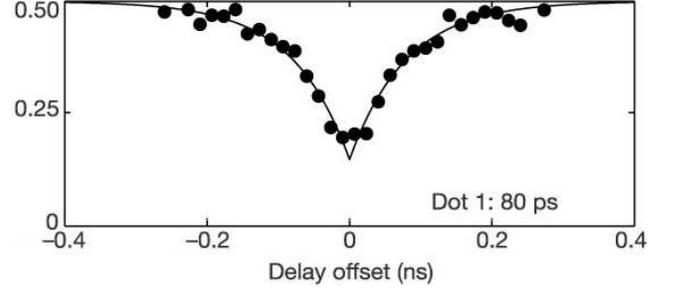, width=8.5cm}
  \end{psfrags}
  \end{center}
 \caption{The Hong-Ou-Mandel effect for one of the InAs quantum dot
   single-photon source of Santori et al.\ (2002b). Reprinted with
   permission from Nature Publishing Group.}  
 \label{fig:hom}
\end{figure}

We now revisit the Hong-Ou-Mandel interferometer introduced in section
\ref{elga} with single photon input states. This example has been considered
by Rohde and Ralph (2005). We label the two sets of modes by the Latin symbols
$a,b$ so, for example, the positive frequency parts of each field are simply
$a(t),b(t)$. The coupling between the modes is described by a beam splitter
matrix connecting the input and output plane waves 
\begin{eqnarray}
 \hat{a}_n^{out}&= &  \sqrt{\upsilon}\,\hat{a}_n+\sqrt{1-\upsilon}\,\hat{b}_n\\
 \hat{b}_n^{out} & = &  \sqrt{\upsilon}\,\hat{b}_n-\sqrt{1-\upsilon}\,\hat{a}_n
\end{eqnarray}
where $0\leq\upsilon\leq 1$. The probability per unit time to find a
coincidence detection of a single photon at each output beam is
proportional to    
\begin{equation}
 C = \overline{\langle \hat{a}^\dagger(t)\hat{b}^\dagger(t) \hat{b}(t)
 \hat{a}(t) \rangle}.
\end{equation}
The over-line represents a time average over a detector response time that is
long compared to the period of the field carrier frequencies.  In this example,
we only need consider the case of one photon in each of the two distinguished
modes, so we take the input state to be 
\begin{equation}
 |1\rangle_a\otimes|1\rangle_b=\sum_{m,n=1}^\infty \alpha_n\beta_m
 \hat{a}^\dagger_n \hat{b}^\dagger_m|0\rangle ,
\end{equation}
where $\alpha_n$ and $\beta_n$ refer to the excitation probability
amplitudes for modes $a_n$ and $b_n$, respectively. This state is
transformed by the unitary transformation $U$ to give
$|\psi\rangle_{out}=U(|1\rangle_a\otimes|1\rangle_b)$.  In the case of a 50:50
beam splitter, for which $\upsilon=0.5$, this is given as ($U|0\rangle =
|0\rangle$) 
\begin{eqnarray*}
|\psi\rangle_{out}& = & \sum_{n,m=1}^\infty\alpha_n\beta_m \left( U a_n^\dagger
 b_m^\dagger U^{\dagger}\right) U |0\rangle\\ 
 & = &\frac{1}{2} \sum_{n,m=1}^\infty
 \alpha_n\beta_m(a_n^\dagger+b_n^\dagger)(b_m^\dagger-a_m^\dagger)|0\rangle\\ 
 & = & \frac{1}{2}\sum_{n,m=1}^\infty\alpha_n\beta_m
 [|1\rangle_{a_n}|1\rangle_{b_m}-|1\rangle_{a_n}|1\rangle_{a_m}|0\rangle_b\\ 
   & & \mbox{}\ \ \ \ \ \ \ \ \ \ \ \ \
 +|1\rangle_{b_n}|1\rangle_{b_m}|0\rangle_a-|1\rangle_{b_n}|1\rangle_{a_m} ]. 
   \end{eqnarray*}
Note that the second and third terms in this sum have no photons in modes $b$
and $a$, respectively. We then have that 
\begin{equation}\label{modematch}
C=\frac{1}{2}-\frac{1}{2}\sum_{n,m=1}^\infty\alpha_n\alpha_m^*\beta_m\beta_n^*.
\end{equation}
If the excitation probability amplitudes at each frequency are identical,
$\alpha_n=\beta_n$ this quantity is zero.  In other words, only if the
two-single photon wave packets are identical do we see a complete cancellation
of the coincidence probability. This is the second essential diagnostic for a
single-photon source. Of course in practice, complete cancellation is
unlikely.  The extent to which the coincidence rate approaches zero is a
measure of the quality of a single-photon source as far as LOQC is
concerned. Whether or not this is the case depends on the nature of the single
photon sources. In Fig.~\ref{fig:hom} we have reproduced the experimental
results for the Hong-Ou-Mandel effect, shown with one of the InAs quantum dot
single-photon sources of Santori et al.\ (2002b).

Broadly speaking, there are currently two main schemes used to realise single
photon sources: (I) conditional spontaneous parametric down conversion, and
(II) cavity-QED Raman schemes.   As discussed by Rohde and Ralph (2005),
type (I) corresponds to a Gaussian 
distribution of $\alpha_n$  as a function of $n$ and thus is the continuum
analogue of the binomial state defined in Eq.~(\ref{binomial}). The second
scheme, type (II), leads to a temporal pulse structure that is the convolution
of the excitation pulse shape and the Lorentzian line shape of a cavity. If
the cavity decay time is the longest time in the dynamics, the
distribution $\alpha_n$ takes the Lorentzian form given in
Eq.~(\ref{lorentzian}).  An early single-photon source based on an optical
emitter in a micro-cavity was proposed and demonstrated by De~Martini
et al.\ (1996). \nocite{demartini96}  

Cavity-based single-photon sources are very complicated experiments in
their own right,
and instead, most single-photon sources used in LOQC experiments are
based on Parametric Down-Conversion (PDC). In PDC a short-wavelength pump
laser generates photon pairs of longer wavelength in a birefringent
crystal. PDC can yield extremely high fidelities ($F>0.99$) because the
data is usually obtained via post-selection: we take only those events
into account that yield the right number of detector coincidents. In
addition, PDC facilitates good mode matching due to energy and
momentum conservation in the down-conversion process. The output of a
non-collinear type-I PDC can be written as   
\begin{equation}
 |\Psi_{\rm PDC}\rangle = \sqrt{1-|\lambda|^2} \sum_{n=0}^{\infty}
  \lambda^n |n,n\rangle\; , 
\end{equation}
where $|n\rangle$ is the $n$-photon Fock state, and $\lambda$ is a
measure for the amount of down-conversion. The probability for
creating $n$ photon pairs is $p(n)=(1-|\lambda|^2) |\lambda|^{2n}$,
which exhibits pair bunching. When $\lambda$ is small, we can make a
probabilistic single-photon gun by detecting one of the two
modes. However, if we use only bucket detectors without single-photon
resolution, then increasing $\lambda$ will also increase the amplitudes
for a two-photon pair and ultimately high-photon pairs to the
output state. Consequently, the single-photon source will deteriorate
badly. A detailed study of the mode structure of the conditional photon pulse
has been undertaken by Grice et al.\ \nocite{grice01} (2001).

Another consideration regarding parametric down-conversion is that the
photons in a pair are typically
highly entangled in frequency and momentum. When we use a bucket
detector that is sensitive over a broad frequency range to herald a
single photon in the freely propagating mode, the lack of frequency
information in the detector read-out will cause the single-photon
state to be mixed. In principle, this can be remedied by embedding the
down-converting material in a micro-cavity such that only certain
frequencies are allowed \cite{raymer05}. The source will then generate
photon pairs with frequencies that match the cavity, and a narrow-band
bucket detector can herald a pure single-photon state with a small
frequency line width.  

Alternatively, we can use the following method of making single-photon
sources \cite{pittman02c,migdall02}: Consider an array of PDCs with
one output mode incident on a photon detector, and the other entering
the quantum circuit. We fire all PDCs simultaneously. Furthermore,
all PDC have small $\lambda$, but if there are approximately
$|\lambda|^{-2}$ of them we still create a single photon on
average. Given that in current PDC configurations $|\lambda|^2 \approx
10^{-4}$, this is quite an inefficient process. Nevertheless, since it
contributes a fixed overhead per single photon to the computational
resources, this technique is strictly speaking scalable. For a
detailed description of parametric down-conversion as a photon source
see U'Ren et al.\ \nocite{uren03} (2003). 

To illustrate the experimental constraints on the generation of single-photon
states, we now review an example of a cavity-QED Raman scheme implemented by
Keller et al.\ \nocite{keller04} (2004). Photon anti-bunching from resonance
fluorescence was demonstrated  long ago. If an atom decays spontaneously from
an excited to a ground state, a single photon is emitted and a second photon
cannot be emitted until the atom is re-excited. Unfortunately the photon is
emitted into a dipole radiation pattern over a complete solid angle. Clearly
we need to engineer the electromagnetic environment with mirrors, dielectrics,
etc., to ensure a preferred mode for emission. However as pointed out by Kiraz
et al.\ \nocite{kiraz04} (2004), this comes at a price. For example, a
carefully engineered cavity around a single dipole emitter can change the free
field spectral density around the emitter such that a photon is indeed emitted
in a preferred direction with an increased rate compared to free space
emission. 

However, single-photon sources based on spontaneous emission are necessarily
compromised by the random nature of spontaneous emission. As demonstrated by
Rohde, Ralph, and Nielsen (2005) \nocite{rohde05c}, single-photon sources that
create Gaussian wave packets are much more robust to mode mismatching than
sources that create Lorentzian wave packets. Spontaneous emission processes
fall in this last category. Clearly, we prefer a {\em stimulated} emission
process yielding a Gaussian wave packet. To this end, a number of schemes
based on stimulated Raman emission into a cavity  mode have been proposed
\cite{hennrich04,maurer04}. As an example, we discuss the experiment by Keller
et al.\ (2004) in some detail.

\begin{figure}[t]
  \begin{center}
  \begin{psfrags}
       \epsfig{file=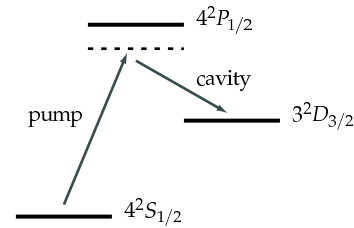, width=4.5cm}
  \end{psfrags}
  \end{center}
 \caption{The Raman process in a three-level atom. A classical pump field
   drives the transition $4^2S_{1/2}\rightarrow 4^2P_{1/2}$ off-resonantly,
   thus generating a photon in the cavity mode. The level $4^2P_{1/2}$ is
   adiabatically eliminated and hence never populated.}
 \label{fig:raman}
\end{figure}

Consider the three-level atomic system in Fig.~\ref{fig:raman}.  The ground
state is coupled to the excited state via a two-photon Raman process mediated
by a well-detuned third level. In the experiment by Keller et al.\ (2004), a
calcium ion $^{40}\mbox{Ca}^+$ was trapped in a cavity via an RF ion
trap. The  cavity 
field is nearly resonant with the $4^2P_{1/2}\rightarrow 3^2D_{3/2}$
transition. Initially there is no photon in the cavity. An external laser is
directed onto the ion and is nearly resonant with the $4^2S_{1/2}\rightarrow
4^2P_{1/2}$ transition. When this laser is switched on, the atom can be
excited to the  $3^2D_{3/2}$ level by absorbing one pump photon and {\em
emitting} one photon in to the cavity. Since this is a stimulated Raman
process, the time of emission of the photon into the cavity is completely
controlled by the temporal structure of the pump pulse. The photon in the
cavity then decays through the end mirror, again as a Poisson process,
at a rate given by the cavity decay rate. This can be made very fast. 

In principle one can now calculate the probability per unit time to detect a
single photon emitted from the cavity. If we assume every photon emitted is
detected, this probability is simply $p_1(t)=\kappa\langle \hat{a}^\dagger(t)
\hat{a}(t)\rangle$ where $\kappa$ is the cavity decay rate and $\hat{a},
\hat{a}^\dagger$ are the annihilation and creation operators for the {\em
  intra-cavity} field and   
\begin{equation}
\langle \hat{a}^\dagger(t) \hat{a}(t)\rangle=\mbox{\rm tr}[\rho(t)
\hat{a}^\dagger \hat{a}] ,
\end{equation}
where $\rho(t)$ is the total density operator for ion-plus-cavity-field
system. This may be obtained by solving a master equation describing the
interaction of the electronic states of the ion and the two fields, one of
which is the time-dependent pump. Of course, for a general time-dependent pump
pulse-shape this can only be done  numerically.  Some typical examples are
quoted by Keller et al. Indeed by carefully controlling the pump pulse shape
considerable control over the temporal structure of the single-photon
detection probability may be achieved. In the experiment the length of the
pump pulse was controlled to optimise the single-photon output rate. The
efficiency of emission was found to be about $8\%$, that  is to say, $92\%$ of
the pump pulses did not lead to a single-photon detection event. This was in
accordance with the theoretical simulations. These photons are probably lost
through the sides of the cavity. It is important to note that this kind of
loss does not affect the temporal mode structure of the emitted (and detected)
photons.   

In a similar way we can compute the second-order correlation function via
\begin{equation}
 G^{(2)}(T)=\kappa^2\mbox{\rm tr}\left[ \hat{a}^\dagger \hat{a}\, e^{{\cal
 L}T}(\hat{a}\rho(t) \hat{a}^\dagger)\right],
\end{equation}
where $e^{{\cal L}T}$ is a formal specification of the solution to the master
equation for a time $T$ after the ``initial'' conditional state
$a\rho(t)a^\dagger$.  Once again, due to the non stationary nature of the
problem, this must be computed numerically.  However, if the pump pulse
duration is very short compared to the cavity decay time, and the
cavity decay time is the fastest decay constant in the system, the probability
amplitude to excite a single photon in a cavity at frequency $\omega$ is
very close to Lorentzian.  The experiment revealed a clear suppression of the
peak at $T=0$ in the normalised correlation function $g^{(2)}(T)$, thus
passing the first test of a good single photon source.  Unfortunately, a
Hong-Ou-Mandel interference experiment was not reported.  

For a practical linear-optical quantum computer, however, we need good
microscopic single-photon sources that can be produced in large
numbers. A recent review on this topic by Lounis and Orrit (2005)
\nocite{lounis05} identifies six types of microscopic sources: i)
Atoms and ions in the gaseous
phase\footnote{\nocite{kuhn02,mckeefer04} Kuhn et al.\ (2002) and
McKeefer et al.\ (2004).}; ii) Organic molecules at low temperature
and room temperature\footnote{\nocite{brunel99} Brunel et al.\ (1999),
\nocite{lounis00a,treussart02,hollars03} Lounis and Moerner (2000),
Treussart et al.\ (2002) and Hollars et al.\ (2003).}; iii)
Chromophoric systems\footnote{\nocite{lee04b} Lee et al.\ (2004b).};
iv) Colour centres in diamond, such  as nitrogen-vacancy\footnote{
\nocite{kurtsiefer00,beverator02,jelezko02} Kurtsiefer et al.\ (2000),
Beveratos et al.\ (2002) and Jelezko et al.\ (2002).} or
nickel-nitrogen\footnote{\nocite{gaebel04} Gaebel et al.\ (2004).}; v)
Semiconductor nano-crystals\footnote{
\nocite{lounis00b,michler00,messin01} Lounis et al.\ (2000), Michler
et al.\ (2000) and Messin et al.\ (2001).}; and vi) self-assembled quantum
dots  and other hetero-structures such as micro-pillars and
micro-mesa\footnote{\nocite{santori02a,pelton02,vuckovic03,gerard02}
Santori et al.\ (2002b), Pelton et al.\ (2002), G\'erard et al.\
(2002) and Vu{\v c}kovi\'c et al.\ (2003).}, quantum dots\footnote{
\nocite{hours03,zwiller03,ward05} Hours et al.\ (2003), Zwiller et
al.\ (2003) and Ward et al.\ (2005).}, and electrically driven
dots\footnote{\nocite{yuan02} Yuan et al.\ (2002).}. The typical
physical mechanisms that reduce the indistinguishability of the
single-photon sources are dephasing of the optical transition,
spectral diffusion, and incoherent pumping. An earlier review on this
topic is given by Greulich and Thiel (2001). \nocite{greulich01}
The subject of single-photon sources using quantum dots was reviewed
by Santori et al.\ \nocite{santori04} (2004).

When single-photon sources are less than ideal, linear optics might be
employed in order to improve the output state. For example, if the source
succeeds with probability $p$, then the output of the source might be $\rho =
p |1\rangle\langle 1| + (1-p) |0\rangle\langle 0|$, where we assumed that the
failure output results in a vacuum state. Using multiple copies of $\rho$,
linear optics, and ideal photon detection, one may increase the probability up
to $p=1/2$, but not higher \cite{berry04}. A general discussion on
improving single-photon sources with linear optical post-processing is
given by Berry et al.\ (2004). \nocite{berry04b}

Single-photon sources must not only create clean single-photon states, in the
sense described above, but all sources must also generate {\em identical}
states in order to achieve good visibility in a Hong-Ou-Mandel test.
Typical experiments demonstrating single-photon sources create
subsequent single-photon states in the {\em same} source and employ a delay
line to interfere the two photons. This way, two-photon quantum
interference effects are demonstrated without having to rely on
identical sources \cite{santori02a}. De Riedmatten {et al}.
demonstrated quantum interference by using identical pulse shapes
triggering different  photon sources \cite{deriedmatten03}. In
applications other than LOQC, such as cryptography, the requirement of
indistinguishable sources may be relaxed. This leads to the concept of
the {\em suitability} of a source for a particular application
\cite{hockney03}.

A variation on single-photon sources is the {\em entangled-photon source}. We
define an ideal entangled-photon source as a source that creates a two-photon
polarisation Bell state. This is an important resource in both the
Browne-Rudolph and the Gilchrist-Hayes-Ralph protocol. It is known that these
states cannot be created deterministically from single-photon sources, linear
optics and destructive photon detection \cite{kok00b}. Nevertheless, such
states are very desirable, since they would dramatically reduce the cost of
linear-optical quantum computing. The same error models for  single-photon
sources apply to entangled-photon sources. Again, a great variety of proposals
for entangled-photon sources exist in the literature, using quantum dots
\nocite{benson00,stace03} (Benson et al.\ 2000; Stace et al.\ 2003) or
parametric down-conversion \cite{sliwa03}. Two-photon states without
entanglement have been created experimentally by Moreau et al.\ (2001), and
\nocite{moreau01,santori02b} Santori et al.\ (2002a), as have entangled photon
pairs \cite{yamamoto03,kuzmich03}.

\subsection{Circuit errors and quantum memories}

\noindent 
In addition to detector errors, and errors in the single-photon
sources, there is a possibility that the optical circuits themselves
acquire errors. Probably the most important circuit error is {\em mode
mismatching}. It occurs when non-identical wave packets are used in an
interferometric setup [e.g., the coefficients $\alpha_n$ and $\beta_n$
in Eq.~(\ref{modematch}) are not identical]. There is a plethora of
reasons why the coefficients $\alpha_n$ and $\beta_n$ might not be
equal. For example, the optical components might not do exactly what
they are supposed to do. More precisely, the interaction Hamiltonian
of the components will differ from its specifications. One
manifestation of this is that there is a finite accuracy in the
parameters in the interaction Hamiltonian of any optical component,
leading to changes in phases, beam splitter transmission coefficients,
and polarisation rotation angles. In addition, unwanted birefringence
in the dielectric media can cause photo-emission and
squeezing. Inaccurate Hamiltonian parameters generally reduce the
level of mode matching, leading, for example, to a reduced
Hong-Ou-Mandel effect and hence inaccurate CZ gates. Indeed, mode
matching is likely to be the main circuit error. Most of this effect
is due to non-identical photon sources, which we discussed in the
previous section. The effect of frequency and temporal mode
mismatching was studied by \nocite{rohde05b} Rohde and Ralph (2005),
and by Rohde, Ralph, and Nielsen (2005).

A second error mechanism is that typically, components such as beam
splitters, half- and quarter-wave plates, etc.\ are made of dielectric
media that have a (small) absorption amplitude. Scheel (2005)
\nocite{scheel05c} showed that there is a lower bound on the absorption
amplitude in physical beam splitters. In addition, imperfect 
impedance matching of the boundaries will scatter photons back to the
source. This amounts to photon loss in the optical circuit. In large
circuits, these losses can become substantial.

A third error mechanism is due to classical errors in the feed-forward
process. This process consists of the read-out of a photon detector, classical
post-processing, and conditional switching of the optical circuit. The
detection errors have been discussed in section \ref{sec:detectors} and
classical computing is virtually error-free due to robust classical error
correction. Optical switches, however, are still quite lossy \cite{thew02}. In
addition, when high-voltage Pockels cells are used, the repetition rate is
slow (on the order of 10 kHz). This may become too slow, as photons need to be
stored in a quantum memory (e.g., a delay loop), which itself may be lossy and
needs feed-forward processing. Feed-forward control for LOQC was demonstrated
by Pittman et al.\ \nocite{pittman02a} (2002a) and  Giacomini et al.\
(2002). \nocite{giacomini02}

\bigskip

\noindent
An important component of linear optical quantum
computing that we have ignored so far is the {\em quantum memory}. When
the probability of a successful (teleported) gate or addition to a
cluster state becomes small, the photons that are part of the circuit
must be stored for a considerable time while the off-line preparation
of entangled photons is taking place. The use of mere fibre loops then
becomes problematic, as these induce photon losses
(0.17~dB$\,$km$^{-1}$ in a standard telecom fibre at 1550~nm). For
example, to store a photon for 100~$\mu$s in a fibre has a loss
probability of $p \approx 0.54$. At present, all linear optical
quantum computer proposals need some kind of quantum memory. This may
be in the form of delay lines with error correction, atomic vapours,
solid-state implementations, etc.

In general, the effect of a quantum memory error boils down to the
inequality of the input state $\rho_{\rm in}$ and the (time-translated) output
state $\rho_{\rm out}$. A good figure of merit is the fidelity $F_{qm}$:
\begin{equation}
 F_{qm} = \left[ \mbox{\rm Tr} \left( \sqrt{\sqrt{\rho_{\rm in}}\, \rho_{\rm
 out}\,  \sqrt{\rho_{\rm in}}} \right) \right]^2 \; .
\end{equation}
The absence of a photon in the output state is an obvious
failure mechanism, but other ways the memory can fail include qubit
decoherence and mode mismatching of the input/output modes. In this
sense, the design specifications of a solid-state based quantum memory
are more stringent than those for solid-state single-photon sources:
Not only does it need to produce a single photon with very high
fidelity, it also needs the ability to couple a photon {\em into} the
device with very high probability. Note that we do not have to couple a
photonic {\em qubit} into a quantum memory: We can use two photon
memories to store one qubit, provided the memory does not retain
information about whether a photon was stored or not.

A proof-of-principle for a free-space delay line was given by
Pittman and Franson \nocite{pittman02} (2002), and quantum memory
delay lines using quantum error correction and QND measurements were
proposed by Gingrich et al.\ \nocite{gingrich03} (2003), and Ralph et
al.\ \nocite{ralph05} (2005). A storage time of 125~$\mu$s for
entangled photons in a telecom fibre was reported by Li et al.\
\nocite{li05} (2005), with subsequent fringe visibilities of
82\%. Using the magnetic sublevels of the ground state of an atomic
ensemble, Julsgaard et al.\ \nocite{julsgaard04} (2004) stored a weak
coherent light pulse for up to 4~ms with a fidelity of 70\%. The
classical limit is 50\%, showing that a true quantum memory was
constructed. Other proposals include dark-state polaritons
\cite{fleischauer02}, and single-photon cavity QED \cite{maitre97}.

\section{General error correction}\label{sec:error}

\noindent
To achieve quantum computing despite inevitable physical
errors in the quantum computer, we have to employ Error
Correction (EC). Typically, an error-correction protocol consists of
a circuit that can correct for one or more types of error. However,
these circuits will in turn introduce errors. For an EC protocol to
be useful, the error in the circuit {\em after} the EC protocol must
be smaller than the error {\em before} the EC protocol. Repeated
nested application of the EC protocol (so-called {\em concatenation})
can then reduce the errors to arbitrarily small levels. In doing so, we
must take care not to sacrifice the scaling behaviour of the quantum
computer. This is captured in the notion of {\em fault tolerance}. The
magnitude of the errors for which fault tolerance breaks down is
called the fault-tolerant {\em threshold}. For more details, see
Nielsen and Chuang (2000). 
 
General fault-tolerant thresholds for quantum computing have been
derived by Steane (2003) and Knill (2005),\nocite{steane03}
\nocite{knill05} and here we address LOQC specific error correction
and fault-tolerant thresholds.  We have seen that the KLM scheme
employs a certain level of error correction in order to turn
high-probability teleported gates into near-deterministic gates, even
though all-optical components are ideal. In this section, we discuss
how an LOQC architecture can be developed with robustness against
component errors. 

Different error models will typically lead to different levels of
robustness. For example, in the cluster-state approach of Browne and
Rudolph, we can relax the condition of perfect photon counting given
ideal photon sources. The type-II fusion operation described in
section \ref{sec:br} must give a coincidence count in the two
detectors. Any other detector signature heralds an error. So if the
photon detectors are lossy, the rate of coincidence counts is
reduced. Since the fusion operation is already probabilistic, a reduced
success rate translates into a larger overhead in the cluster-state
generation. However, if the photon sources are not ideal and if there
is a substantial number of dark counts in the detectors, then we
rapidly lose quantum information. This raises two important questions:
(1) Given a certain error model, what is the error correcting
capability for a given LOQC architecture? (2) What is the realistic
error model? This last question depends on the available photon
sources, detectors and memories, as well as the architecture of the
optical quantum computer. Currently, theoretical research in LOQC is
concentrating on these questions.

The three main errors that need to be coded against are inefficient
detectors, noisy photon sources, and unfaithful quantum
memories. There are other error mechanisms as well (see section
\ref{sec:real}), and these will become important in concatenated error
correction. In order to find the fault-tolerant level for a given
architecture, these other errors must be taken into account. In the next
section, we discuss how photon loss can be corrected in both the cluster-state
model and the circuit-based model. In section \ref{sec:ft} we discuss
fault-tolerant quantum computing in the cluster-state model.

\subsection{Correcting for photon loss}

\noindent 
We first consider photon loss. Its effect on the original
teleportation component in the KLM protocol was studied by Glancy et
al.\ (2002), \nocite{glancy02} who found that in the KLM scheme a gate
teleportation fidelity better than 99\% requires detectors with an
efficiency $\eta > 0.999\, 987$. Using the seven qubit CSS quantum
code, the photon loss $\epsilon$ in the KLM scheme is allowed to be as
large as $1.78\% \leq \epsilon \leq 11.5 \%$, depending on the
construction of the entangling gates \cite{silva05}. Using the type-I
and  type-II fusion gates in creating entanglement, the photon loss
can be much higher: In the Browne-Rudolph protocol, a low detection
efficiency merely reduces the rate with which the cluster state is
created, whereas in the circuit-based model a low detection efficiency
requires a higher level of encoding.

However, we not only need the ability to grow the cluster or the parity
encoding efficiently, we also need to do the single-qubit measurements. Since
in LOQC the single-qubit measurements amount to photon detection, we have a
problem: Failing to measure a photon is also a single-qubit
failure. Therefore, every logical qubit must be constructed with multiple
photons, such that photon loss can be recovered from. In particular, this
means that we can no longer straightforwardly remove redundant qubits in the
cluster-state model if they are not properly encoded. In this section, we show
how cluster states can be protected from photon loss by ``planting trees'' in
the cluster \cite{varnava05}, and we will describe how an
extra layer of encoding protects the circuit model of Ralph, Hayes, and
Gilchrist \nocite{ralph05} (2005) from detection inefficiency, probabilistic
sources, and memory loss.

\begin{figure}[t]
  \begin{center}
  \begin{psfrags}
       \epsfig{file=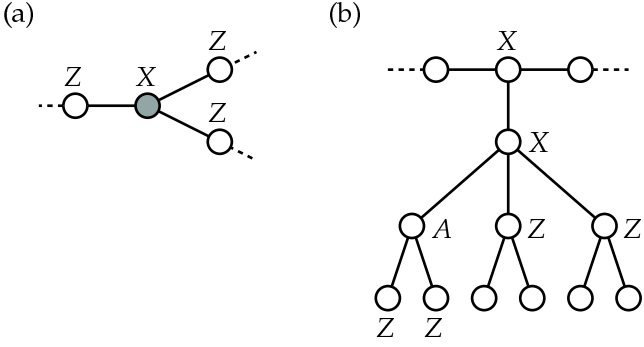}
  \end{psfrags}
  \end{center}
  \caption{Photon-loss tolerant cluster states. (a) We can measure the
  Pauli operator $X$ on the shaded (lost) qubit by measuring all the
  adjacent qubits in the $Z$ basis. (b) Planting a cluster tree
  using two adjacent $X$ measurements in order to do a single-qubit
  measurement in the basis $A$.} 
  \label{fig:horticulture} 
\end{figure}

Varnava, Browne, and Rudolph \nocite{varnava05} (2005) introduced a
code exploiting the property that a cluster state is an eigenstate of
every stabiliser generator, and that the eigenvalue of each is known
beforehand (we will assume that all eigenvalues are $+1$). This allows
us to measure the value of a lost qubit as follows: Suppose we
wish to measure a qubit in the computational basis, that is, we
require a $Z$ measurement. If that qubit is no longer present, we can
choose $S_i = X_i \prod_{j\in n(i)} Z_j$ such that our lost qubit is
in the neighbourhood $n(i)$ of the $i^{\rm th}$ qubit. If we successfully measure
$X_i$ and all $Z_j$ except for the lost qubit, we can multiply the
eigenvalues to find either $+1$ or $-1$. Since the stabiliser
generator has eigenvalue $+1$, this determines the $Z$ eigenvalue, and
therefore the $Z$ eigenstate of our lost qubit. In figure
\ref{fig:horticulture}a, we show how an $X$ measurement can be
performed on a lost qubit by $Z$ measurements on the adjacent qubits.

In cluster-state quantum computing, we need the ability to do
single-qubit measurements in an arbitrary basis; $A = \cos\phi\, X +
\sin\phi\, Y$. To this end, we use the cluster-state property that two
adjacent $X$ measurements remove the qubits from the cluster and
transfer the bonds. This way, we can plant the qubit labelled $A$
into the cluster (see Fig.~\ref{fig:horticulture}b). Instead of doing
the $A$ measurement on the in-line qubit, we perform the measurement
on a qubit in the third (horizontal) level. When this measurement succeeds, we
have to break the bonds with all the other qubits in the tree. Therefore,
we measure all the remaining qubits in the third level as well as the
qubits in the fourth level that are connected to the $A$ qubit in the
$Z$ basis. 

Sometimes the photon detection that constitutes the qubit measurement
$A$ will fail due to the detector inefficiency. In that case, we can 
attempt the $A$ measurement on a second qubit in the third
level. Again, the remaining qubits and the fourth-level qubits
connected to the $A$ qubit must be measured in the $Z$ basis. Whenever
such a $Z$ measurement fails (as is the case for the qubit that failed
the $A$ measurement), we need to do an {\em indirect} $Z$
measurement according to the method outlined above: When a photonic
qubit is lost, we need to choose a stabiliser generator for which that photon
was represented by a $Z$ operator. The tree structure ensures that
such an operator can always be found. We then measure all the photons
in this stabiliser generator to establish the $Z$ eigenvalue for the lost
photon. In the case of additional photon loss, we repeat this process until we
succeed. 

When the success probability of the measurement of a logical qubit is given by
$p$, then the number of qubits in a tree $n$ is given by
\begin{equation}
 n = \mathrm{polylog} \left( \frac{1}{1-p} \right) .  
\end{equation}
Numerical simulations indicate that a detector loss of up to 50\% can be
corrected \cite{varnava05}. Moreover, if more than 50\% of the photons
were allowed to be lost, then we can imagine that all the lost photons
are collected by a third party who can perform a measurement
complementary to $A$ on the same qubit. Since this would violate
various no-cloning bounds, such a strategy must be ruled out. Hence,
a detection efficiency of 50\% is the absolute minimum
\cite{seanprivate}.  

\bigskip

\noindent
In the circuit-based model by Gilchrist, Hayes, and Ralph, the lowest
level of encoding consists of a 
polarised photon such that $|0\rangle \equiv |H\rangle$ and
$|1\rangle \equiv |V\rangle$. The second level of encoding is the
parity code $\{ |0\rangle^{(n)}, |1\rangle^{(n)} \}$ of
Eq.~(\ref{eq:rhg}), which allows us to use the probabilistic fusion 
gates in a deterministic manner. The third level of encoding is a {\em
redundant encoding} such that a logical qubit is encoded in a GHZ
state of parity-encoded qubits \cite{ralph05}: 
\begin{equation}
 |\psi\rangle_L \equiv \alpha |0\rangle^{(n)}_1\ldots|0\rangle^{(n)}_q
  + \beta |1\rangle^{(n)}_1 \ldots|1\rangle^{(n)}_q.
\end{equation}
To demonstrate that this code protects quantum information from photon 
loss, we show that heralded photon loss merely yields a recoverable
error, and that we can perform a universal set of deterministic
quantum gates on these logical qubits. 

First of all, we note that every pair of optical modes that constitutes
the lowest level qubit encoding must contain exactly one photon, and
no high number counting is required (contrary to the KLM scheme). This is also
true for the Browne-Rudolph protocol. We
therefore assume that we have bucket detectors with a certain
detection efficiency $\eta$ and a negligible dark count rate. The
type-I and type-II fusion gates are then no longer (1,1) and (2,1)
strategies, respectively. The type-I fusion gate ceases to yield pure
output states, while the type-II gate yields a pure output state only
if a detector coincidence is found (we assume perfect sources). A
failure not only removes the mode that was involved with the PBS, but
we should also measure one mode in order to purify the cluster. Hence,
the type-II fusion gate with bucket detectors is a (2,2)-strategy, and
the growth requirement is $m>2/p$.

Suppose we wish to measure the value of the logical qubit
$|\psi\rangle_L$ in the computational basis (as we discuss below, any
other measurement can be performed by first applying a single-qubit
rotation to $|\psi\rangle_L$). Since the logical qubit is a GHZ state,
it is sufficient to measure only one parity qubit, e.g., the first
one. Physically, this measurement constitutes a counting of
horizontally and vertically polarised photons: if the number of
vertically polarised photons is even, then the value of the parity
qubit is $|0\rangle_1^{(n)}$, and if it is odd, then its value is
$|1\rangle_1^{(n)}$. In order to successfully establish the parity, we
therefore need to detect {\em all} $n$ photons. 

When we include photon loss in this measurement, there are three
possible measurement outcomes for every optical mode: ``horizontal'',
``vertical'', or ``detector failure''. In the language of POVMs, this
can be written as 
\begin{eqnarray}
 \hat{E}^{(H)} &=& \eta |H\rangle \langle H| \; ; \cr
 \hat{E}^{(V)} &=& \eta |V\rangle \langle V| \; ; \cr
 \hat{E}^{(0)} &=& (1-\eta) \left( |H\rangle \langle H| +
 |V\rangle \langle V| \right) \; .
\end{eqnarray}
These POVMs add up to unity in the subspace spanned by $|H\rangle$ and
$|V\rangle$, as required. A particular measurement
outcome on $n$ modes can then be written as a string of $n$ outcomes
$s = (s_1,\ldots,s_n)$, where every $s_i \in \{ H,V,0\}$. The optical
state after finding a particular measurement outcome $s$ is then
\begin{equation}
 \rho_{2\ldots q} = \mathrm{Tr}_1 \left[ \hat{E}_1^{(s)}
 |\psi\rangle_L\langle\psi| \right] \; .
\end{equation}
When all photons are detected, the qubit is projected onto its logical
value. However, when one or more qubits are lost it is no longer
possible to establish the parity. Therefore, as soon as a photon is lost the
next photon is measured in a diagonal basis, thus disentangling the parity
qubit from the other parity qubits. A few lines of algebra shows that
the remaining $q-1$ parity qubits are in the state 
\begin{equation}
 |\psi\rangle_L' = \alpha |0\rangle^{(n)}_2\ldots|0\rangle^{(n)}_q
  + \beta |1\rangle^{(n)}_2 \ldots|1\rangle^{(n)}_q.
\end{equation}
In other words, the encoding has become smaller but the quantum
information {\em has not been erased}. We can therefore retry the
measurement of the qubit $q$ times.

Next, we have to show that we can perform deterministic one- and
two-qubit gates using this redundant encoding. To this end, recall how
deterministic gates were implemented in the parity encoding: A
universal set of gates is $\{ X_{\theta}^{(p)}, Z^{(p)},
Z_{\pi/2}^{(p)}, \mathrm{CNOT}^{(p)} \}$. We added a superscript $(p)$
to indicate that these gates act on the parity qubit. As we have seen,
the gates $Z_{\pi/2}^{(p)}$ and CNOT$^{(p)}$ cannot be implemented
deterministically, and have to be built using fusion gates.  

How can these gates be used to form a universal set on the redundantly
encoded (logical) qubit? First, the single-qubit gate $Z$ is still 
implemented deterministically: $Z = Z^{(p)} = n \sigma_z$. Therefore,
in order to apply a $Z$ gate, a $\sigma_z$ operation must be applied
to all $n$ photons in {\em one and only one} parity qubit. Secondly,
the gate $Z_{\pi/2}$ is diagonal in the computational basis, and can
therefore be implemented using one $Z_{\pi/2}^{(p)}$. Thirdly, the
$X_{\theta}$ gate on the redundantly encoded qubit is somewhat
problematic, since the gate transforms separable states of parity
qubits into highly entangled GHZ states. However, if we apply 
$(q-1)$ CNOT$^{(p)}$ gates, we can decode the qubit such that its
state is $(\alpha |0\rangle_1^{(n)} + \beta |1\rangle_1^{(n)}) \otimes
|0\rangle_2^{(n)}\ldots|0\rangle_q^{(n)}$. We can then apply the
deterministic gate
$X_{\theta}^{(p)}$ to the first parity qubit, and use another set of
$(q-1)$ CNOT$^{(p)}$ gates to re-encode the qubit. Therefore, the gate
$X_{\theta}$ ``costs'' $2(q-1)$ CNOT$^{(p)}$ gates. Finally, the CNOT
gate on the redundantly encoded qubit can be implemented using $q$
CNOT$^{(p)}$ gates. 

Since every single-qubit operation can be constructed from
$X_{\theta}$ and $Z_{\pi/2}$, we can perform arbitrary single-qubit
measurements. We now have a universal set of gates on our logical
qubit, together with computational-basis read out and an efficient
encoding mechanism. Numerical simulations indicate that this method
allows for combined detector, source and memory efficiencies of $\eta > 55\%$
\cite{ralph05}.  

Note that there seem to be conflicting requirements in this code: In 
order to execute successful fusion gates, we want $n$ to be reasonably
large. On the other hand, we want $n$ to be as small as possible such
that the probability of measuring all $n$ photons $p=\eta^n$ is not too
small. We assumed that every parity qubit is encoded with the same
number of photons $n$, but this is not necessary. In principle, this
method works when different parity qubits have different-sized
encodings. However, some care should be taken to choose every $n_i$ as
close to the optimal value as possible.

\subsection{General error correction in LOQC}\label{sec:ft}

\noindent
As we mentioned before, photon loss is not the only error in LOQC, and
creating large cluster trees or a sizable redundant encoding in the circuit
model will actually amplify other  errors, such as dephasing. A truly
fault-tolerant quantum computer architecture  must therefore be able to handle
the actual physical noise that will be present.  Given a certain noise model
and error correcting codes, we can derive  fault-tolerant {\em thresholds}:
The errors must be smaller than the threshold value  for concatenated error
correction to eliminate them all. Knill et al.\ (2000) considered a
combination of photon loss and dephasing in the original KLM proposal
and found that the accuracy threshold for the optical components in
that scheme was higher than 99\%. 

Dawson, Haselgrove, and Nielsen (2006a,b) \nocite{dawson05,dawson06}
performed an extensive numerical study of fault-tolerant thresholds
for linear optical cluster state quantum computing. The computational
model they adopted is Nielsen's micro-cluster approach, described in
section \ref{sec:nielsen}, with type-I fusion gates instead of
KLM-type CZ gates. The physical operations in this model are
Bell-state preparation, single-photon gates and memories, type-I fusion
gates, and photon measurements. The computation proceeds in time steps,
with exactly one operation at each step. Furthermore, it is assumed
that any two single-photon qubits in the computation can serve as the
inputs of the fusion gate. In other words, we have a {\em direct
interaction} between qubits. In addition, parallel operations are
allowed to speed up the computation and minimise the use of quantum
memories. Finally, the classical computation needed to control the
cluster state computing is taken to be sufficiently fast.

The noise model adopted by Dawson et al.\ consists of the inherent
probabilistic nature of the fusion gates, as well as photon loss and
depolarisation at every time step in the computation. The photon loss
is characterised by a uniform loss probability $\gamma$, and the
depolarisation comes in two flavours: Single-qubit operations have a
probability $\epsilon/3$ of undergoing a Pauli operation $X$, $Y$, or
$Z$. After the Bell-state preparation and before the fusion gate
input, the two photons undergo a correlated depolarising noise: With
probability $(1-\epsilon)$ nothing happens to the qubits, while with
probability $\epsilon/15$ any of the remaining 15 two-qubit Pauli
operators are applied. This is a completely general model for the
noise that can affect optical cluster state quantum computing, and the
resulting fault-tolerance simulation gives  an accuracy threshold
region on $\gamma$ and $\epsilon$. Thresholds were obtained for both a
seven-qubit CSS error correction code and a 23-qubit Golay error
correction code. The study shows that scalable quantum computing with
the 23-qubit code is possible for a maximum loss probability of
$\gamma < 3\cdot 10^{-3}$ and a maximum depolarising probability of
$\epsilon < 10^{-4}$.

Even though this noise model accounts for general noise, and the
fault-tolerant threshold puts a bound on its magnitude, it is clearly a
simplification of the physical noise that is expected in cluster state
LOQC. It is argued that the difference between correlated two-qubit
noise and independent noise does not change the threshold
much. Similarly, using one parameter to describe both photon
absorption and detector efficiency will not have a dramatic effect on
the threshold \cite{nielsen06}. The next milestone for establishing
fault-tolerance thresholds is to adopt a noise model in which the
parameters are measurable quantities, such as the visibility in a
Hong-Ou-Mandel experiment and photon loss probabilities.

When various parameters in a noise model differ significantly, it
might be beneficial to diversify the error correction codes. EC codes
that correct specific errors such as photon loss or depolarisation may
be smaller than generic EC codes, and therefore introduce less
noise. A round of special error correction might be used to reduce
large errors, and subsequent generic error correction will further
reduce the errors below the fault-tolerant threshold.  

In addition, certain types of errors or noise might be naturally
suppressed by a suitable alteration in the architecture. For example,
there is a way to create high-fidelity four-photon GHZ states with
lossy bucket detectors and inefficient sources
\cite{gilchrist05b}. Assume that the Bell-pair source creates a state
of the form $p_s |0\rangle\langle 0| +
(1-p_s)|\Psi^-\rangle\langle\Psi^-|$, where $|\Psi^-\rangle$ is the
two-photon polarization singlet state. This is a reasonable error
model when the source obeys selection rules that prevent single-photon
components to contribute to the output state (c.f., Benson et al.\
2000). In order to make a three-photon GHZ state using these sources
we use a type-I fusion gate and post-select on a single detector
click. The detector click indicates that at least one source created a
photon pair. However, if only one photon pair was created, the output
mode of the type-I fusion gate must necessarily empty. By taking the
output modes of two type-I fusion gates in two separate three-photon
GHZ creation attempts, and leading them into a type-II fusion gate, we
can post-select on finding two detector clicks. As a result,
high-fidelity four-photon GHZ states are produced.

Several other specialised circuits have been proposed that either detect
errors or correct them. For example, Ralph (2003) \nocite{ralph03}
proposed a simple demonstration circuit that detects and corrects bit
flip errors on a single qubit
using the encoded qubit state $\alpha|0\rangle |0\rangle + \beta
|1\rangle |1\rangle$ and an ancilla qubit $|0\rangle$. However, since
this is a probabilistic protocol, this circuit cannot naively be
inserted in a quantum computing circuit. If we assume the availability
of perfectly efficient detectors (not necessarily photon-number
resolving), deterministic polarisation-flip detection for distributing
entanglement can be achieved \cite{kalamidas04}. In a similar fashion,
a single-qubit error correction circuit can be constructed with
polarising beam splitters, half-wave plates, and Pockels cells
\cite{kalamidas05}. Here, we assume that these passive optical
elements do not induce additional noise. A full analysis would have to
take this noise into account.

\section{Outlook: beyond linear optics}\label{sec:outlook}

\noindent  
We have seen in this review that it is possible to construct a quantum
computer with linear optics, single-photon sources, and photon
detection alone. Knill, Laflamme, and Milburn (2001) overturned the
conventional wisdom that a lack of direct photon-photon interactions
prohibits scalability. Since KLM, several groups have proposed
modifications to building a linear optical quantum computer with
reduced resources and realistic (noisy) components.

The basic principles of LOQC have all been demonstrated
experimentally, predominantly using parametric down-conversion (PDC)
and bucket photon detection. Due to the small efficiency of PDC photon
sources, however, these techniques cannot be considered scalable in a
practical sense. Currently, there is a concerted effort to build the
necessary single-photon sources, photon detectors, and quantum
memories for a scalable linear optical quantum computer. On the
theoretical front, there is an ongoing effort to design more efficient
architectures and effective error correction codes tailored to the
noise model that is relevant to LOQC.

Nevertheless, constructing the necessary components and using
fault-tolerant encoding is hard, and several extensions to LOQC have
been proposed. In this last section we sketch a few additions to the
linear optical toolbox that can make quantum computing a little bit
easier. 

First, we have seen in section \ref{sec:kerr} that a cross-Kerr nonlinearity
can be used to induce a photon-photon interaction, and how two-qubit
quantum gates can be constructed using such a
nonlinearity. Unfortunately, natural Kerr nonlinearities are extremely
small, and this is not a practical method for creating optical
gates. However, recently it was suggested that a small nonlinearity
might still be used for quantum computing. It was shown by Munro et
al.\ (2005) how such nonlinearities can make a number-resolving QND
detector, and Barrett et al.\ \nocite{barrett05b} (2005) showed how a
small cross-Kerr nonlinearity can be used to perform complete Bell
measurements without destroying the photons. Subsequently, it was
realized by Nemoto and Munro (2004) \nocite{nemoto04} that this
technique can also be used to create a deterministic CNOT gate on
photonic qubits. Recent work in electromagnetically induced
transparencies by Lukin and Imamo{\v g}lu (2000, 2001)
\nocite{lukin00} \nocite{lukin01} suggests that the small-but-not-tiny
nonlinearities needed for this method are on the threshold of becoming
practical. Alternatively, relatively large nonlinearities can be
obtained in photonic band-gap materials \cite{friedler04}.  

Secondly, if we have high-fidelity single-photon sources
and memories, it might become beneficial to engineer these systems
such that they support coherent single-qubit operations. This way, we
can redefine our qubits as isolated static systems, and we have
circumvented the problem of qubit loss. When these matter qubits emit
a qubit-dependent photon, they can in turn be entangled using
techniques from linear optical quantum computing. It was shown by
Barrett and Kok (2005) \nocite{barrett05} that such an architecture
can support scalable quantum computing, even with current realistic
components. Independently, Lim et al.\ (2005) \nocite{lim05} showed
how a similar setup can be used to implement deterministic two-qubit
quantum gates. Recently, these two methods were combined in a fault
tolerant, near-deterministic quantum computer architecture
\cite{lim05b}.   

Thirdly, Franson, Jacobs, and Pittman proposed the implementation of a
two-qubit $\sqrt{\mbox{\sc swap}}$ gate using the quantum
Zeno effect: Two optical fibres are fused and split again, such that
the input modes overlap in a small section of the fibre. This acts as
a beam splitter on the modes in the input and output fibres. At
regular intervals inside the joint fibre we place atoms with a
two-photon transition. This transition acts as a two-photon
measurement, while single-photon wave-packets propagate through the
fibre undisturbed. Furthermore, the single-photon wave-packets
maintain coherence. The effect of this repeated two-photon measurement
is to suppress the Hong-Ou-Mandel effect via the quantum Zeno
effect. In this way, two single-photon qubits in the input modes are
transformed into two single-photon qubits in the output modes and
undergo a $\sqrt{\mbox{\sc swap}}$ gate.

Finally, an alternative approach to linear optical quantum computing
involves encoding qubits in squeezed or 
coherent states of light \cite{gottesman01,ralph03b}. Linear elements
take on a new capability in these implementations. For example Bell measurements
and fan-out gates become deterministic elements
\cite{vanenk02,jeong01}. The downside is that it is difficult to
produce the superposition states that are required as resources in
such schemes, although considerable theoretical and experimental
progress has been made recently  \cite{lund04,wenger04}. If this
problem is solved, considerable savings in resources could result from
adopting such implementations.

Whatever the ultimate architecture of quantum computers will be, there
will always remain a task for (linear) optical quantum information
processing: In order to distribute quantum information over a network
of quantum computers, the qubit of choice will most likely be
optical. We therefore believe that the techniques reviewed here are an
important step towards full-scale distributed quantum computing ---
the Quantum Internet.

\section*{Acknowledgements}

\noindent 
We would like to thank James Franson, Andrew White, Geoff Pryde, Philip
Walther, and their co-workers for providing us with the experimental
data of their respective CNOT gates, and Charles Santori for allowing
us to reproduce his figures. PK wishes to thank Sean Barrett and Dan
Browne for stimulating discussions, Radu Ionicioiu, Michael Raymer,
and Colin Williams for valuable comments, and Michael Nielsen for
extensive correspondence on fault-tolerance in LOQC. 
This work was supported by
ARO,
the Australian Centre for Quantum Computer Technology,
DTO,
the Hearne Institute,
JSPS,
LSU BOR-LINK,
MIC,
NSA,
the QIPIRC,
and the EU RAMBOQ Project.

\end{document}